\documentclass[pdftex,twocolumn,epjc3]{svjour3}          

\RequirePackage[T1]{fontenc}

\smartqed  

\RequirePackage{subfigure}

\RequirePackage{graphicx}
\RequirePackage{mathptmx}      
\RequirePackage{flushend}
\RequirePackage[numbers,sort&compress,merge]{natbib}
\RequirePackage[colorlinks,citecolor=blue,urlcolor=blue,linkcolor=blue]{hyperref}
\RequirePackage{wasysym}

\newcommand{\bea}{\begin{eqnarray}}
\newcommand{\eea}{\end{eqnarray}}
\newcommand{\be}{\begin{equation}}
\newcommand{\ee}{\end{equation}}
\newcommand{\mw}   {\mbox{$M_W$}}
\newcommand{\mwsq}   {\mbox{$M_W^2$}}

\newcommand{\mz}   {\mbox{$M_Z$}}
\newcommand{\mzsq}   {M_Z^2}

\newcommand{\seffl}{\sin^2\theta_{eff}^{l}\,}

\journalname{Eur. Phys. J. C}

\begin{document}

\title{On electroweak corrections to neutral current Drell--Yan
  with the POWHEG BOX}

\author{Mauro Chiesa\thanksref{e1}
\and
Clara Lavinia Del Pio\thanksref{e2}
\and
Fulvio Piccinini\thanksref{e3}
}
\author{Mauro Chiesa\thanksref{e1,addr1}
\and
Clara Lavinia Del Pio\thanksref{e2,addr2}
\and
Fulvio Piccinini\thanksref{e3,addr1}
}
\thankstext{e1}{e-mail: mauro.chiesa@pv.infn.it}
\thankstext{e2}{e-mail: claralavinia.delpio01@universitadipavia.it}
\thankstext{e3}{e-mail: fulvio.piccinini@pv.infn.it}

\institute{INFN, Sezione di Pavia,  Via A. Bassi 6, 27100 Pavia, Italy
\label{addr1}
\and
INFN, Sezione di Pavia and Dipartimento di Fisica, Universit\`a di Pavia, 
Via A. Bassi 6, 27100 Pavia, Italy\label{addr2}
}

\date{Received: date / Accepted: date}

\maketitle

\begin{abstract}
Motivated by the requirement of a refined and flexible treatment of
electroweak corrections to the neutral current Drell-Yan process, we
report on recent developments on various input parameter/renormalization 
schemes for the calculation of fully differential cross sections,
including both on-shell and $\overline{ \rm MS}$ schemes.
The latter are particularly interesting for direct determinations of running
couplings at the highest LHC energies. The calculations feature
next-to-leading order precision with additional higher order contributions
from universal corrections such as $\Delta \alpha$ and $\Delta \rho$. 
All the discussed input parameter/renormalization scheme options are
implemented in the package of {\tt POWHEG-BOX-V2} dedicated to the neutral
current Drell-Yan simulation, i.e. {\tt Z\_ew-BMNNPV}, which is used to
obtain the presented numerical results. In particular, a comprehensive
analysis on physical observables calculated with different input
parameter/renormali\-zation $\quad$ schemes is presented, addressing the 
$Z$ peak invariant mass region as well as the high energy window. 
We take the opportunity of reporting
also on additional improvements and options introduced in the package 
{\tt Z\_ew\--BMNNPV} after svn revision 3376, such as different
options for the treatment of the
hadronic contribution to the running of the electromagnetic coupling
and for the handling of the unstable $Z$ resonance. 

\keywords{Hadron colliders \and Drell-Yan \and QCD \and Electroweak corrections}
\PACS{12.15-y \and 12.15.Lk \and 12.38-t \and 12.38.Bx \and 13.85.-t}
\end{abstract}

\section{Introduction}
\label{section:introduction}
The neutral current Drell-Yan (NC DY) process plays a particular role in the
precision physics programme of the LHC. In fact, considering its large
cross section and clean experimental signature, together with the high
precision measurement of the $Z$-boson mass at LEP, this process is
a standard candle that can be used for different general purposes
such as detector calibration, Parton Distribution Functions (PDFs)
constraining and tuning of non-perturbative parameters
in the general purpose Monte Carlo event generators. 
Moreover, in the high tail of the transverse momentum and 
invariant mass distributions of the produced leptons,
the NC DY is one of the main irreducible backgrounds to the searches
for New Physics at the LHC. 
Recently an impressive precision at the sub-percent level has
been reached by the experimental analysis in large regions of
the dilepton phase space. 

In addition to the above general aspects,
it has to be stressed that the NC DY process 
also allows to perform precision tests of the Standard Model electroweak (SM EW) parameters
through the direct determination of the $W$-boson
mass~\cite{CDF:2013dpa,ATLAS:2017rzl,CDF:2022hxs,ATLAS:2023fsi}
and the weak mixing angle in hadronic collisions~\cite{CMS:2011utm,ATLAS:2015ihy,LHCb:2015jyu,CDF:2018cnj,CMS:2018ktx}.
While in the former case the NC DY observables enter only indirectly, 
the latter can be determined directly, without reference to the
charged current DY process. Another interesting possibility is the direct
determination of the running of the coupling $\sin^2\theta$,
defined in the $\overline{ \rm MS}$ scheme,
at the highest available LHC energies, in order to check its consistency
with low energy and $Z$-peak measurements within the SM framework~\cite{Amoroso:2023uux}. 

For all the above reasons, many efforts have been dedicated to 
the calculation of DY observables with the most advanced calculation 
techniques. On the perturbative side, the differential cross section,
fully exclusive in the leptonic kinematical variables,
is known with next-to-next-to-leading order (NNLO) accuracy in
QCD~\cite{Hamberg:1990np,Anastasiou:2003ds,Melnikov:2006kv,Melnikov:2006di,Catani:2009sm,Gavin:2010az,Gavin:2012sy}
and next-to-leading order (NLO) accuracy
in the electroweak sector (EW) of the Standard Model
(SM)~\cite{Baur:1997wa,Zykunov:2001mn,Baur:2001ze,Dittmaier:2001ay,Baur:2004ig,Arbuzov:2005dd,CarloniCalame:2006zq,Zykunov:2005tc,CarloniCalame:2007cd,Arbuzov:2007db,Brensing:2007qm,Dittmaier:2009cr,Boughezal:2013cwa}.
Threshold effects at next-to-next-to-next-to-leading order (N$^3$LO) in QCD
have been studied in Refs.~\cite{Ahmed:2014cla,Catani:2014uta}.
First calculations of inclusive N$^3$LO QCD corrections
have been calculated for charged current DY in Ref.~\cite{Duhr:2020sdp}
and for the production of a lepton pair mediated by photon
exchange in Ref.~\cite{Duhr:2020seh}. The N$^3$LO QCD corrections
to the inclusive NC DY process have been calculated
in Refs.~\cite{Duhr:2021vwj,Baglio:2022wzu} while N$^3$LO QCD corrections
to single-differential distributions appeared in
Refs.~\cite{Chen:2021vtu,Chen:2022lwc}
for neutral and charged current DY, respectively,
and the N$^3$LO QCD corrections to fiducial
cross sections have been calculated
in Refs.~\cite{Chen:2022cgv,Neumann:2022lft}. 
The resummation of large logarithms due to soft gluon emission at
small transverse momenta has been investigated in
Refs.~\cite{Neumann:2022lft,Guzzi:2013aja,Kulesza:2001jc,Catani:2015vma,Balazs:1997xd,Landry:2002ix,Cieri:2015rqa,Bozzi:2010xn,Mantry:2010mk,Becher:2011xn,Bizon:2018foh,Bizon:2019zgf,Re:2021con,Ju:2021lah,Camarda:2021ict,Camarda:2023dqn,Isaacson:2023iui}, while 
the combined effect of QED and QCD transverse momentum resummation for
$W$ and $Z$ production has been addressed in
Ref.~\cite{Cieri:2018sfk,Autieri:2023xme}. Relevant effects due to 
multiphoton emission have been studied in
Refs.~\cite{Placzek:2003zg,CarloniCalame:2003ux,CarloniCalame:2007cd,Brensing:2007qm,Dittmaier:2009cr}. 
The mixed NNLO contributions of
${\cal O}(\alpha_s \alpha)$ have been evaluated 
in pole approximation for neutral and charged current
DY~\cite{Dittmaier:2014qza,Dittmaier:2015rxo,Dittmaier:2024row}
while the complete corrections of 
${\cal O}(N_f \alpha_s \alpha)$ have been presented
in Ref.~\cite{Dittmaier:2020vra}. Several
results on ${\cal O}(\alpha_s \alpha)$ corrections to DY are
available in narrow width approximation: 
QCD$\times$QED \cite{deFlorian:2018wcj}, 
QCD$\times$EW~\cite{Bonciani:2019nuy,Bonciani:2020tvf,Bonciani:2021iis}
corrections to inclusive $Z$ production
and QCD$\times$EW corrections to fully exclusive 
$Z$~\cite{Delto:2019ewv,Buccioni:2020cfi,Cieri:2020ikq}
and $W$ production~\cite{Behring:2020cqi}. 
More recently important progress has been achieved
on the necessary ingredients for the calculation of
${\cal O}(\alpha_s \alpha)$ corrections to the off-shell
DY processes \cite{Bonciani:2016ypc,Heller:2019gkq,Hasan:2020vwn,Heller:2020owb}. 
The complete mixed ${\cal O}(\alpha_s \alpha)$ NNLO 
corrections to the NC DY process have been reported for the first time in
Refs.~\cite{Bonciani:2021zzf,Armadillo:2022bgm,Armadillo:2022ugh,Buccioni:2022kgy},
while, for the charged current DY process, they have been calculated
in approximate form, i.e. neglecting the exact NNLO virtual contribution, in
Ref.~\cite{Buonocore:2021rxx}.

Several fixed-order simulation tools have been developed, dedicated to 
NNLO QCD corrections like {\tt DYNNLO}~\cite{Catani:2009sm}, 
{\tt DY\-TURBO} \cite{Camarda:2019zyx}, {\tt FEWZ}~\cite{Melnikov:2006kv,Gavin:2010az,Gavin:2012sy},
{\tt MATRIX}~\cite{Grazzini:2017mhc} and {\tt MCFM}~\cite{Boughezal:2016wmq} 
and to NLO EW corrections, like
{\tt HORACE}~\cite{CarloniCalame:2006zq,CarloniCalame:2007cd} and
{\tt ZGRAD}/ {\tt ZGRAD2}~\cite{Baur:1997wa,Baur:2001ze}, 
while the {\tt SANC} framework~\cite{Arbuzov:2007db,Andonov:2008ga,Bardin:2012jk,Bondarenko:2013nu,Arbuzov:2015yja,Arbuzov:2016wfy,Arbuzov:2022sep} 
and the {\tt RADY} code~\cite{Dittmaier:2009cr} allow
to evaluate both QCD and EW NLO corrections. 
NNLO QCD corrections were combined with NLO EW contributions
in {\tt FEWZ}~\cite{Li:2012wna}. The state of the art of electroweak corrections
in the form of form factors (featuring (N)NLO accuracy (at)/off the
$Z$-resonance) is available in the libraries {\tt DIZET}~\cite{Arbuzov:2023afc}
and {\tt GRIFFIN}~\cite{Chen:2022dow}. 

A fundamental role in simulations for collider pheno\-menology is played by Monte Carlo event generators capable
to consistently match fixed order calculations to parton showers (PS)
simulating multiple soft/collinear radiations.
The {\tt MC@NLO}~\cite{Frixione:2002ik} and
{\tt POWHEG}~\cite{Nason:2004rx,Frixione:2007vw} algorithms have been developed for
the matching of NLO QCD computations to QCD PS and
implemented in the public softwares
{\tt MadGraph5\-\_aMC@NLO}~\cite{Alwall:2014hca}
and {\tt POWHEG-BOX}~\cite{Alioli:2010xd,Jezo:2015aia}.
Alternative formulations of the above algorithms
are used in event generators like
{\tt Sherpa}~\cite{Hoeche:2011fd,Sherpa:2019gpd} and
{\tt HERWIG}~\cite{Bellm:2015jjp,Platzer:2011bc}. More recently,
algorithms like
NNLOPS~\cite{Hamilton:2013fea} (based on reweighting of the
MiNLO$'$~\cite{Hamilton:2012rf} merging strategy),
U\-NNLOPS~\cite{Hoche:2014uhw,Hoche:2014dla},
{\tt Geneva}~\cite{Alioli:2013hqa},
and MiNNLO$_{PS}$~\cite{Monni:2019whf,Monni:2020nks}
have been proposed for the matching of NNLO QCD calculations to QCD PS.
Though several studies appeared on the approximated inclusion of EW corrections in event generators including higher-order
QCD corrections (see, for instance
Refs. \cite{Gutschow:2018tuk,Brauer:2020kfv,Lindert:2022qdd}),
event generators including both NLO QCD and NLO EW corrections consistently
matched to both QCD and QED parton showers are only available for a
limited number of processes, namely:
charged and neutral current
Drell-Yan~\cite{Barze:2012tt,Bernaciak:2012hj,Barze:2013fru,Muck:2016pko,Chiesa:2019nqb},
$HV+0/1$ jet~\cite{Chiesa:2019nqb}, diboson production~\cite{Chiesa:2020ttl}, and electroweak $H+2$ jets
production~\cite{Jager:2022acp}\footnote{In Ref.~\cite{Chiesa:2019ulk} an event generator for same-sign $WW$ scattering
  at the LHC at NLO EW accuracy matched to QED PS and supplemented with QCD PS was presented.}. QED correction exclusive exponentiation for DY processes
within YFS framework is realized within the event generator
{\tt KKMC-hh}~\cite{Jadach:2016zsp,Jadach:2017sqv,Yost:2019bmz,Yost:2020jin,Yost:2022kxg}. Very recently, the resummation of EW and mixed QCD-EW effects up to
next-to-leading logarithmic accuracy has been presented in
Ref.~\cite{Buonocore:2024xmy} for charged and neutral current DY.

The first implementation of QCD and EW NLO corrections and their interplay
in a unique simulation framework have been given
in Ref.~\cite{Barze:2012tt} for charged current DY~\footnote{An independent
  implementation of EW corrections in the {\tt POWHEG-BOX-V2} framework
was presented in Ref.~\cite{Bernaciak:2012hj}.} and in
Ref.~\cite{Barze:2013fru} for NC DY. 
An important improvement in the matching of QED radiation
in the presence of a resonance, following the ideas presented in 
Ref.~\cite{Jezo:2015aia}, has been discussed in
Ref.~\cite{CarloniCalame:2016ouw} for the
charged current DY process ({\tt W\_ew-BMNNP} svn revision 
3375) and extended to the NC one ({\tt Z\_ew-BMNNPV}
svn revision 3376)~\footnote{An independent realization of 
resonance-improved parton-shower matching for DY processes including 
EW corrections has been presented in Ref.~\cite{Muck:2016pko}.
}.
After the above mentioned release, additional improvements
and options have been introduced, in particular for the NC DY package
{\tt Z\_ew\--BMNNPV}, motivated by the need of a refined and flexible
treatment of EW corrections that allows a consistent internal estimate 
of the uncertainties affecting the theoretical predictions.
They can be schematically enumerated as follows:
\begin{itemize}
\item input parameter/renormalization schemes 
\item introduction of known higher order corrections
\item treatment of the hadronic contribution to the running of
  the electromagnetic coupling $\Delta \alpha_{\rm had}$
\item scheme for the treatment of the unstable resonance
\end{itemize}
In the following we give a detailed description of the various 
input parameter schemes and the related higher order corrections. 
A key ingredient of the latter is given by the running
of the electromagnetic coupling $\alpha$ between different scales.
In particular, when low scales are involved in the running,
the hadronic contribution is intrinsically non-perturbative and
different parametrizations have been developed in the literature
relying on low-energy experimental data, which we properly include
in our formulation of the electroweak corrections.

The formulae relevant for the various input parameters/ renormalization
schemes are presented in a complete and self-contained form, so 
that they can be implemented in any simulation tool. 

Though the {\tt Z\_ew-BMNNPV} package allows to simulate NC DY production
at NLO QCD$+$NLO EW accuracy with consistent matching to the QED and QCD parton showers
provided by {\tt PYTHIA8}~\cite{Sjostrand:2006za,Sjostrand:2014zea} and/or 
{\tt Photos}~\cite{Golonka:2005pn,Barberio:1990ms,Barberio:1993qi}, in the present paper we are 
mainly interested in various aspects of the fixed-order calculation
(NLO EW plus universal higher orders).
For this reason we show numerical results obtained at fixed-order and 
including only the weak corrections, since the QED contributions
are not affected by the choice of the input parameter scheme and
are a gauge-invariant subset of the EW corrections for the NC DY process.

The layout of the paper is the following: 
Section~\ref{sect:inputschemesintro} provides an introduction
to the input parameter schemes available in the code, while general
considerations on higher order universal corrections, common to all
the schemes, are presented in Section~\ref{sect:HO}. 
A detailed account of various input/renormalization schemes
at NLO accuracy and the related higher order corrections
is given in Section~\ref{sect:input-schemes-detail}, while
a numerical analysis of the features of the schemes,
with reference to cross section and forward-backward asymmetry,
as functions of the dilepton invariant mass, 
is presented in Section~\ref{sect:input-schemes-numerics},
together with a discussion on the main parametric
uncertainties in Section~\ref{sec:parametric}. The treatment of
the hadronic contribution to the running of $\alpha$ is discussed
in Section~\ref{sect:dahad}, while 
Section~\ref{sect:widths} is devoted to the description of the improvement
of the code with respect to the treatment of the unstable $Z$ resonance.
In Section~\ref{sec:high-energy} we analyse the effect of
the various input/renormalization schemes in the high energy regimes,
which will be accessible to the HL-LHC phase and future FCC-hh. 
A brief summary is given in Section~\ref{sec:conclusions}. 
The list of the default parameter values~\footnote{The same parameter values 
are also used to obtain the presented numerical results.} is contained
in~\ref{appendix:numerical-param},
while the list of the flags activating the available
options is given in~\ref{appendix:flags}.

\section{Input parameter schemes: general considerations}
\label{sect:inputschemesintro}
The input parameter schemes available in the {\tt Z\_ew-BMNNPV} package
of {\tt POWHEG-BOX-V2} can be divided in three categories:
the ones including both the $W$ and the $Z$
boson masses among the independent parameters,
the $(\alpha_0$, $G_{\mu}$, $M_Z)$ scheme, where we use the notation $\alpha_0$ for the QED coupling constant at $Q^2=0$, and the schemes
with $M_Z$ and the sine of the weak mixing angle as free parameters.
The latter class includes the schemes that use
$\sin \theta_{eff}^l$ as input parameter, where $\theta_{eff}^l$ is the effective
weak mixing angle, 
and an hybrid $\overline{ \rm MS}$ scheme where the independent
quantities are $\alpha_{\overline{\rm MS}}(\mu^2)$, $\sin^2 \theta_{\overline{\rm MS}}(\mu^2)$ and $M_Z$, with the couplings renormalized in the $\overline{\rm MS}$ scheme and $M_Z$ with the usual on-shell prescription.

The first class of input parameter schemes, namely $(\alpha_i$, $M_W$, $M_Z)$
with $\alpha_i=\alpha_0$, $\alpha(M_Z^2)$, $G_\mu$,
is widely used for the calculation of the EW corrections for processes
of interest at the LHC. On the one hand, the fact that
the $W$ boson mass is a free parameter is a useful feature, in particular, 
in view of the experimental determination of $M_W$
from charged current Drell-Yan production using template fit methods;
on the other hand, the predictions obtained 
in these schemes can suffer from relatively large parametric uncertainties
related to the current experimental precision
on $M_W$. This drawback is overcome, for instance,
in the $(\alpha_0$, $G_\mu$, $M_Z)$ scheme used for the
calculation of the EW corrections in the context of LEP physics,
where all the input parameters are experimentally known with high precision.
The third class of input parameter schemes uses the sine of the weak mixing
angle as a free parameter.
In the {\tt Z\_ew-BMNNPV} package the 
$(\alpha_i$, $\sin \theta_{eff}^l$, $M_Z)$ schemes, 
with $\alpha_i=\alpha_0$, $\alpha(M_Z^2)$, $G_\mu$, and the
$(\alpha_{\overline{\rm MS}} (\mu)$, $\sin^2 \theta_{\overline{\rm MS}}(\mu^2)$, $M_Z)$ one are implemented. 
The schemes where $\sin \theta_{eff}^l$
$(\sin^2 \theta_{\overline{\rm MS}}(\mu^2))$ is a free parameter are particularly useful
in the context of the experimental
determination of $\sin \theta_{eff}^l$
$(\sin^2 \theta_{{\overline{\rm MS}}}(\mu^2))$ from NC DY production
at the LHC using template fits~\cite{Chiesa:2019nqb,Amoroso:2023uux}.

The predictions for NC DY production obtained in the schemes
that use $\alpha(M_Z^2)$ or $G_\mu$ 
as inputs show a better convergence of the perturbative series
compared to the corresponding results
from the schemes with $\alpha_0$ as free parameter~\footnote{This is true 
if we consider only the LO contribution $q\overline{q}\to l^+l^-$.
If we include also the $\gamma\gamma\to l^+l^-$ LO process in the theoretical
predictions, the EW radiative corrections to the latter are minimized
with $\alpha_0$ as input parameter.}.
This is a consequence of the fact that, 
when using $\alpha(M_Z^2)$ or $G_\mu$ as independent variables,
large parts of the radiative corrections
related to the running of $\alpha(Q^2)$ from $Q^2=0$ to the electroweak scale
are reabsorbed in the LO predictions.
On the contrary, the EW corrections in the schemes with $\alpha_0$ as
input tend to be larger, because 
the running of $\alpha(Q^2)$ involves logarithmic corrections of
the form $\log (m^2/Q^2)$, where $m$
stands for the light-fermion masses (we refer to Sect.~\ref{sect:dahad}
for the treatment of the
light-quark contributions to the running of $\alpha(Q^2)$) and
$Q^2$ is the typical large mass scale of the process.

From a technical point of view, the calculation of the one-loop electroweak
corrections in the above-mentioned input parameter schemes differs in the
renormalization prescriptions used for the computation, while the bare part
of the Drell-Yan amplitudes remains formally the same and it is just
evaluated with different numerical values of the input parameters.
For each choice of input scheme, the renormalization is performed as follows:
first the electroweak parameters are expressed as a function of the three
selected independent quantities, then the counterterms corresponding to these parameters are fixed
by imposing some renormalization condition, and finally the counterterms for the derived electroweak parameters
are written in terms of the ones corresponding to the input parameters.

The fact that the counterterm part of the Drell-Yan amplitude differs
in the considered input parameter schemes implies that also the expression
of the universal fermionic corrections changes, as these
corrections at NLO can be related to the counterterm amplitude. In fact,
they can be computed at $\mathcal{O}(\alpha^2)$ 
taking the square of the fermionic universal contributions at NLO
(after subtracting the
$\mathcal{O}(\alpha)$ terms already included in the NLO calculation). 
We refer to Sect.~\ref{sect:HO} for details.

The input parameter schemes described in the following sections are formally equivalent for a given order
in perturbation theory; however, the numerical results obtained 
differ because of the truncation of the perturbative expansion. Although there is some arbitrariness in the choice of the input
parameter scheme to be used for the calculations, there can be phenomenological motivations to prefer one
scheme to the others, depending on the observables under consideration and on the role played by the
theory predictions with respect to the interpretation of the experimental measurements\footnote{
  We refer to~\cite{Brivio:2021yjb} for a discussion on the choice of input parameter schemes in the context
of the Standard Model effective field theory.}. For instance, in the context of cross section
or distribution measurements, where the theory predictions are used as
a benchmark for the experimental results but do not provide input for
parameter determination, those input parameter schemes should be
preferred that involve independent quantities known experimentally with high precision in order to
minimize the corresponding parametric uncertainties. One should also try to minimize the parametric uncertainties from quantities that enter
the calculation only at loop level (such as, for instance,
the top quark mass in DY processes). Another aspect that should be taken into account when choosing
an input parameter scheme is the convergence of the perturbative expansion in the predictions for
the observables of interest, which is mainly related to the possibility of reabsorbing large parts
of the radiative corrections in the definition of the coupling at LO. 

A different situation is the direct determination of electroweak parameters
using template fit methods, as done for example for the $W$ boson mass at Tevatron
and LHC. In this case, the theory predictions enter the
interpretation of the measurement
(with the Monte Carlo templates) and the theory uncertainties become part of the total systematic 
error on the quantity under consideration: it is thus important to use an input parameter scheme
where the quantity to be measured is a free parameter that can be varied independently not only
at LO, but also at higher orders in perturbation theory.

\section{Higher-order corrections}
\label{sect:HO}

At moderate energies, the leading corrections to NC DY production are related
to the logarithms of the light fermion masses and to terms proportional
to the top quark mass squared. 
These contributions can be traced back to the running of $\alpha(Q^2)$
(i.e. to $\Delta \alpha$) and to $\Delta \rho$
and are thus related to the counterterm amplitude for the process under
consideration. Following
Refs.~\cite{Consoli:1989fg,Denner:1991kt,Diener:2005me,Dittmaier:2009cr},
these effects can be taken into account at $\mathcal{O}(\alpha^2)$
by taking the square of the part of the counterterm
amplitude proportional to $\Delta \alpha$ and $\Delta \rho$.
They can thus be combined to the full NLO 
calculation after subtracting the part of linear terms in $\Delta \alpha$ and $\Delta \rho$
appearing in the square of the counterterm amplitude that are already present in the NLO computation.

The numerical results for the fermionic higher-order corrections presented in the following are
obtained using the one-loop expression for $\Delta \alpha$ (even though the two-loop leptonic
corrections are also available in the {\tt Z\_ew-BMNNPV} package and can be activated with the
flag {\tt dalpha\_lep\_2loop}), while for $\Delta \rho$ we include the leading Yukawa corrections
up to $\mathcal{O}(\alpha_S^2)$, $\mathcal{O}(\alpha_S x_t^2)$, and $\mathcal{O}(x_t^3)$, with
$x_t=\sqrt{2}G_\mu M_{\rm top}^2/16\pi^2$. More precisely, the expression used for $\Delta \rho$
is
\bea
\Delta \rho^{HO} & = 3 x_t (1 + x_t \Delta \rho^{(2)})(1+\frac{\alpha_S}{\pi}
\delta^{(2)}_{QCD}+\frac{\alpha_S^2}{\pi^2} \delta^{(3)}_{QCD}) \nonumber \\
&+x_t^3 \Delta \rho^{x_t^3}+\frac{\alpha_S}{\pi} x_t^2 \Delta \rho^{x_t^2 \alpha_S}
-3 x_t^2 \Delta \rho^{(2)} \frac{\alpha_S}{\pi}\delta^{(2)}_{QCD}, 
  \label{eq:drho}
\eea
where  $\Delta \rho^{(2)}$ is the two-loop heavy-top corrections
to the $\rho$ parameter~\cite{Veltman:1977kh,Fleischer:1993ub,Fleischer:1994cb}, $\delta^{(2)}_{QCD}$ and
$\delta^{(3)}_{QCD}$ are the two and three-loop QCD corrections~\cite{Djouadi:1987gn,Djouadi:1987di,Kniehl:1989yc,Avdeev:1994db},
while the three-loop contributions $\Delta \rho^{x_t^3}$ and $\Delta \rho^{x_t^2 \alpha_S}$ are taken
from Ref.~\cite{Faisst:2003px}. The last term in Eq.~(\ref{eq:drho}) is introduced in order to avoid
the double counting of the $\mathcal{O}(x_t^2 \alpha_S^2)$ contribution already present in factorized
approximation in the product of the QCD corrections and the Yukawa corrections at two loops.
The four-loop QCD corrections to the $\rho$ parameter~\cite{Chetyrkin:2006bj}
are not included. By inspection of the numerical impact of three-loop
QCD corrections (cf. Figs.~\ref{fig:mv_ho_deltarho3qcd}
and~\ref{fig:afb_ho_deltarho3qcd}), the phenomenological impact of
four-loop QCD corrections to the $\rho$ parameter is
expected to be negligible at the LHC. 
For the numerical studies presented in the following, the scale for the $\alpha_S$ factors entering
the QCD corrections to $\Delta \rho$ is set to the invariant mass of the dilepton pair.

\section{Input parameter schemes: detailed description}
\label{sect:input-schemes-detail}
In the following subsections we present a detailed account of the available
input parameter schemes at NLO weak accuracy and the related universal
higher order corrections (in what follows, the label NLO+HO refers to NLO plus higher-order accuracy). In the last subsection we present a comparison 
of the radiative corrections obtained with the different parameter schemes 
for two relevant differential observables (cross section and forward-backward
asymmetry as functions of the dilepton invariant mass $M_{ll}$) of the NC DY
process at the LHC with $\sqrt{s}=13$~TeV, considering $\mu^+\mu^-$ final states. In the following, for the sake of simplicity of notation,
whenever the complex mass scheme (CMS in the following) is used for the
treatment of the unstable gauge bosons, the symbol $M^2_V$, with $V=W, Z$,
represents the quantity $\mu^2_V = M^2_V - i \Gamma_V M_V$.

\subsection{\it The $(\alpha_0/\alpha(M_Z^2)/G_\mu$, $M_W$, $M_Z)$ schemes}
\label{sect:amwmz}

In these schemes, the input parameters are the $W$ and $Z$ boson masses and
$\alpha_i=\alpha_0$, $\alpha(M_Z^2)$, $G_\mu$. The counterterms for the independent parameters are defined as
\bea
M_{Z,b}^2 &=& \mzsq + \delta\mzsq
\label{eq:amwmzctmz}\\
M_{W,b}^2 &=& \mwsq + \delta\mwsq
\label{eq:amwmzctsin}\\
e_b &=& e (1 + \delta Z_e) \, ,
\label{eq:amwmzcte}
\eea
where the subscript $b$ denotes the bare parameter. 
The expression for $\delta Z_e$ is fixed by imposing that the NLO EW
corrections to the $\gamma e^+ e^-$ vertex 
vanish in the Thomson limit, while $\delta \mwsq$ and $\delta \mzsq$ are
obtained by requiring that the gauge-boson
masses do not receive radiative corrections.

The analytic expression of the counterterms can be found in
Ref.~\cite{Denner:1991kt} (and in
Refs.~\cite{Denner:1999gp,Denner:2005fg,Denner:2006ic} 
if the complex-mass scheme is used). In the following, 
for the self energies and the counterterms we will use the notation
of Ref.~\cite{Denner:1991kt}.

In the schemes with $M_W$ and $M_Z$ as independent parameters, the sine of the weak-mixing angle is a derived
quantity defined as
\begin{equation}
  s_W^2=1-\frac{M_W^2}{M_Z^2},
  \label{eq:swos}
\end{equation}
and the corresponding counterterm reads:
\begin{equation}
  \frac{\delta s_W}{s_W}=-\frac{1}{2}\frac{c_W^2}{s_W^2}
  \Big( \frac{\delta M_W^2}{M_W^2}- \frac{\delta M_Z^2}{M_Z^2}\Big).
  \label{eq:swctos}
\end{equation}
When $\alpha(M_Z^2)$ or $G_\mu$ are used as input parameters,
the calculation of the $\mathcal{O}(\alpha)$ corrections is
formally the same one as in the $(\alpha_0$, $M_W$, $M_Z)$
scheme but with the replacements
$\delta Z_e \to \delta Z_e -\Delta \alpha(M_Z^2)/2$ and
$\delta Z_e \to \delta Z_e -\Delta r/2$, respectively, that take into
account the running of $\alpha(Q^2)$ from $Q^2=0$ to the weak scale
which is absorbed in the LO coupling ($\alpha(M_Z^2)$ or $G_\mu$).
It is worth noticing that these replacements
remove the logarithmically enhanced fermionic corrections coming from
$\Delta \alpha$.
The factor $\Delta r$ represents the full one-loop electroweak corrections to the muon decay
in the scheme $(\alpha_0$, $M_W$, $M_Z)$  after the subtraction of the QED effects in the Fermi theory and reads:
\be
\Delta r =  \Delta \alpha - \frac{c_W^2}{s_W^2} \Delta \rho^{self} + \Delta r^{self}_{rem}=\Delta \alpha - \frac{c_W^2}{s_W^2} \Delta \rho + \Delta r_{rem},
\label{eq:dr}
\ee
with
\begin{eqnarray}
  \Delta \rho^{self} &=&\frac{\Sigma_T^{ZZ}(0)}{M_Z^2}-\frac{\Sigma_T^{WW}(0)}{M_W^2}=\frac{\alpha}{4\pi} \frac{3}{4s_W^2}\frac{M_{\rm top}^2}{M_W^2}+\cdots \nonumber \\
  & =& \Delta \rho^{1-loop}|_{\alpha} +{\rm non-enhanced \; terms},
  \label{eq:drh0-prec}
\end{eqnarray}
(clearly $\Delta \rho^{1-loop}|_{\alpha(G_\mu)}=3x_t$) and 
\begin{eqnarray}
 & & \Delta r_{rem}^{self} =
  \frac{{\rm Re}\Sigma^{AA}_T(s)}{s} 
   - \frac{c_W^2}{s_W^2}\left( \frac{\delta M_Z^2}{M_Z^2}
   - \frac{\Sigma_T^{ZZ}(0)}{M_Z^2} \right) \nonumber\\
   &+& \frac{c_W^2 - s_W^2}{s_W^2}\left( \frac{\delta M_W^2 }{M_W^2} - \frac{\Sigma_T^{WW}(0)}{M_W^2} \right)
   +2 \frac{c_W}{s_W}\frac{\Sigma_T^{AZ}(0)}{M_Z^2}    \nonumber\\
   &+& \frac{\alpha}{4\pi s_W^2} \Big( 6+ \frac{7-4s_W^2}{2 s_W^2} {\rm log} ( c_W^2 ) \Big) .
   \label{eq:dr-remn}
\end{eqnarray}

From the expression of the counterterms, one can notice that the leading fermionic corrections at NLO EW
are related to
\begin{equation}
\delta Z_e \sim \frac{\Delta \alpha}{2} , ~~~ \frac{\delta s_W}{s_W} \sim \frac{1}{2} \frac{c_W^2}{s_W^2} \Delta \rho, 
~~~\Delta r \sim \Delta \alpha -\frac{c_W^2}{s_W^2} \Delta \rho. 
\label{eq:amwmzHO}
\end{equation}
In the {\tt Z\_ew-BMNNPV} package, we implemented a slightly modified version of Eqs.~(3.45)-(3.49) of Ref.~\cite{Dittmaier:2009cr}
for the computation of the leading fermionic corrections to neutral-current Drell-Yan up to $\mathcal{O}(\alpha^2)$. More precisely,
those equations are modified in such a way to be valid also in the complex-mass scheme. 
As discussed in Sect.~\ref{sect:HO}, in order to combine these higher-order fermionic
corrections with the NLO EW results, it is mandatory to subtract those effects that are included in the full one-loop
calculation to avoid double counting. In particular, this implies the replacement $\Delta \rho \to (\Delta \rho -\Delta \rho^{\rm 1-loop})$
in the linear terms of the fermionic corrections up to $\mathcal{O}(\alpha^2)$: 
if we use, optionally by means of the flag 
{\tt a2a0-for-QED-only}~\footnote{This is the setting used for the numerical results 
  obtained in the present study.}, for the overall weak-loop factors the same value of $\alpha_i$
used in the LO couplings, $\Delta \rho^{\rm 1-loop}$ is computed as a function of
$\alpha_0$, $\alpha(M_Z^2)$, or $G_\mu$ for the $(\alpha_0$, $M_W$, $M_Z)$, $(\alpha(M_Z^2)$, $M_W$, $M_Z)$, and
$(G_\mu$, $M_W$, $M_Z)$ schemes, respectively.
If instead we use $\alpha_0$ for the overall weak-loop factors,
we subtract the quantity $\Delta \rho^{\rm 1-loop}|^{\alpha_0}$ computed in the
$\alpha_0$ scheme, regardless of the value of $\alpha_i$ used as independent parameter.

\subsection{\it The $(\alpha_0/\alpha(M_Z^2)/G_\mu$, $\seffl$, $M_Z)$
  schemes}
\label{sect:amzsw}

In the $(\alpha_i$, $\seffl$, $M_Z)$ schemes (where $\alpha_i=\alpha_0$, $\alpha(M_Z^2)$, $G_\mu$)
the sine of the effective weak mixing angle is used as input
parameter~\footnote{The bulk of the results presented in this subsection
    is contained in Ref.~\cite{Chiesa:2019nqb}.
    We report them here for the sake of completeness.}.
This quantity is defined
from the ratio of the vectorial and axial-vectorial couplings of the $Z$ boson to the leptons
$g_V^l$ and $g_A^l$, or, equivalently, in terms of the chiral $Zll$ couplings
$g_L^l$ and $g_R^l$, measured at the $Z$ resonance and reads
\be
\seffl
~\equiv~\frac{I_3^{l}}{Q_l}\,{\rm Re}\left(\frac{-g_R^l(\mzsq)}{g_L^l(\mzsq)-g_R^l(\mzsq)} \right)\, ,
\label{eq:def-sinthetaeff0}
\ee
where ${I_3^{l}}$ is the third component of the weak isospin for left-handed leptons\footnote{In the keys of the plots of Sec.~\ref{sect:input-schemes-numerics}, the short-hand notation $s^2_{eff}$ is used in place of $\seffl$.}.
Since $\seffl$ is used as an independent parameter, this scheme is particularly useful in the context
of the direct extraction of $\seffl$ from NC DY at the LHC using template fit
methods at NLO EW accuracy.

The counterterms corresponding to the input parameters are defined as
\bea
M_{Z,b}^2 &=& \mzsq + \delta\mzsq
\label{eq:ctmz}\\
\sin^2\theta_b&=&\seffl + \delta\seffl
\label{eq:ctsin}\\
e_b &=& e (1 + \delta Z_e) \, .
\label{eq:cte}
\eea
The expressions of $\delta Z_e$ and $\delta\mzsq$ are determined as in Sect.~\ref{sect:amwmz}, while the
expression of $\delta\seffl$ is fixed by requiring that the definition in Eq.~(\ref{eq:def-sinthetaeff0})
holds to all orders in perturbation theory. More precisely, we write Eq.~(\ref{eq:def-sinthetaeff0})
at one loop as
\bea
& & \seffl
~\equiv~\frac{I_3^{l}}{Q_l}\,{\rm Re}\left(\frac{-{\cal G}_R^l(\mzsq)}{{\cal G}_L^l(\mzsq)-{\cal G}_R^l(\mzsq)} \right)\, \nonumber \\
&=&\frac{1}{2} {\rm Re} \frac{-g_R^l}{g_L^l-g_R^l}+\frac{1}{2} {\rm Re} \frac{g_L^l g_R^l}{(g_L^l - g_R^l)^2} \Big( \frac{\delta g_L^l}{g_L^l}-\frac{\delta g_R^l}{g_R^l}\Big).
\label{eq:def-sinthetaeff1}
\eea
where $\mathcal{G}_{L(R)}^l(M_Z^2)=g_{L(R)}^l+\delta g_{L(R)}^l(M_Z^2)$ represent the $Zl^Ll^L$ and  $Zl^Rl^R$ form
factors computed at one loop accuracy at the scale $M_Z^2$ and we impose the condition:
\be
   {\rm Re}\left(\frac{-{\cal G}_R^l(\mzsq)}{{\cal G}_L^l(\mzsq)-{\cal G}_R^l(\mzsq)} \right)=
   {\rm Re}\left(\frac{-g_R^l}{g_L^l-g_R^l} \right),
\label{eq:def-sinthetaeff2}
\ee
which implies
\be
   {\rm Re} \left(
   \frac{\delta g_L^l}{g_L^l}-
   \frac{\delta g_R^l}{g_R^l}
   \right)\, ~=~ 0.
\label{eq:deltasintheta_0}
\ee
The $\delta g_{L(R)}^l$ factors contain both bare vertices and counterterms and since they are functions of $\delta\seffl$,
Eq.~(\ref{eq:deltasintheta_0}) can be used to compute the counterterm corresponding to the effective weak mixing
angle. By inserting the expressions for $\delta g_{L(R)}^l$ one gets
\bea
&&\frac{\delta \seffl}{\seffl}
= {\rm Re} \Big\{-\frac{1}{2} \frac{\cos \theta_{eff}^l}{\sin \theta_{eff}^l} \delta Z_{AZ} \label{eq:deltasintheta_1} \\
                 && + 
                  \left( 1 - \frac{Q_l}{I_3^{l}} \seffl  \right)
                  \left[ \delta Z^l_L + \delta V^L
                    - \delta Z^l_R - \delta V^R\right] \Big\}, \nonumber
\eea
where $\delta Z^l_{L(R)}$ are the pure weak parts of the wave function renormalization counterterms
for the leptons and $\delta V^{L(R)}$ are the one-loop weak corrections to the left/right $Zll$ vertices defined as
\bea
\delta V^L &=& \left( g^l_L \right)^2 \frac{\alpha}{4 \pi} {\cal V}_a\left( 0, M_Z^2, 0, M_Z, 0, 0\right) \nonumber \\
                 &+& \frac{1}{2 s_W^2} \frac{g^{\nu}_L}{g^{l}_L} \frac{\alpha}{4 \pi} {\cal V}_a\left( 0, M_Z^2, 0, M_W, 0, 0\right) \nonumber \\
     &-& \frac{c_W}{s_W}\frac{1}{2 s_W^2} \frac{1}{g^{l}_L} \frac{\alpha}{4 \pi} {\cal V}_b^-\left( 0, M_Z^2, 0, 0, M_W, M_W\right) \nonumber \\
\delta V^R &=& \left( g^l_R \right)^2 \frac{\alpha}{4 \pi} {\cal V}_a\left( 0, M_Z^2, 0, M_Z, 0, 0\right) 
\label{eq:vertex-v2}
\eea
and the vertex functions ${\cal V}_a$ and ${\cal V}_b^-$ are given in Eqs.~(C.1) and (C.2)
of Ref.~\cite{Denner:1991kt}, respectively.
No QED correction is included in Eq.~(\ref{eq:deltasintheta_1}), since the QED contributions to the $Zll$
vertex are the same for left or right-handed fermions and cancel in Eq.~(\ref{eq:deltasintheta_0}).
When the complex mass scheme is used, the input value for $\seffl$ remains real: this implies
that $g_{R/L}^{\rm LO}$ remain real and the condition~(\ref{eq:def-sinthetaeff2}) still reduces to~(\ref{eq:deltasintheta_0}). 
As a consequence, the definition in Eq.~(\ref{eq:deltasintheta_1}) remains valid in the complex-mass scheme, provided
that the CMS expressions for $\delta Z^l_{L(R)}$  and $\delta Z_{AZ}$ are used.
Note that the vertex functions ${\cal V}_a$ and ${\cal V}_b$ are computed for a real scale $M_Z^2$,
while the gauge boson masses appearing in the loop diagrams are promoted to complex.
If one instead uses a complex-valued $s_W^2$ (see the discussion on fermionic higher-order effects in Sect.~\ref{sect:agmumz}),
the condition in~(\ref{eq:deltasintheta_0}) can still be used but without taking the real part.

As already discussed in Sect.~\ref{sect:amwmz}, the counterterms in the $(\alpha(M_Z^2)$, $\seffl$, $M_Z)$
and $(G_\mu$, $\seffl$, $M_Z)$ schemes can be obtained from the ones in the $(\alpha_0$, $\seffl$, $M_Z)$
scheme performing the replacements $\delta Z_e \to \delta Z_e -\Delta \alpha(M_Z^2)/2$ and
$\delta Z_e \to \delta Z_e -\Delta \tilde{r}/2$, respectively, where $\Delta \tilde{r}$ represents
the one-loop electroweak corrections to the muon decay (after subtracting the QED effects in the Fermi
theory) in the scheme $(\alpha_0$, $\seffl$, $M_Z)$ and reads
\bea
& & \Delta \tilde r = 2 \delta Z_e -\Big( 1-\frac{s_W^2}{c_W^2} \Big) \frac{\delta \seffl}{s^2_W} -\frac{\delta M_Z^2}{M_Z^2}
+ \frac{\Sigma^{WW}_T (0)}{M_W^2}   \nonumber \\
&+& \frac{2}{c_W s_W}  \frac{\Sigma^{AZ}_T(0)}{M_Z^2} + \frac{\alpha}{4\pi s_W^2}\Big( 6 +\frac{7-4s_W^2}{2 s_W^2} \log{c_W^2} \Big)
\label{eq:drtildelong}
\eea
that can be written also as
\be
\Delta\tilde r =  \Delta \alpha - \Delta \rho^{self} + \Delta\tilde r^{self}_{rem}=  \Delta \alpha - \Delta \rho + \Delta\tilde r_{rem},
\label{eq:drtilde}
\ee
with
\begin{eqnarray}
\Delta\tilde r^{self}_{rem} &=&
  \frac{{\rm Re}\Sigma^{AA}_T(s)}{s} 
   - \left( \frac{\delta M_Z^2}{M_Z^2}
   - \frac{\Sigma_T^{ZZ}(0)}{M_Z^2} \right) \nonumber\\
   &+& \frac{s_W^2 - c_W^2}{c_W^2}\frac{\delta \seffl}{s_W^2}  +2 \frac{c_W}{s_W}\frac{\Sigma_T^{AZ}(0)}{M_Z^2}    \nonumber\\
   &+& \frac{\alpha}{4\pi s_W^2} \Big( 6+ \frac{7-4s_W^2}{2 s_W^2} {\rm log} ( c_W^2 ) \Big) ,
   \label{eq:drtilde-remn}
\end{eqnarray}
where we used the short-hand notation $s_W=\sin\theta_{eff}^{l}$ and $c_W=\cos\theta_{eff}^{l}$.

From Eqs.~(\ref{eq:deltasintheta_1})-(\ref{eq:drtilde}) it is clear that the leading fermionic corrections in the
schemes with $\seffl$ as input parameter are only related to $\delta Z_e \sim \frac{\Delta \alpha}{2}$ and
$\Delta \tilde{r} \sim \Delta \alpha -\Delta \rho$, while the counterterm of $\seffl$ does not contain terms
proportional to the logarithms of the light-fermion masses or to the square of the top quark mass.
As a result, the fermionic higher-order corrections in these schemes (after the subtraction of the
effects already included in the $\mathcal{O}(\alpha)$ calculation) are just overall factors that multiply
the LO matrix element squared and read:
\begin{eqnarray}
  |M_{\rm LO}|^2 \Big( \frac{1}{(1-\Delta \alpha(M_Z^2))^2} -1- 2 \Delta \alpha(M_Z^2) \Big) ,
   \label{eq:amzsw-HO-a0}
\end{eqnarray}
and
\begin{eqnarray}
  |M_{\rm LO}|^2\Big( \frac{1}{(1-\Delta \rho)^2} -1- 2 \Delta \rho^{(1)} \Big)  ,
   \label{eq:amzsw-HO-gmu}
\end{eqnarray}
for the schemes with $\alpha_0$ and $G_\mu$ as input parameters, respectively, while these
corrections are zero when $\alpha(M_Z^2)$ is used as independent parameter.
In equation~(\ref{eq:amzsw-HO-a0}), a resummation of the logarithms of the light-fermion masses was performed,
while the overall factor in~(\ref{eq:amzsw-HO-gmu}) comes from the relation between $\alpha$ and $G_\mu$ at NLO plus higher orders,
namely $\alpha=G_\mu s_W^2 c_W^2 M_Z^2 \sqrt{2} (1+\Delta \tilde r - \Delta \alpha)/\pi$\footnote{In the LO matrix element,
the value of $\alpha$ used is derived from $G_\mu$ at LO accuracy: $\alpha=G_\mu s_W^2 c_W^2 M_Z^2\sqrt{2}/\pi$.}.

\subsection{\it The $(\alpha_0$, $G_\mu$, $M_Z)$ scheme}
\label{sect:agmumz}

In the $(\alpha_0$, $G_\mu$, $M_Z)$ scheme, the input parameters are $\alpha_0$, $G_\mu$, and the mass of the $Z$ boson.
The main advantage of using this scheme is that all the independent parameters are experimentally
known with high precision and the corresponding parametric uncertainties are small (in particular, compared to
the schemes in Sect.~\ref{sect:amwmz}, it is independent of the uncertainties
related to the experimental knowledge of $M_W$).

In the scheme under consideration, the sine of the weak mixing angle and the $W$ boson mass 
are derived quantities. At the lowest order in perturbation theory they can be computed using the relations
\begin{eqnarray}
  & (1-s_W^2)s_W^2=\frac{\pi \alpha}{\sqrt{2} G_\mu M_Z^2}, \\
  & c_W^2=1-s_W^2, \qquad c_W^2=\frac{M_W^2}{M_Z^2}, \nonumber \\
  & s_W^2=\frac{1}{2}-\sqrt{\frac{1}{4}-\frac{\pi \alpha}{\sqrt{2} G_\mu M_Z^2}}, \quad M_W^2=\frac{\pi \alpha}{\sqrt{2} G_\mu s_W^2}.
  \label{eq:agfmzLOcpl}
\end{eqnarray}
In terms of Eqs.~(\ref{eq:agfmzLOcpl}), it is possible to write the LO amplitude for NC DY as the
sum of the photon exchange amplitude proportional to $\alpha$ and the
$Z$ exchange amplitude proportional to $G_\mu M_Z^2$, namely:
\begin{equation}
  M^{\sigma,\tau}=A^{\sigma,\tau}\Big[-\frac{4\pi \alpha}{s}Q_qQ_l-\frac{4\sqrt{2}G_\mu M_Z^2}{s} G_q^{\sigma}G_l^{\tau} \chi_Z(s)\Big],
  \label{eq:LOagmz}
\end{equation}
where $A^{\sigma,\tau}$ is the part of the amplitude containing the
$\gamma$ matrices and the external fermions spinors ($\sigma,\tau=L,R$) and  
\begin{equation}
  G_{q(l)}^{\sigma}=I_3^{\sigma  \, q(l)}-s_W^2 Q_{q(l)},~~ \chi_Z(s)=\frac{s}{s-M_Z^2+i\Gamma_Z M_Z},
  \label{eq:LOagmzparts}
\end{equation}
$Q_{q(l)}$ and  $I_3^{\sigma \, q(l)}$ being the quark (lepton) charges and third components of the weak isospin.
In the complex-mass scheme, the definition of $\chi_Z$ is $s/(s-M_Z^2)$. 
Clearly the $Z$-boson exchange diagram contains a residual dependence on $\alpha$ from $s_W$ in Eq.~(\ref{eq:LOagmzparts}).
Two different realizations of the $(\alpha_0$, $G_\mu$, $M_Z)$ scheme are available in the {\tt Z\_ew-BMNNPV}
package: users can select a specific one through the {\tt azinscheme4} flag in the {\tt powehg.input} file. 
If the flag is absent or negative, in Eqs.~(\ref{eq:agfmzLOcpl}) $\alpha=\alpha_0$ 
in such a way that the $\gamma f \bar{f}$ interaction
is evaluated at low scale while the $Z f \bar{f}$
couplings are computed at the weak scale. If {\tt azinscheme4} is positive, $\alpha=\alpha_0/(1-\Delta \alpha(M_Z^2))$:
this way also the photon part of the amplitude is evaluated at the weak scale. Note that we compute $\Delta \alpha(M_Z^2)$
form $\alpha_0$ rather that taking $\alpha(M_Z^2)$ as an independent parameter. For dilepton invariant masses
in the resonance region or larger, the latter running mode allows to reabsorb in the couplings the 
mass logarithms originating from the running of $\alpha$ from $q^2=0$ to the weak scale.
If not otherwise stated, the numerical results presented for the $(\alpha_0$, $G_\mu$, $M_Z)$ scheme are
obtained with {\tt azinscheme4}$=1$.

The counterterms for the independent quantities are defined as:
\bea
M_{Z,b}^2 &=& \mzsq + \delta\mzsq
\label{eq:agmzCTmz}\\
G_{\mu,b}&=&G_\mu + \delta G_\mu
\label{eq:agmzCT0g}\\
e_b &=& e (1 + \delta Z_e).
\label{eq:agmzCTe}
\eea
The expression of the $\delta Z_e$ and $\delta\mzsq$ counterterms is fixed as in Sect.~\ref{sect:amwmz} (if {\tt azinscheme4}$=1$, there
is the additional shift $\delta Z_e \to \delta Z_e -\Delta \alpha(M_Z^2)/2$), while
$\delta G_\mu$ is determined by requiring that the muon decay computed in the $(\alpha_0$, $G_\mu$, $M_Z)$ scheme
does not receive any correction at NLO (after the subtraction of the QED effects in the Fermi theory), namely:
\bea
  \frac{\delta G_\mu}{G_\mu} &=&  -\frac{2}{s_Wc_W}\frac{\Sigma^{AZ}_T(0)}{M_Z^2} -\frac{\Sigma^{WW}_T(0)}{M_W^2}\nonumber\\
  &&-\frac{\alpha}{4\pi s_W^2}[6 +\frac{7-4s_W^2}{2s_W^2}{\rm log} c_W^2],
\label{eq:agmzCT1g}
\eea
where $\alpha$ in the last term correspond to the loop factor governed by the flag {\tt a2a0-for-QED-only}. 
The counterterms for the dependent quantities read
\begin{eqnarray}
     \frac{\delta M_W^2}{M_W^2} & = & \frac{s_W^2}{c_W^2-s_W^2}\Big( - 2 \delta \tilde Z_e
  + \frac{\delta G_\mu}{G_\mu} + \frac{c_W^2}{s_W^2}\frac{\delta M_Z^2}{M_Z^2} \Big),  \\  
  \label{eq:agfmzCTmw}
   \frac{\delta s_W}{s_W} & = & \frac{1}{2} \frac{c_W^2}{c_W^2-s_W^2} \Big( 2 \delta \tilde  Z_e
   - \frac{\delta G_\mu}{G_\mu} -\frac{\delta M_Z^2}{M_Z^2} \Big),\\
   \label{eq:agfmzCTsw}
   \delta \tilde Z_e & = & \delta Z_e -\frac{1}{2} \delta_{{\rm opt}\, ,1} \Delta \alpha(M_Z^2)
    \label{eq:agfmzCTze}
\end{eqnarray}
where $\delta_{{\rm opt}\, ,1}$ is one if the {\tt azinscheme4} flag is active and zero otherwise.

By looking at the expressions of the counterterms in equations~(\ref{eq:agmzCT1g})--(\ref{eq:agfmzCTze}),
it is clear that at NLO the leading fermionic corrections to the photon exchange amplitude are related to $\delta \tilde Z_e$,
while for the Z exchange amplitude they come from the counterterms of the overall factor $G_\mu M_Z^2$ and from $\delta s_W/s_W$,
with
\bea
& & \delta \tilde Z_e \sim (1-\delta_{{\rm opt}\, ,1}) \frac{\Delta \alpha}{2} , \qquad \frac{\delta G_\mu}{G_\mu} + \frac{\delta M_Z^2}{M_Z^2} \sim \Delta \rho, \nonumber\\
& & \delta s_W^2 \sim \frac{s_W^2 c_W^2}{c_W^2-s_W^2} \Big( (1-\delta_{{\rm opt}\, ,1})\Delta \alpha - \Delta \rho \Big).
\label{eq:agmzHO}
\eea
In order to include these effects beyond $\mathcal{O}(\alpha)$, we follow the strategy described, for instance, in~\cite{Bardin:1999ak}:
the fermionic higher-order corrections are written as a Born-improved amplitude written in terms of effective couplings $\alpha=\alpha_0/(1-\Delta \alpha(M_Z^2))$
and $\overline{\sin}^2 \theta^l_{eff}$ (computed as a function of $\alpha_0$, $M_Z$, $G_\mu$) after subtracting those parts of the corrections already present in the NLO result. The sine of effective leptonic weak-mixing angle in the $(\alpha_0$, $G_\mu$, $M_Z)$
can be computed at NLO using Eq.~(\ref{eq:def-sinthetaeff1}): after noticing that the second term in the last line
goes like $\delta s_W^2$ and it would be zero if the $s_W$ counterterm was $\delta \seffl$ (i.e. it had the expression
derived according to Eq.~(\ref{eq:def-sinthetaeff2}) but with a numerical value of $s_W$ fixed by Eq.~(\ref{eq:agfmzLOcpl})),
by adding and subtracting $\delta \seffl$ Eq.~(\ref{eq:def-sinthetaeff1}) boils down to
\begin{eqnarray}
  & &\overline{\sin}^2 \theta^l_{eff} = s_W^2 \Big( 1+\frac{\delta s_W^2}{s_W^2}-\frac{\delta \seffl}{s_W^2} \Big)
  =s_W^2 +\frac{s_W^2 c_W^2}{c_W^2-s_W^2}\Delta \tilde r  \nonumber \\
  & & = \frac{1}{2}-\sqrt{\frac{1}{4}-\frac{\pi \alpha}{\sqrt{2} G_\mu M_Z^2}} + \frac{1}{2}\frac{\frac{\pi \alpha}{\sqrt{2} G_\mu M_Z^2}}{\sqrt{\frac{1}{4}-\frac{\pi \alpha}{\sqrt{2} G_\mu M_Z^2}}} \Delta \tilde r,
  \label{eq:seffsc4NLO}
\end{eqnarray}
where in the second equality the explicit expression of the counterterms
$\delta s_W^2$~(\ref{eq:agfmzCTsw}) and $\delta G_\mu$~(\ref{eq:agmzCT1g}) was used
and compared to the explicit expression of $\Delta \tilde r$ (if the flag {\tt azinscheme4}
is on, $\Delta \tilde r$ must be computed in terms of $\delta \tilde Z_e$ rather than $\delta Z_e$).
Equation~(\ref{eq:seffsc4NLO}) is the NLO expansion of
\begin{equation}
  \overline{\sin}^{2, \, {\rm HO}}  \theta^l_{eff} =\frac{1}{2}-\sqrt{\frac{1}{4}-\frac{\pi \alpha}{\sqrt{2} G_\mu M_Z^2} \Big(1+\Delta \tilde r_{\rm HO} \Big)},
  \label{eq:seffsc4HO}
\end{equation}
where $\Delta \tilde r_{\rm HO}$ is obtained from $\Delta \tilde r$ by adding
to the $\Delta \rho$ term in ~(\ref{eq:drtilde}) the higher-order corrections 
in Eq.~(\ref{eq:drho}). Note that $\Delta \tilde r_{\rm HO}$ depends on $\Delta \rho$, but not on $\Delta \alpha$:
in fact, if {\tt azinscheme4} is equal to one, $\Delta \tilde r$ is function of $\delta \tilde Z_e$,
while for {\tt azinscheme4} equal to zero the $\Delta \alpha$ factor originally present in $\Delta \tilde r$
is subtracted and resummed in the $\alpha=\alpha_0/(1-\Delta \alpha)$ factor under the square root in Eq.~(\ref{eq:seffsc4HO}).

To summarize, the fermionic higher-order effects are computed
in terms of a LO matrix-element squared computed as a function of the effective parameters $\alpha_0/(1-\Delta \alpha(M_Z^2))$
and $\overline{\sin}^{2, \, {\rm HO}}  \theta^l_{eff}$ and the removal of the double-counting of the $\mathcal{O}(\alpha)$ correction
is achieved by subtracting its first-order expansion in $\Delta \tilde r$ (and $\Delta \alpha$ if the {\tt azinscheme4} flag is off).
If the complex-mass scheme is used, $\Delta \tilde r$ in Eq.~(\ref{eq:seffsc4NLO}) becomes complex,
but we decided to include in Eq.~(\ref{eq:seffsc4HO}) $-$ and thus, effectively, resum $-$ only its real part
in order to minimize the spurious effects introduced by the CMS prescription.

As a conclusive remark, we recall that $\alpha_0$, $M_Z$, and $G_\mu$ are the input parameters
used for the theory predictions/tools \cite{Hollik:1988ii,Consoli:1989pc,Burgers:1989bh,Altarelli:1989hv,Bardin:1990fu,
  Bardin:1990de,Bardin:1990xe,Bardin:1992jc,Novikov:1993vn,Novikov:1994wk,Novikov:1994qi,
  Montagna:1993py,Montagna:1993ai} developed for the precise determination of the $Z$-boson
properties at LEP1 (see for instance~\cite{Bardin:1997xq} for a tuned comparison).
The realizations of the $(\alpha_0$, $G_\mu$, $M_Z)$ scheme described in this section differ 
from the ones used in the above-mentioned references, even though they are equivalent
at the perturbative order under consideration. In fact, a typical strategy in the literature was
to perform the calculation in a given scheme, for instance $(G_\mu$, $M_W$, $M_Z)$, using
the formulae for the NLO (or NLO plus fermionic higher-order corrections) derived in that scheme
but computing the numerical value of $M_W$ and $s_W$ (at the same perturbative accuracy)
from $(\alpha_0$, $G_\mu$, $M_Z)$ through the expression of $\Delta r$, namely:
\begin{equation}
  s_W^2=\frac{1}{2}-\sqrt{\frac{1}{4}-\frac{\pi \alpha_0}{\sqrt{2} G_\mu M_Z^2 (1-\Delta \alpha)}
    \frac{1+\Delta r_{remn}}{1+\frac{c_W^2}{s_W^2}\Delta \rho^{\rm HO}}},
  \label{eq:tuningOS}
\end{equation}
where the leading fermionic effects related to the running of the parameters $\alpha$ and $s_W$ in $\Delta r$ have been
resummed~\cite{Consoli:1989fg} and $\Delta \rho^{\rm HO}$ includes also higher-order corrections. In Eq.~(\ref{eq:tuningOS}) $s_W^2$ is a short-hand notation
for the quantity $1 - \frac{M_W^2}{M_Z^2}$, $M_W$ being the actual input
parameter in the $(G_\mu$, $M_W$, $M_Z)$ scheme. 
Equation~(\ref{eq:tuningOS}) is solved iteratively as $\Delta r_{remn}$ is a function of $s_W$.
In Ref.~\cite{Bardin:1997xq} a slightly modified version of Eq.~(\ref{eq:tuningOS}) was used,
with $\Delta \rho^{HO}$ promoted to $\Delta \rho^{HO}+\Delta \rho_X$ and $\Delta r_{remn}$
changed accordingly to $\Delta r_{remn}+\frac{c_W^2}{s_W^2}\Delta \rho_X$, with
\begin{equation}
  \Delta \rho_X= \Big( \frac{\Sigma_T^{ZZ}(M_Z^2)}{M_Z^2}-\frac{\Sigma_T^{WW}(M_W^2)}{M_W^2} \Big)_{\overline{\rm MS}} -\Delta \rho^{1-loop},
  \label{eq_drx}
\end{equation}
where the notation $\overline{\rm MS}$ just means that, within the brackets, UV poles have been removed and the mass scale $\mu_{\rm Dim}$ was replaced with $M_Z$.
In the following, the LEP1-like tuned comparison will be performed with the convention from Ref.~\cite{Bardin:1997xq}.
A similar strategy could be followed for the $(\alpha$, $\seffl$, $M_Z)$ schemes, where the
value of $\seffl$ can be obtained from the iterative solution of
\begin{equation}
  s_W^2=\frac{1}{2}-\sqrt{\frac{1}{4}-\frac{\pi \alpha_0}{\sqrt{2} G_\mu M_Z^2 (1-\Delta \alpha)}
    \Big(1+\Delta \tilde r_{\rm HO}\Big)}.
  \label{eq:tuningSEFF}
\end{equation}

\subsection{\it The $(\alpha_{\overline{\rm MS}}$, $s_{W \,\overline{\rm MS}}^2$, $M_Z)$ scheme and its decoupling variants}
\label{sect:swms}

In the $(\alpha_{\overline{\rm MS}}$, $s_{W \,\overline{\rm MS}}^2$, $M_Z)$ scheme, the independent parameters are the
$\overline{\rm MS}$ running couplings $\alpha_{\overline{\rm MS}}$ and $s_{W \,\overline{\rm MS}}^2$ and the $Z$-boson mass. 
More precisely, the input parameters are the numerical values of $\alpha_{\overline{\rm MS}}(\mu_0^2)$ and
$s_{W \,\overline{\rm MS}}^2(\mu_0^2)$ for a given $\overline{\rm MS}$ renormalization scale $\mu_0$ selected by the user
and the on-shell $Z$ mass (internally converted to the corresponding pole value). The values of $\alpha_{\overline{\rm MS}}(\mu_0^2)$ and
$s_{W \,\overline{\rm MS}}^2(\mu_0^2)$ are then evolved to $\alpha_{\overline{\rm MS}}(\mu^2)$ and
$s_{W \,\overline{\rm MS}}^2(\mu^2)$, where $\mu$ is the $\overline{\rm MS}$ renormalization scale selected for the calculation.
Both fixed and dynamical renormalization scale choices are implemented in the code.  The numerical results presented in the following are
obtained with a dynamical renormalization scales,
$\mu$ being set to the dilepton invariant mass.

The calculation of the tree-level and bare one-loop amplitudes in the
$(\alpha_{\overline{\rm MS}}$, $s_{W \,\overline{\rm MS}}^2$, $M_Z)$ scheme proceeds in the very same way as in the other schemes
described above, with the only difference that the electric charge and the sine of the weak-mixing angle are set to
$e_{\overline{\rm MS}}(\mu^2)$  and $s_{W \,\overline{\rm MS}}^2(\mu^2)$, respectively. The additional factor of $\alpha$
coming from the virtual and real QED corrections is always set to $\alpha_0$.
In the numerical studies presented in the next sections, the value of $\alpha$ used in the loop factor
in the virtual weak loops corresponds to $\alpha_{\overline{\rm MS}}(\mu^2)$,
as the flag {\tt a2a0-for-QED-only} is active, but the code allows the use of $\alpha_0$ as well.

The renormalization in the weak sector is performed in a hybrid scheme: the $Z$-boson mass counterterm
as well as the external-fermions wave-function counterterms are derived in the on-shell scheme (with the modifications
related to the complex-mass scheme choice), while the electric charge and the sine of the weak-mixing angle are
renormalized in the $\overline{\rm MS}$ scheme (possibly supplemented with $W$-boson and top-quark decoupling).

The electric charge counterterm in the $(\alpha_{\overline{\rm MS}}$, $s_{W \,\overline{\rm MS}}^2$, $M_Z)$ scheme
reads:
\begin{eqnarray}
  & & \delta Z_{e \; \overline{\rm MS}}(\mu^2) = - \frac{\alpha}{4\pi} \Big\{
  \sum_{f=l,q} \frac{N_C^f 2 Q_f^2}{3}\Big[ -\Delta_{\rm UV} +\log \frac{\mu^2}{\mu^2_{\rm Dim}} \Big]\nonumber \\
  & & + \frac{7}{2} \Big( \Delta_{\rm UV} - \log \frac{\mu^2}{\mu^2_{\rm Dim}} \Big)
  + \delta_{\rm D, \; \rm top} \frac{8}{9} \log \frac{M_{\rm top}^2}{\mu^2}\theta(M_{\rm top}^2 -\mu^2) \nonumber \\
  & & + \delta_{\rm D, \; W} \Big[ -\frac{7}{2} \log \frac{M_{W, \,\rm thr.}^2}{\mu^2}+\frac{1}{3} \Big] \theta(M_{W, \,\rm thr.}^2 -\mu^2) \Big\}, 
  \label{eq:dzewmsbar}
\end{eqnarray}
where $\mu_{\rm Dim}$ is the unphysical dimensional scale introduced with dimensional regularization and
cancels in the sum of bare and counterterm amplitudes.
The last two terms in Eq. (\ref{eq:dzewmsbar}) implement
the top and $W$ decoupling: if $\mu$ is greater than $M_{\rm top}$ ($M_{W, \,\rm thr.}$), only the part of the top-quark ($W$) loop
proportional to the combination $\Delta_{\rm UV} - \log \frac{\mu^2}{\mu^2_{\rm Dim}}$ contributes to the electric-charge
counterterm, while for $\mu < M_{\rm top}$ ($M_{W, \,\rm thr.}$) the full top-quark ($W$) loop enters the counterterm expression.
Note that the discontinuity at $\mathcal{O}(\alpha)$ on the $W$ threshold cancels the corresponding discontinuity
in the running of $\alpha_{\overline{\rm MS}}(\mu^2)$.
In equation~\ref{eq:dzewmsbar}, $\delta_{\rm D, \; W}$ ($\delta_{\rm D, \; \rm top}$) is equal to one if the $W$ (top)
decoupling is enabled together with the threshold corrections and zero otherwise (flags {\tt decouplemtOFF}, {\tt decouplemwOFF},
{\tt OFFthreshcorrs}). 

The counterterm corresponding to the sine of the weak-mixing angle reads:
\begin{eqnarray}
  & & \frac{\delta s_{W \,\overline{\rm MS}}^2}{s_{W \,\overline{\rm MS}}^2}(\mu^2)=
  \frac{c_{W \,\overline{\rm MS}}}{2 s_{W \,\overline{\rm MS}}}
  \Big( \delta Z_{ZA\,\overline{\rm MS}} - \delta Z_{AZ\,\overline{\rm MS}} \Big) \nonumber \\
  & & + \delta_{\rm D, \; W} \frac{\alpha}{6 \pi} \frac{c_{W \,\overline{\rm MS}}^2}{  s_{W \,\overline{\rm MS}}^2}\theta(M_{W, \,\rm thr.}^2 -\mu^2),
  \label{eq:dswmsbar}
\end{eqnarray}
where $\delta Z_{ZA\,\overline{\rm MS}}$ and $\delta Z_{AZ\,\overline{\rm MS}}$ have the usual expression of
$\delta Z_{ZA}$ and $\delta Z_{AZ}$ in the on-shell scheme upon the replacement $\Sigma^{AZ}_T \to \Sigma^{AZ,\overline{\rm MS}}_T$,
with:
\begin{eqnarray}
  & & -\frac{4 \pi}{\alpha} \Sigma^{AZ,\overline{\rm MS}}_T (q^2,\mu^2) = \Big( \Delta_{\rm UV} - \log \frac{\mu^2}{\mu^2_{\rm Dim}}\Big) \nonumber \\
  & &  \Big\{ -q^2 \sum_{f=l,q} \frac{2}{3} N_C^f Q_f(g_f^+ + g_f^-)
   +\frac{( 9 c_{W \,\overline{\rm MS}}^2 +\frac{1}{2}) q^2 +6 M_W^2}{3 s_{W \,\overline{\rm MS}} c_{W \,\overline{\rm MS}}} \Big\} \nonumber \\
   & & - \delta_{\rm D, \; W} \frac{( 9 c_{W \,\overline{\rm MS}}^2 +\frac{1}{2}) q^2 +6 M_W^2}{3 s_{W \,\overline{\rm MS}} c_{W \,\overline{\rm MS}}} \log \frac{M_{W, \,\rm thr.}^2}{\mu^2} \theta(M_{W, \,\rm thr.}^2 -\mu^2) \nonumber \\
  & & + \frac{4 \delta_{\rm D, \; \rm top}}{3} q^2 (g_{\rm top}^+ + g_{\rm top}^-) \log \frac{M_{\rm top}^2}{\mu^2}\theta(M_{\rm top}^2 -\mu^2).   
   \label{eq:SAZmsbar}
\end{eqnarray}
Note that $M_W^2$ in Eq.~(\ref{eq:SAZmsbar}) is computed as $(1-s_{W \,\overline{\rm MS}}^2) M_Z^2$ and does not necessarily coincide
with $M_{W, \,\rm thr.}^2$. When the decoupling is active, 
the $\mathcal{O}(\alpha)$ threshold correction for $\mu=M_{W, \,\rm thr.}$ in the running of $\alpha_{\overline{\rm MS}}$
induces a similar discontinuity in the running of $s_{W \,\overline{\rm MS}}^2$: the last term in Eq.~(\ref{eq:dswmsbar})
cancels this discontinuity at the $W$ threshold at $\mathcal{O}(\alpha)$.

The running of $\alpha_{\overline{\rm MS}}$ from the scale $\mu_0$ to the scale $\mu$ is taken
from Eqs.~(9)--(13) of Ref.~\cite{Erler:1998sy},
which contain QED and QCD corrections to the fermionic contributions to the $\beta$ function up to $\mathcal{O}(\alpha)$ and
$\mathcal{O}(\alpha_s^3)$~\cite{Gorishnii:1988bc,Surguladze:1990tg,Larin:1994va,Chetyrkin:1996ez}, respectively.
When the calculation is performed in the
decoupling scheme, the threshold corrections corresponding to the $W$ and the top-quark thresholds are also implemented: while the former
are $\mathcal{O}(\alpha)$ effects, the latter are included at $\mathcal{O}(\alpha^2)$, $\mathcal{O}(\alpha\alpha_S)$, and
$\mathcal{O}(\alpha \alpha_S^2)$~\cite{Chetyrkin:1996cf,Erler:1998sy}. In the code, the running of $\alpha_{\overline{\rm MS}}$
is only computed between scales $\mu_0$ and $\mu$ well within the perturbative regime ($\mu_0^2$, $\mu^2$ $\gg 4 m_b^2$):
non-pertur\-ba\-tive QCD effects are effectively
included through the numerical value of 
$\alpha_{\overline{\rm MS}}(\mu_0^2)$ selected
by the user (see~\ref{appendix:flags} for the corresponding default value and related discussion).
The running of $s_{W \,\overline{\rm MS}}$ is taken from Eq.~(25) of Ref.~\cite{Erler:2004in} (see also~\cite{Erler:2017knj}),
which contains $\mathcal{O}(\alpha^2)$, $\mathcal{O}(\alpha\alpha_S)$, $\mathcal{O}(\alpha\alpha_S^2)$, and
$\mathcal{O}(\alpha \alpha_S^3)$ corrections to the fermionic part of the $\beta$ function
~\cite{Gorishnii:1988bc,Surguladze:1990tg,Larin:1994va}.
As in the case of $\alpha_{\overline{\rm MS}}$, when the decoupling is active,
the corrections associated with the crossing of the $W$ and the top-quark 
thresholds at $\mathcal{O}(\alpha)$ and at $\mathcal{O}(\alpha^2)$, $\mathcal{O}(\alpha\alpha_S)$, and
$\mathcal{O}(\alpha \alpha_S^2)$, respectively, are also computed.
For some of the results presented below, the running is only performed at NLO (flag {\tt excludeHOrun}$=1$).

Similarly to the $(\alpha_0$, $\seffl$, $M_Z)$ scheme, where the fermionic higher-order corrections
effectively account for the running of $\alpha$ from the Thomson limit to the weak scale, in the
$(\alpha_{\overline{\rm MS}}$, $s_{W \,\overline{\rm MS}}^2$, $M_Z)$ scheme the universal higher-order
effects  are included through the running of the couplings.

\begin{figure}
  \includegraphics[width=0.48\textwidth]{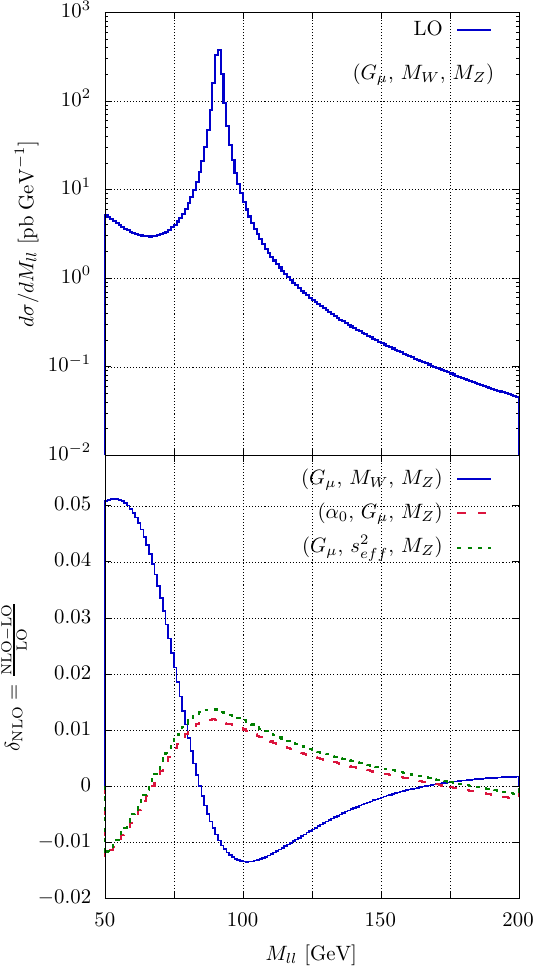}
  \caption{Upper panel: cross section distribution as a function of the
    leptonic invariant mass, at leading order in the $(G_\mu, M_W, M_Z)$
    scheme. Lower panel: relative difference between the NLO cross section
    and the LO one in the three renormalization schemes $(G_\mu, M_W, M_Z)$
    (solid blue), $(\alpha_0, G_\mu, M_Z)$ (dashed red),
    and $(G_\mu, \sin^2 \theta^l_{eff}, M_Z)$ (dotted green).
\label{fig:mv_2p_nlo}}
\end{figure}

\begin{figure}
  \includegraphics[width=0.48\textwidth]{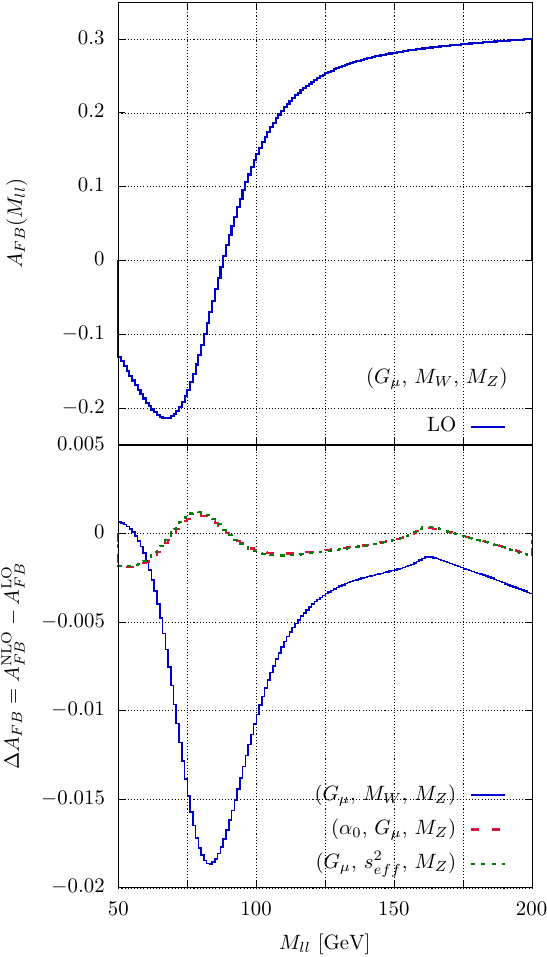}
  \caption{Upper panel: forward-backward asymmetry distribution as a function
    of the leptonic invariant mass, at leading order in the
    $(G_\mu, M_W, M_Z)$ scheme. Lower panel: absolute difference between NLO
    and LO asymmetry in the three renormalization schemes $(G_\mu, M_W, M_Z)$
    (solid blue), $(\alpha_0, G_\mu, M_Z)$ (dashed red), and
    $(G_\mu, \sin^2 \theta^l_{eff}, M_Z)$ (dotted green).
    \label{fig:afb_2p_nlo}}
\end{figure}

In the {\tt Z\_ew-BMNNPV} package, the choice of leaving $\alpha_{\overline{\rm MS}}(\mu_0^2)$ and
$s_{W \,\overline{\rm MS}}^2(\mu_0^2)$ as free parameters is motivated by the possibility of measuring
$s_{W \,\overline{\rm MS}}^2$ at the LHC and future hadron colliders from neutral-current Drell-Yan through
a template fit approach, as investigated in Ref.~\cite{Amoroso:2023uux}. Such measurements would require
the generation of Monte Carlo templates for different vales of $s_{W \,\overline{\rm MS}}^2(\mu_0^2)$
(and possibly $\alpha_{\overline{\rm MS}}(\mu_0^2)$) to be fitted to the data. While the present study
is focused on fixed-order results and in particular on weak corrections,
the {\tt Z\-\_ew-BMNNPV} can generate the required templates at NLO QCD+NLO EW accuracy with the consistent
matching to QCD and QED parton showers. Another possibility could be to use the $\overline{\rm MS}$ scheme
for a precise prediction (rather than determination) of $s_W^2$ at the weak scale as done in
Ref.~\cite{Degrassi:1990tu,Degrassi:1990ec,Sirlin:2012mh,Degrassi:1996ps,Gambino:1993dd,Ferroglia:2002rg,Ferroglia:2001cr}
up to full $\mathcal{O}(\alpha)$ accuracy (plus higher order corrections to the running of $\alpha$ and $\Delta \rho$).
In this approach $\alpha_{\overline{\rm MS}}(\mu_0^2)$ and $s_{W \,\overline{\rm MS}}^2(\mu_0^2)$ are derived quantities
computed as functions of other input parameters: typically the calculation is performed in the $(\alpha_0$, $G_\mu$, $M_Z)$
scheme given the high accuracy at which these parameters are measured. For what concerns $\alpha_{\overline{\rm MS}}$,
it can be computed from $\alpha_0$ via the relation
\begin{equation}
  \alpha_{\overline{\rm MS}}(\mu^2) = \frac{\alpha_0}{1-\Delta \hat{\alpha}(\mu^2)},
  \label{eq:alphaMSfroma0}
\end{equation}
where $\Delta \hat{\alpha}(\mu^2)=2[\delta Z_e -\delta Z_{e \; \overline{\rm MS}}(\mu^2)]$ at $\mathcal{O}(\alpha)$
(in the code, leptonic corrections at $\mathcal{O}(\alpha^2)$ as well as  QCD corrections of order $\alpha_S$ and $\alpha_S^2$ are
also available). At order $\alpha$, $s_{W \,\overline{\rm MS}}^2(\mu^2)$ can be computed from
\begin{equation}
  \sin^2 \theta_b= s_{W \,\overline{\rm MS}}^2+\delta s_{W \,\overline{\rm MS}}^2=s_{W}^2+\delta s_{W}^2,
  \label{eq:swMSfroma0gmu1}
\end{equation}
where the (renormalized) sine of the weak-mixing angle and the corresponding counterterm on the right-hand side of the second
equality are computed in the $(\alpha_0$, $G_\mu$, $M_Z)$ scheme. From equation~(\ref{eq:swMSfroma0gmu1}) it follows that,
at $\mathcal{O}(\alpha)$:
\begin{eqnarray}
  & & s_{W \,\overline{\rm MS}}^2(\mu^2)=s_{W}^2+\delta s_{W}^2-\delta s_{W \,\overline{\rm MS}}^2 + \delta \seffl - \delta \seffl \nonumber \\
  & = & s_{W}^2 +\frac{c_W^2s_W^2}{c_W^2-s_W^2} \Delta \tilde r+\delta \seffl-\delta s_{W \,\overline{\rm MS}}^2 \nonumber \\
  & = & s_{W}^2 +\frac{c_W^2s_W^2}{c_W^2-s_W^2} \Big[ \Delta r_{\overline{\rm MS}} + 2\Big( \delta Z_e -\delta Z_{e \; \overline{\rm MS}}(\mu^2)  \Big) \Big],
  \label{eq:swMSfroma0gmu2}
\end{eqnarray}
$\Delta r_{\overline{\rm MS}}$ being formally identical to $\Delta \tilde r$, but with $\delta Z_e$ and $\seffl$ replaced with
$\delta Z_{e \; \overline{\rm MS}}(\mu^2)$ and $\delta s_{W \,\overline{\rm MS}}^2$.
As in the case of Eqs.~(\ref{eq:seffsc4NLO}) and~(\ref{eq:seffsc4HO}), we can consider Eq.~(\ref{eq:swMSfroma0gmu2}) as the NLO expansion of
\begin{eqnarray}
  && s_{W \,\overline{\rm MS}}^2(M_Z^2) =  \nonumber\\
  && \frac{1}{2}-\sqrt{\frac{1}{4}-\frac{\pi \alpha_0}{\sqrt{2} G_\mu M_Z^2 (1-\Delta \hat{\alpha}(M_Z^2))} \Big(1+\Delta r_{\overline{\rm MS},\, {\rm HO}} \Big)}. 
  \label{eq:swMSfroma0gmu3}
\end{eqnarray}
In equation~(\ref{eq:swMSfroma0gmu3}) the renormalization scale was identified with $M_Z^2$, given the input-parameter set used,
and $\Delta r_{\overline{\rm MS},\, {\rm HO}}$ is obtained from $\Delta r_{\overline{\rm MS}}$ by replacing the $\mathcal{O}(\alpha)$
expression of $\Delta \rho$ with the one including the higher-order corrections discussed in Sect.~\ref{sect:HO}.

A last comment is in order concerning the decoupling procedure. We decouple the top-quark and the $W$ boson
in the $\alpha_{\overline{\rm MS}}$ and $s_{W \,\overline{\rm MS}}^2$ running to make contact with Refs.~\cite{Erler:2004in,Erler:2017knj},
mainly motivated by the huge impact of the $W$ decoupling. However we adopt a minimal (and simplified) approach where
the top and the $W$ are integrated out only in the renormalization-group equations for
$\alpha_{\overline{\rm MS}}$ and $s_{W \,\overline{\rm MS}}^2$
and in the expression of the NLO counterterms for $\delta Z_{e \; \overline{\rm MS}}$ and 
$\delta s_{W \,\overline{\rm MS}}^2 / s_{W \,\overline{\rm MS}}^2$, which are closely related to the
evolution equations. The heavy degrees of freedom are not integrated out in the calculation of the relevant matrix elements.

\section{Input parameter schemes: numerical results}
\label{sect:input-schemes-numerics}
In this section we investigate the numerical impact of the
radiative corrections to differential observables of the NC DY process at the LHC, according to the above described input parameter schemes.
In particular, we focus on the dilepton invariant mass distribution
$d\sigma / dM_{ll}$ and on the forward-backward asymmetry $A_{FB}(M_{ll})$,
defined as:
\begin{eqnarray}
  A_{FB}(M_{ll})&=&\frac{d\sigma^F/dM_{ll} - d\sigma^B/dM_{ll}}
 {d\sigma^F/dM_{ll} + d\sigma^B/dM_{ll}}, 
  \label{eq:afb} \\
  d\sigma^F/dM_{ll} &=& \int_0^1 dc\,  \frac{d\sigma}{dM_{ll}\, dc}\, ,
  \label{eq:sigma_F} \\
  d\sigma^B/dM_{ll} &=& \int_{-1}^0 dc\, \frac{d\sigma}{dM_{ll}\, dc}\, ,
  \label{eq:sigma_B}
\end{eqnarray}
where $c$ is the cosine of the lepton scattering angle in the Collins-Soper
frame, as a function of the invariant mass $M_{ll}$. 
We consider the $\mu^+ \mu^-$ final state, with $\sqrt{s}=13$~TeV. 
All results are obtained for an inclusive setup, where no cuts are imposed
on the final-state leptons except for an invariant mass cut 
$M_{ll} \geq 50$~GeV.
The numerical values of the relevant parameters are specified
in~\ref{appendix:numerical-param}
and the default values for the higher order options, for the hadronic
contributions to $\Delta \alpha$ as well as for the $W/Z$ boson width options
are adopted.

The upper panels of Figs.~\ref{fig:mv_2p_nlo} and~\ref{fig:afb_2p_nlo}
show the LO predictions obtained in the $\left( G_\mu, M_W, M_Z \right)$
scheme for the differential cross section and $A_{FB}$ distribution computed
as functions of the dilepton invariant mass $M_{ll}$
in the window $50$~GeV $\leq M_{ll} \leq 200$~GeV, without
additional kinematical cuts on the leptons.
While the invariant mass distribution has a Breit-Wigner peak for
$M_{ll}$ equal to $Z$ mass,
the asymmetry crosses zero and changes sign in the resonance region: because
of this behaviour, in the following we quantify the impact of EW corrections
or the differences among predictions obtained in different input-parameter
schemes in terms of absolute (rather than relative) differences for the
$A_{FB}$ distribution.

To analyse the main features of the NLO weak corrections and the higher-order
effects discussed in Sects.~\ref{sect:HO}
and~\ref{sect:input-schemes-detail}, we consider the
$(\alpha_0$, $G_\mu$, $M_Z)$ scheme together with a representative of the
class of schemes using $M_W$ as independent parameter and a representative
for the class using $\seffl$ as input. We choose schemes with couplings
defined at the weak scale, namely $\left( G_\mu, M_W, M_Z \right)$,
$\left( G_\mu, \seffl, M_Z \right)$, and  $(\alpha_0$, $G_\mu$, $M_Z)$
with the flag {\tt azinscheme4} equal to one (i.e. using
$\alpha=\alpha_0/(1-\Delta \alpha)$ in the calculation).
Other schemes where couplings are defined at low-energy,
like the ones involving $\alpha_0$, lead to larger corrections with
respect to the LO because of the running of the parameters up to the
weak scale: such effects tend to reduce the differences w.r.t. to
the predictions of schemes relying on $\alpha(M_Z^2)$ or $G_\mu$ when moving
from LO to NLO and NLO+HO accuracy
(see Figs.~\ref{fig:mll_ratio_schemes}-\ref{fig:afb_diff_schemes}).
In the $\overline{\rm MS}$ scheme of Sect.~\ref{sect:swms}, the running of
$\alpha_{\overline{\rm MS}}(\mu^2)$ and $s_{W \,\overline{\rm MS}}^2(\mu^2)$ 
reabsorbs large part of the corrections in the Born matrix elements: for
this reason, the relative corrections with respect to the LO are not shown
for the $(\alpha_{\overline{\rm MS}}$, $s_{W \,\overline{\rm MS}}^2$, $M_Z)$
scheme and we only show the $\overline{\rm MS}$ results at NLO with the best
predictions (i.e. NLO+HO) in the other schemes
(Figs.~\ref{fig:mll_ratio_schemes}-\ref{fig:afb_diff_schemes}). 

The lower panel of Fig.~\ref{fig:mv_2p_nlo} shows the NLO relative correction
to $d\sigma / dM_{ll}$ w.r.t. the LO prediction, for three input
schemes: $( G_\mu, \sin^2 \theta_{eff}^l, M_Z)$ (dotted green line),
$\left( \alpha_0, G_\mu, M_Z \right)$ (dashed red line),
and $(G_\mu, M_W, M_Z)$ (solid blue line). The corrections in the first
two schemes are very similar, 
ranging from $-1\%$ to about $+1\%$, with the line corresponding to the
$( G_\mu, \sin^2 \theta_{eff}^l, M_Z)$ scheme
slightly above the one for the  $\left( \alpha_0, G_\mu, M_Z \right)$ scheme.
When using $M_W$ as an independent parameter,
the corrections have different shape and are in general larger, ranging form
$+5\%$ at 40~GeV to $-1\%$ around 100~GeV.
The picture could be understood as follows. The analytic expression of the
one-loop matrix element in the three schemes
is identical once the counterterms are expressed in terms of $\delta Z_e$
(or $\delta \tilde Z_e$) and $\delta s_W^2$,
the only differences being the actual form of the counterterms ($\delta Z_e$
and $\delta s_W^2$) and the $\Delta r$ or $\Delta \tilde r$ subtraction
terms that factorize on a the tree-level matrix-element in the scheme
$(G_\mu, M_W, M_Z)$ or $( G_\mu, \sin^2 \theta_{eff}^l, M_Z)$ schemes,
respectively\footnote{More precisely, $\Delta r -\Delta \alpha$ or 
  $\Delta \tilde r -\Delta \alpha$, since in all the three schemes there is a
  term $-\Delta \alpha M_{LO}$.}. 
If one replaces the counterterm $\delta s_W^2$ with
$\delta s_W^2 + \delta \seffl - \delta \seffl$, one can split the one-loop
matrix-element in a term that corresponds to the one-loop amplitude in the
scheme $( G_\mu, \sin^2 \theta_{eff}^l, M_Z)$
(up to the above-mentioned subtraction terms which however appear as
constant shifts in the relative corrections) plus a reminder that might
be written as $\Delta s^2_W \partial M_{\rm LO} / \partial s^2_W$ which
represents the change in the LO matrix-element when the numerical value
of $s^2_W$ is shifted by a factor $\Delta s^2_W=\delta s_W^2 -\delta \seffl$.
In the $\left( \alpha_0, G_\mu, M_Z \right)$ scheme, $\Delta s^2_W$ is about
$2.7 \times 10^{-4}$ and the corresponding impact is hardly visible on the
scale of the plot, while in the $(G_\mu, M_W, M_Z)$ $\Delta s^2_W$ is much
larger, of order $1 \times 10^{-2}$, and it is the main responsible for the
shape and the size of the effects shown in Fig.~\ref{fig:mv_2p_nlo}. The
relative corrections at NLO in the schemes with $\alpha_0$ or $\alpha(M_Z^2)$
as input together with $M_W$ ($\seffl$) can be obtained from the ones shown
in Fig.~\ref{fig:mv_2p_nlo} by removing the constant term $-2\Delta r$
($-2\Delta \tilde r$) or replacing it with $-2 \Delta \alpha(M_Z^2)$,
respectively. The lower panel of Fig.~\ref{fig:afb_2p_nlo} shows the
NLO correction to the asymmetry, defined as the absolute difference
$\Delta A_{FB} = A_{FB}^{\rm NLO} - A_{FB}^{\rm LO}$. Similarly to what happens
for the cross section, the NLO weak corrections with the schemes
$(G_\mu, \seffl, M_Z)$ and $\left( \alpha_0, G_\mu, M_Z \right)$ are 
very close and in general smaller, falling in the range $\pm 0.002$,
while the corrections in the $(G_\mu, M_W, M_Z)$ are larger, reaching
the value of $-0.018$ at about 80~GeV. The results for the asymmetry and
the ones for the dilepton invariant mass basically share the same
interpretation detailed above, with the main difference that the effect
of the overall subtraction terms like $\Delta r$ and $\Delta \tilde r$
largely cancel in $A_{FB}^{\rm NLO}$.
\begin{figure}
  \includegraphics[width=0.48\textwidth]{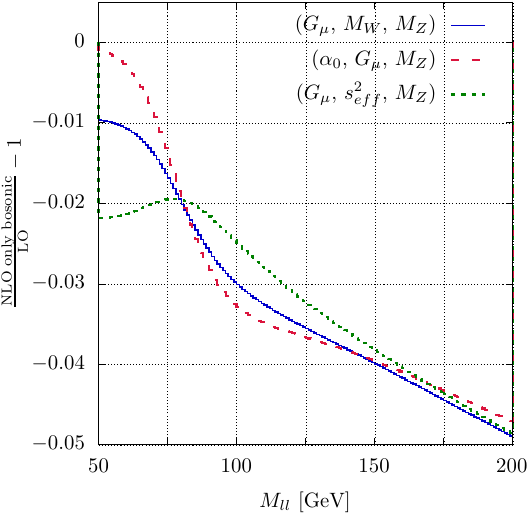}
  \caption{Relative corrections to the invariant mass cross section
    distribution obtained by considering only NLO bosonic contributions.
  \label{fig:mv_bosonly}}
\end{figure}
\begin{figure}
  \includegraphics[width=0.48\textwidth]{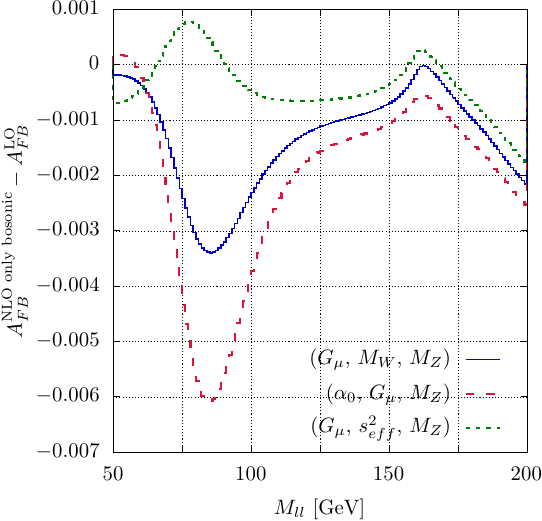}
  \caption{Absolute corrections to the forward-backward asymmetry obtained
    by considering only NLO bosonic contributions. \label{fig:afb_bosonly}}
\end{figure}
\begin{figure}
  \includegraphics[width=0.48\textwidth]{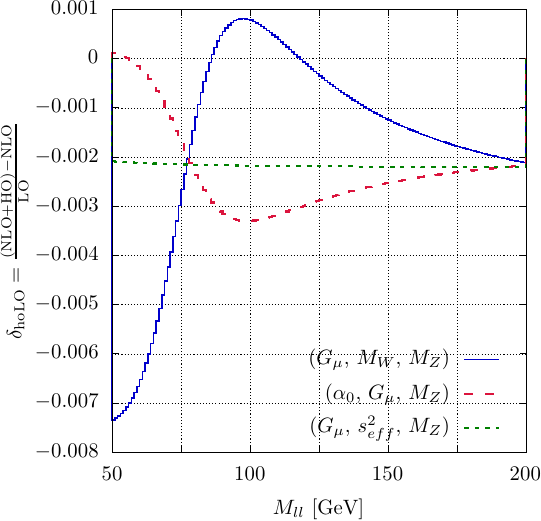}
  \caption{Higher-order correction to the cross section distribution as a
    function of the leptonic invariant mass. The three curves correspond to
    the three different choices of renormalization scheme discussed.
    \label{fig:mv_ho}}
\end{figure}
\begin{figure}
  \includegraphics[width=0.48\textwidth]{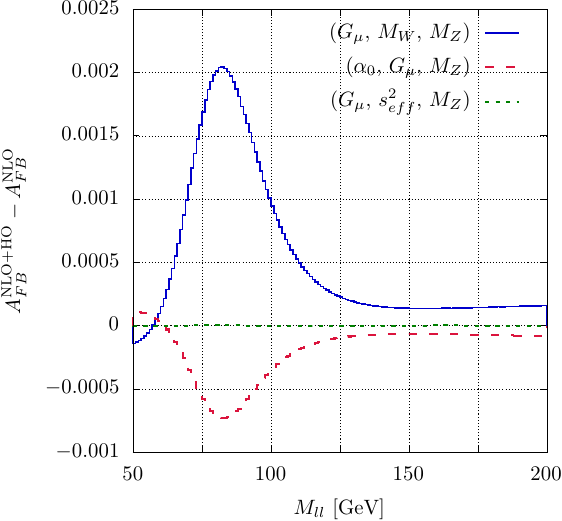}
  \caption{Higher-order correction to the forward-backward asymmetry
    distribution as a function of the leptonic invariant mass. The three
    curves correspond to the three different choices of renormalization
    scheme discussed.\label{fig:afb_ho}}
\end{figure}

Figures~\ref{fig:mv_bosonly} and~\ref{fig:afb_bosonly} show the relative (absolute)
NLO weak corrections to the cross section (forward-backward asymmetry) if only
the gauge-invariant subset of the bosonic loops is included. At low dilepton
invariant masses, large part of these corrections come from the bosonic
contribution to $\Delta r$ and $\Delta \tilde r$ entering the calculation in
the $(G_\mu, M_W, M_Z)$ and in the $( G_\mu, \sin^2 \theta_{eff}^l, M_Z)$
schemes, respectively. For larger $M_{ll}$ values, on the right part of the plot,
the bosonic corrections are relatively large (order $-5\%$), while the full
NLO weak corrections are at the permille level, pointing out a strong cancellation
between bosonic and fermionic corrections.  
The contribution from $\Delta r$ and $\Delta \tilde r$ essentially
cancels in $A_{FB}$
and the asymmetry difference in Fig.~\ref{fig:afb_bosonly} is dominated
by the shift induced in the effective $s_W^2$  by the bosonic part of the $\mathcal{O}(\alpha)$
corrections in the $( \alpha_0, G_\mu, M_Z)$ and $(G_\mu, M_W, M_Z)$ schemes
($\delta s_W^2 \sim 3 \times10^{-3}$ and $\delta s_W^2 \sim 2.5 \times10^{-3}$, respectively).
By comparing Figs.~\ref{fig:afb_2p_nlo} and~\ref{fig:afb_bosonly}, one notices that 
a large cancellation between bosonic and fermionic effects is still there in the $( \alpha_0, G_\mu, M_Z)$ and $(G_\mu, M_W, M_Z)$ schemes, while in the
$( G_\mu, \sin^2 \theta_{eff}^l, M_Z)$ bosonic corrections dominate over the fer\-mio\-nic ones
and the lines corresponding to this scheme in Figs.~\ref{fig:afb_2p_nlo} and~\ref{fig:afb_bosonly}
are almost the same in the scale of the plot.

Figs.~\ref{fig:mv_ho} and~\ref{fig:afb_ho} show the higher-order universal
corrections (i.e. beyond NLO) defined in Sect.~\ref{sect:HO}, to 
the cross section invariant mass distribution (normalized to
the LO predictions) and the forward-backward asymmetry, 
respectively, in the three renormalization schemes. As for the NLO case,
the plots display the relative corrections for the cross section distribution
and the absolute correction for $A_{FB}(M_{ll})$.
In the $( G_\mu, \sin^2 \theta_{eff}^l, M_Z)$ scheme the corrections
are small (order $0.2\%$) and essentially flat: this is because the corrections
in Eq.~(\ref{eq:amzsw-HO-gmu}) factorize on the LO matrix-element squared and
the only dependence on $M_{ll}$ comes from the running of $\alpha_S$
in the QCD corrections to $\Delta \rho$. 
The corrections in the $(\alpha_0$, $G_\mu$, $M_Z)$ scheme fall in the range
$[-0.3\%,0]$, being basically zero for low dilepton invariant masses and
reaching their maximum around the $Z$ peak: the shape of the corrections is
determined by the additional shift $\Delta s_W^{\rm HO}$ on top of the NLO one
($\sim 5 \times 10^{-5}$) which only affects the $Z$-boson exchange
amplitude, while for small invariant masses the dominant contribution is the
$\gamma$ exchange. The impact of the fermionic higher-order effects in the
$(G_\mu, M_W, M_Z)$ scheme is larger than for the other choices of input
parameters, ranging from about $-0.7\%$ at $M_{ll}=50$~GeV to
about $+0.1\%$ at the $Z$-peak: in this scheme, the corrections come from
the interplay of the 
shift to $s^2_W$ ($-9 \times 10^{-4}$ in addition to the NLO shift) and the
overall factor
$2(\Delta \rho -\Delta \rho^{(\alpha)}) c_W/s_w +\Delta \rho^2 c_W^2/s_W^2$
coming from the relation between $\alpha$ and $G_\mu$. The latter effect
enters also the $\gamma$-exchange diagram and thus affects also the
low-invariant mass region of the plot. When considering the asymmetry
(Fig.~\ref{fig:afb_ho}), any overall term common to numerator and denominator
of Eq.~(\ref{eq:afb}) cancels: this is almost the case for the 
higher-order corrections in the $( G_\mu, \sin^2 \theta_{eff}^l, M_Z)$
scheme, where the factorization of the higher-order terms is only approximate,
due to the presence of the NLO corrections, leading to a negligible residual 
effect of order $10^{-6}$ on the asymmetry difference not visible within the 
resolution of the plot. The impact in the  $(G_\mu, M_W, M_Z)$ scheme is
larger, with a maximum of $2 \times 10^{-3}$ for $M_{ll}$ around
$80$~GeV: the behaviour is essentially determined by the above-mentioned
shift in $s_W^2$ on top of the $\mathcal{O}(\alpha)$ one.
In the $(\alpha_0$, $G_\mu$, $M_Z)$ scheme the corrections 
are negative, reaching the value of about $-7 \times 10^{-4}$ again
at about $80$~GeV, and driven by the higher-order shift on $s_W^2$. 
\begin{figure}
  \includegraphics[width=0.48\textwidth]{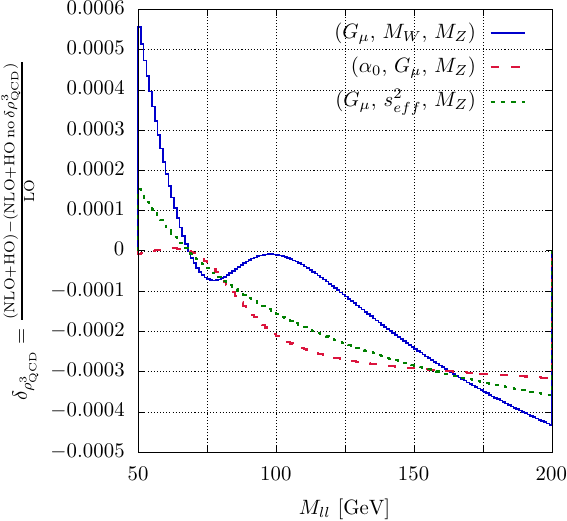}
  \caption{Impact of the three-loop QCD contribution to $\Delta \rho$
    on the cross section distribution $d\sigma / dM_{ll}$.
    The three curves correspond to the three different renormalization
    schemes. \label{fig:mv_ho_deltarho3qcd}}
\end{figure}
\begin{figure}
  \includegraphics[width=0.48\textwidth]{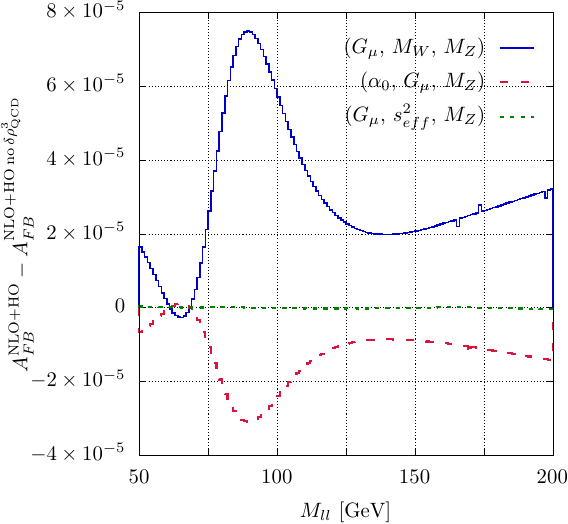}
  \caption{The same as in Fig.~\ref{fig:mv_ho_deltarho3qcd} for
    $A_{FB}$. The absolute difference between the three-loop QCD contribution
    to $\Delta \rho$ and the two-loop case is shown.
    \label{fig:afb_ho_deltarho3qcd}}
\end{figure}
Figs.~\ref{fig:mv_ho_deltarho3qcd} and~\ref{fig:afb_ho_deltarho3qcd}
show the impact of the three-loop QCD correction to $\Delta \rho$, 
$\delta_{QCD}^{(3)}$ in Eq.~(\ref{eq:drho}), normalized to the LO predictions,
to the invariant mass cross section distribution and the forward-backward
asymmetry, respectively, in the three renormalization schemes.
The leading $\delta_{QCD}^{(3)}$ contribution comes from the replacement of
the $\Delta \rho^{(\alpha)}$
terms in the NLO calculation with the expression of $\Delta \rho$ in
Eq.~(\ref{eq:drho}). In the $( G_\mu$, $\sin^2 \theta_{eff}^l$, $M_Z)$ scheme,
$\Delta \rho$ comes from the relation between $\alpha$ and $G_\mu$ and
simply multiplies the LO matrix-element squared: the corresponding line
in Fig.~\ref{fig:mv_ho_deltarho3qcd} is twice the factor
$3 x_t (1+x_t\Delta \rho^{(2)})\frac{\alpha_S^2}{\pi^2}\delta_{QCD}^{(3)}$ and the dependence on $M_{ll}$ is the residual scale
dependence of the QCD correction. 
In the $(\alpha_0$, $G_\mu$, $M_Z)$ scheme, the linear term in $\Delta \rho$
comes from the corrections to the $G_\mu M_Z^2$ factor in the $Z$-boson
exchange diagram: as a consequence, in the low dilepton invariant mass region, 
where the dominant contribution is from $\gamma$ exchange, the correction
tends to vanish, while for larger values of  $M_{ll}$ it does not
factorize on the LO matrix-element. In the $(G_\mu, M_W, M_Z)$ scheme, linear
terms in $\Delta \rho$ come both from the $\delta s_W^2$ counterterm and
from  the $\Delta r$ relating $\alpha$ and $G_\mu$ at NLO
($\sim c_W^2 /s_W^2 \, \Delta \rho$). Only the latter contribution 
factorizes on the Born result and it is the only one affecting the
$\gamma$-exchange which dominates the cross section in the low-invariant
mass limit, where the effect is basically three times ($\sim c_W^2/s_W^2$) 
larger than the  one observed in the
$( G_\mu$, $\sin^2 \theta_{eff}^l$, $M_Z)$ scheme.
Moving to the asymmetry, the impact of $\delta_{QCD}^{(3)}$ in the
$( G_\mu, \sin^2 \theta_{eff}^l, M_Z)$ scheme is not visible
in Fig.~\ref{fig:afb_ho_deltarho3qcd}, since it largely cancels between
numerator and denominator in $A_{FB}$. 
Also for the other two schemes the effect is tiny, of the order of $10^{-5}$.
The four-loop QCD corrections to $\Delta \rho$
(not included in {\tt Z\_ew-BMNNPV} code) computed at the scale $M_{\rm top}$
should be about five times smaller than the three loop
ones~\cite{Chetyrkin:2006bj}, but with
reduced scale dependence, so that the numerical impact on the
$M_{ll}$ and $A_{FB}$ distributions
is negligible compared to the other effects discussed in the following.

We close this subsection presenting in Figs.~\ref{fig:mll_ratio_schemes}
and~\ref{fig:afb_diff_schemes} the predictions for different
schemes referred to the ones obtained in the $( \alpha(M_Z^2)$,
$\sin^2 \theta_{eff}^l$, $M_Z)$ scheme,
for different levels of perturbative accuracy: LO, NLO, and  NLO+HO
(relative differences for $d\sigma / dM_{ll}$
and absolute differences for $A_{FB}(M_{ll})$). 
The lower panels, referring to the NLO+HO predictions, contain
also the results in the hybrid $\overline{ \rm MS}$ scheme
discussed in Section~\ref{sect:swms}, 
$(\alpha_{\overline{\rm MS}}$, $s_{W \,\overline{\rm MS}}^2$, $M_Z)$.
The choice of the reference scheme is motivated by the fact that, in the
$( \alpha(M_Z^2)$, $\sin^2 \theta_{eff}^l$, $M_Z)$ scheme, the corrections
do not involve $\Delta \alpha$- or $\Delta \rho$-enhanced terms
and thus the higher-order corrections discussed in
Sect.~\ref{sect:HO} are absent. On the other hand, in the other schemes,
the corrections can be split in a non-enhanced part $-$ which is formally
the same one as in the $( \alpha(M_Z^2)$, $\sin^2 \theta_{eff}^l$, $M_Z)$
scheme with a different numerical value for $\alpha$ and $s_W^2$ $-$
plus a shift in $s_W^2$ from Eq.~(\ref{eq:seffsc4NLO})
and an overall effect coming from the running of $\alpha$ or from
the corrections to the relation between $\alpha$ and $G_\mu$
when $\alpha_0$ or $G_\mu$ are used as input, respectively.
When going beyond NLO, the latter effects can have a non-trivial 
interplay leading, for instance, to mixed contributions of the form
$\Delta \alpha \Delta \rho$. 
As a general comment, the spread of the predictions for the differential
cross section based on different input parameter schemes tends to shrink
from order $20\%$ at LO to $2\%$ at NLO and few $0.1\%$ 
with the inclusion of universal additional corrections. 
The absolute differences for $A_{FB}$ are at the level $0.02$ at
LO and become of the order of $10^{-3}$ ($10^{-4}$) 
when the NLO (NLO plus fermionic higher-order) corrections are included.
In the low invariant mass region, dominated by the $\gamma$-exchange diagram,
the cross section ratios computed at LO (upper panel of 
Fig.~\ref{fig:mll_ratio_schemes}) reduce to ratios of the values of $\alpha$
used in the numerator and in the denominator (squared). For the schemes
employing $\alpha(M_Z^2)$, including $(\alpha_0$, $G_\mu$, $M_Z)$ since the
{\tt azinscheme4} flag is active, the ratios tend to one at low
$M_{ll}$. The same holds for the $(G_\mu$, $\seffl$, $M_Z)$ scheme, 
since the value of $\alpha$ computed from $G_\mu$ and $\seffl$ at LO is
pretty close to $\alpha(M_Z^2)$. For the schemes based on $\alpha_0$ the
ratios are about $12\%$ smaller, while for the $(G_\mu$, $M_W$, $M_Z)$ scheme 
the corresponding ratio is about $6\%$ smaller than the one for the
$\alpha(M_Z^2)$-based schemes.  At the LO, the only difference in the
predictions for the cross section computed in schemes using $\seffl$ 
as input comes from the value of $\alpha$ used in the LO couplings:
this explains the horizontal lines corresponding to the $(\alpha_0$,
$\seffl$, $M_Z)$ and  and $(G_\mu$, $\seffl$, $M_Z)$ schemes.
When using the schemes with $M_W$ as input or $(\alpha_0$, $G_\mu$, $M_Z)$, 
not only the value of $\alpha$ used for the couplings changes with respect
to the one used in the denominator, but also $s_W^2$ is different:
since a variation of this parameter affects in a different way the $Z$- and
$\gamma$-exchange amplitudes (the latter only when $\alpha$ is derived from
$G_\mu$) which are weighted by the factors $\frac{1}{s-M_Z^2+i\Gamma_ZM_Z}$
and $1/s$, respectively, the ratios corresponding to the $M_W$-based
schemes in the upper panel of Fig.~\ref{fig:mll_ratio_schemes} have a non
trivial shape as a function of $M_{ll}$. For the $(\alpha_0$,
$G_\mu$, $M_Z)$ scheme, the ratio is still close to one since the value
of $s_W^2$ computed at LO in this scheme is close to the value of $\seffl$
used in the denominator ($0.2308$ versus $0.2315$, to be compared with
$0.2228$ in the $M_W$-related schemes). 
As any overall constant factor cancels between numerator and denominator in
$A_{FB}$, the LO asymmetry difference with respect to the
$(\alpha(M_Z^2)$, $\seffl$, $M_Z)$ scheme is zero when $(\alpha_0$, $\seffl$,
$M_Z)$ or $(G_\mu$, $\seffl$, $M_Z)$ are used as input parameters 
(upper panel of Fig.~\ref{fig:afb_diff_schemes}). For the other schemes
one only sees the impact of the different value of $s_W^2$ used:
the three lines for the schemes based on $M_W$ overlap, while the one for
the $(\alpha_0$, $G_\mu$, $M_Z)$ is again closer
to the one of the $\seffl$-based schemes.

At NLO, the spread of the cross section ratios is considerably reduced.
In the low $M_{ll}$ region, the ratio stays one for the schemes
based on $\alpha(M_Z^2)$, while for the $\alpha_0$-related schemes it is much
closer to one compared to the LO results. 
This is because the NLO corrections in these schemes develop an overall
factor $2\Delta \alpha$ that, once added to the LO term,
leads to a sort of LO-improved prediction proportional to the effective
coupling $\alpha_0^2 (1+2\Delta \alpha)$
which is just the first-order expansion of
$\alpha(M_Z^2)^2 = \alpha_0^2/(1-\Delta \alpha)^2$.
Something similar happens for the $(G_\mu$, $M_W$, $M_Z)$ scheme including
the one-loop corrections to the $\alpha$-$G_\mu$ 
relation contained in $\Delta r$. Besides changing the effective value of
$\alpha$ used in the calculation, the one-loop corrections also change the
value of the effective $s_W^2$ used for the $M_W$-based
and the $(\alpha_0$, $G_\mu$, $M_Z)$ schemes: the latter effect is the main
responsible for the shape differences in the plots
and in particular it explains the change of trend moving from LO to NLO when
$M_W$ is used as input parameter
($s_{W,\,\rm LO}^{2\, eff}(M_W)< \seffl$ while $s_{W,\, \rm NLO}^{2\, eff}(M_W)> \seffl$)\footnote{
  Since in our simulations the flag {\tt a2a0-for-QED-only} is switched on,
  the loop factors for the
  schemes $(\alpha_0$, $M_W$, $M_Z)$, $(\alpha(M_Z^2)$, $M_W$, $M_Z)$,
  and $(G_\mu$, $M_W$, $M_Z)$ are not the same, leading to slightly
  different values of $s_{W,\, \rm NLO}^{2\, eff}(M_W)$.}.

Including the higher-order corrections goes in the direction of further
reducing the differences among the predictions in different schemes
(lower panel of Fig.~\ref{fig:mll_ratio_schemes}), as this class of
corrections is basically obtained in terms of Born-improved
matrix elements squared written as functions of \emph{effective} couplings
$\alpha$ and $s_W^2$ reabsorbing the leading part 
of the fermionic corrections up to the scale $M_Z$ and numerically close to
$\alpha(M_Z^2)$ and $\seffl$. It is worth noticing that 
this sort of redefinition of the couplings in the LO matrix-element does not 
affect the part of the one-loop result that 
is not enhanced by large fermionic corrections and in particular does not
apply to the bosonic part of the $\mathcal{O}(\alpha)$ result: the different
couplings entering this part of the corrections are the main responsible 
for the residual deviations from one in the lower panel
of Fig.~\ref{fig:mll_ratio_schemes}. As an example, one can take the predictions
in the $(\alpha_0$, $\seffl$, $M_Z)$ scheme: according to
Eq.~(\ref{eq:amzsw-HO-a0}),  the expression for the NLO+HO corrections is
identical to the one in the $(\alpha(M_Z^2)$, $\seffl$, $M_Z)$ scheme,
when $\alpha(M_Z^2)$ is obtained from $\alpha_0/(1-\Delta\alpha$), 
and the only difference is the non-enhanced part of the $\mathcal{O}(\alpha)$
result, that is proportional to $\alpha_0^3$ in the numerator and
to $\alpha(M_Z^2)^3$ in the denominator of the ratio in
Fig.~\ref{fig:mll_ratio_schemes}. This difference alone leads to an effect
of order $\pm 0.2\%$ as shown in the plot.

The lower panel of Fig.~\ref{fig:mll_ratio_schemes} also shows the ratio of
the ${\overline{\rm MS}}$ predictions with respect to the 
ones in the $(\alpha(M_Z^2)$, $\seffl$, $M_Z)$ scheme. In the
{\tt Z\_ew-BMNNPV} package, we implemented the expressions 
of Refs.~\cite{Erler:1998sy} and~\cite{Erler:2004in,Erler:2017knj} for the
running of $\alpha_{\overline{\rm MS}}$ and $s_{W \,\overline{\rm MS}}^2$ 
from a scale $\mu_0^2$ to a scale $\mu^2$\footnote{Under the assumption that
  both $\mu_0^2$ and $\mu^2$ are in the region above $4m_b^2$.} leaving both
$\mu_0^2$ and the actual values of $\alpha_{\overline{\rm MS}}(\mu_0^2)$ and
$s_{W \,\overline{\rm MS}}^2(\mu_0^2)$ as free parameters, since we had in
mind the determination of $s_{W \,\overline{\rm MS}}^2$ from neutral-current
Drell-Yan at the LHC and future hadron-colliders by means of template fits,
as in Ref.~\cite{Amoroso:2023uux}: the ${\overline{\rm MS}}$ results thus 
depend on $\alpha_{\overline{\rm MS}}(\mu_0^2)$ and
$s_{W \,\overline{\rm MS}}^2(\mu_0^2)$ as input parameters.
The solid black line shows the ratio of the ${\overline{\rm MS}}$ prediction
over the one in the $(\alpha(M_Z^2)$, $\seffl$, $M_Z)$ scheme
setting $\alpha_{\overline{\rm MS}}(M_Z^2)$ and
$s_{W \,\overline{\rm MS}}^2(M_Z^2)$ ($\mu_0^2=M_Z^2$) to the values quoted
by the PDG~\cite{Workman:2022ynf} (see also~\ref{appendix:numerical-param}).
While the numbers fall in the same ballpark as the ones obtained in the other
schemes, the discrepancy tends to be a little larger. The source of the
differences is twofold: on the one hand, the values in
Ref.~\cite{Workman:2022ynf} are computed with a theoretical accuracy that
is not matched by the rest of the calculation in {\tt Z\_ew-BMNNPV} and, 
on the other hand, the parameters used in their computation and the ones
employed in the present study are not tuned.
The dashed black line corresponds to the ${\overline{\rm MS}}$ predictions
for a tuned choice of $\alpha_{\overline{\rm MS}}(M_Z^2)$ and
$s_{W \,\overline{\rm MS}}^2(M_Z^2)$: 
$\alpha_{\overline{\rm MS}}(M_Z^2)$ is consistently computed from $\alpha_0$
using the same parameters as in the rest of the calculation,
while $s_{W \,\overline{\rm MS}}^2(M_Z^2)$ is derived from the input
parameters $(\alpha_0$, $G_\mu$, $M_Z)$ as described in
Sect.~\ref{sect:swms}. 

The interpretation of Fig.~\ref{fig:afb_diff_schemes} for the asymmetry
difference follows closely the one for the dilepton invariant mass
cross section distribution, with the main difference that the corrections
connected to $\Delta \alpha$ and $\Delta r$ largely cancel between numerator
and denominator in $A_{FB}$ (though not exactly, leading for instance to small
deviations from zero in the low invariant mass region for the $\seffl$-based
schemes at NLO), and the spread of the predictions in the considered schemes
is mainly due to the different values of $s_W^2$ effectively employed.

The numerical results in Figs.~\ref{fig:mll_ratio_schemes}--\ref{fig:afb_diff_schemes} are obtained
under the assumption that the input parameters are actually free parameters to be set
to the corresponding experimental values (or to be used as variables in template fit analyses)
and no attempt was made to tune the input parameters for the different schemes
(with the only exceptions of $\alpha(M_Z^2)$ $-$ computed from $\alpha_0$ $-$
and $s_{W \,\overline{\rm MS}}^2(\mu_0^2)$ in the tuned $\overline{\rm MS}$ calculation).
Another possibility, closer to the strategy used for the numerical predictions for LEP1 studies
mentioned in Sect.~\ref{sect:agmumz}, would be to take a reference input scheme, say $(\alpha_0$, $G_\mu$, $M_Z)$,
and perform the calculation in other schemes, like the $(G_\mu$, $M_W$, $M_Z)$, $(G_\mu$, $\seffl$, $M_Z)$, or
$(\alpha_{\overline{\rm MS}}$, $s_{W \,\overline{\rm MS}}^2$, $M_Z)$ ones,
but deriving the numerical values of $M_W$, $\seffl$, and $s_{W \,\overline{\rm MS}}^2(\mu_0^2)$, 
from the parameters $\alpha_0$, $G_\mu$, $M_Z^2$ using the quantity $\Delta r$ ($\Delta \tilde r$)
as in Eqs.~(\ref{eq:tuningOS}),~(\ref{eq:tuningSEFF}), and (\ref{eq:swMSfroma0gmu3}).
Clearly the tuning procedure reduces all the tuned schemes to the reference one at the
considered theoretical accuracy (in our case, NLO plus leading fermionic corrections of
order $\Delta \alpha^2$, $\Delta \rho^2$, $\Delta \alpha \Delta \rho$) and it is expected to reduce the
spread of the predictions in the peak region (where the tuning is actually performed)
but not necessarily away from the resonance. 
The effect of tuning is shown in Fig.~\ref{fig:tuned} for the dilepton invariant mass cross section distribution (upper panel)
and for the forward-backward asymmetry (lower panel), which basically correspond to the lower panels of 
Figs.~\ref{fig:mll_ratio_schemes}--\ref{fig:afb_diff_schemes} but taking as reference the $(\alpha_0$, $G_\mu$, $M_Z)$ scheme.
The maximum spread of the cross section ratios as a function of $M_{ll}$ is of about $0.025$\%, while the one for the
asymmetry difference is of the order of $0.005$\%. As a technical remark, the plots are obtained in the pole scheme in order
to minimize the spurious $\mathcal{O}(\alpha^2)$ effects induced by the CMS, and the fermionic HO corrections
in the $(G_\mu$, $\seffl$, $M_Z)$ scheme are obtained with a modified version of Eq.~(\ref{eq:amzsw-HO-gmu})
where $\Delta \rho$ is replaced with $\Delta \tilde r$: the expressions are equivalent at the considered theoretical
accuracy, differing by terms at most of order $\Delta \tilde r_{\rm remn} \Delta \rho$, but this way the effective
couplings entering the $Zf\overline{f}$ vertex in the calculation of the fermionic higher-orders in the
$(G_\mu$, $\seffl$, $M_Z)$ and the  $(\alpha_0$, $G_\mu$, $M_Z)$ schemes become identical.

\begin{figure}
  \includegraphics[width=0.48\textwidth]{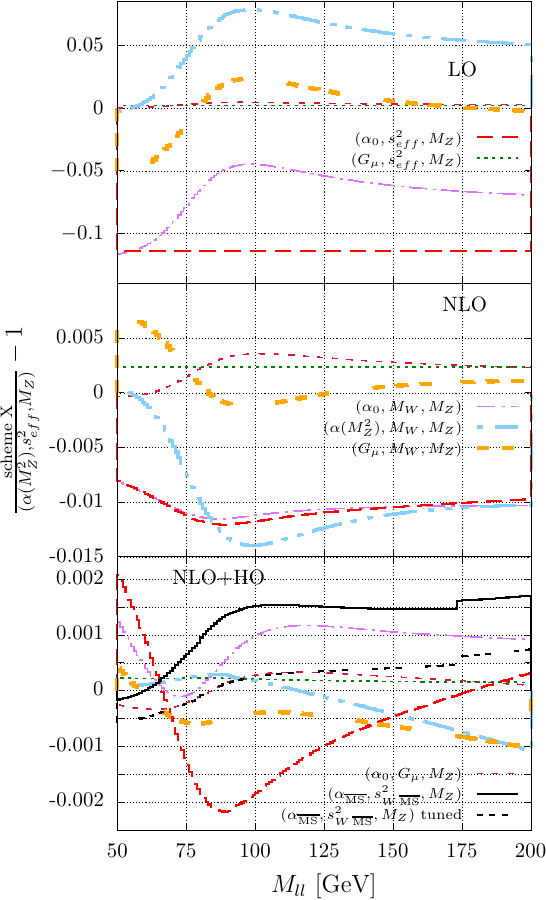}
  \caption{Relative difference of the predictions for the dilepton invariant mass cross section distribution 
    at LO (upper panel), NLO (middle panel), NLO+HO (lower panel). The calculation is
    performed in the CMS. The $\alpha$ values used in the loop factors corresponds to the
    ones used for the LO couplings. In the $(\alpha_0$, $G_\mu$, $M_Z)$ the actual value of
    $\alpha$ is $\alpha_0/(1-\Delta \alpha)$.
  \label{fig:mll_ratio_schemes}}
\end{figure}

\begin{figure}
  \includegraphics[width=0.48\textwidth]{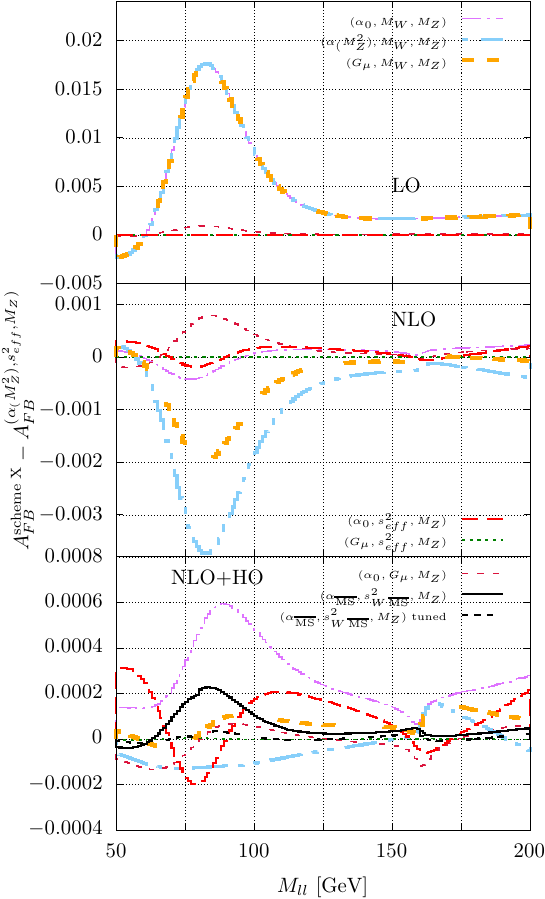}
  \caption{Absolute difference of the predictions for the forward-backward asymmetry
    as a function of the dilepton invariant mass 
    at LO (upper panel), NLO (middle panel), NLO+HO (lower panel). The calculation is
    performed in the CMS. The $\alpha$ values used in the loop factors corresponds to the
    ones used for the LO couplings. In the $(\alpha_0$, $G_\mu$, $M_Z)$ the actual value of
    $\alpha$ is $\alpha_0/(1-\Delta \alpha)$.
  \label{fig:afb_diff_schemes}}
\end{figure}

\begin{figure}
  \includegraphics[width=0.48\textwidth]{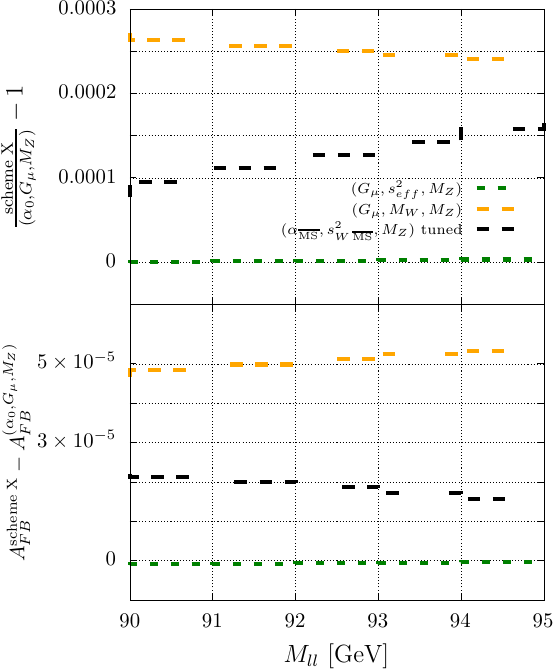}
  \caption{Relative difference (absolute difference) of the predictions for the cross section
    (forward-backward asymmetry) as a function of the dilepton invariant mass at NLO+HO accuracy.
    The calculation is performed in the pole scheme. The $\alpha$ values used in the loop factors corresponds to the
    ones used for the LO couplings. 
    In the $(\alpha_0$, $G_\mu$, $M_Z)$ the actual value of $\alpha$ is $\alpha_0/(1-\Delta \alpha)$.
    The value of $M_W$ used in the $(G_\mu$, $M_W$, $M_Z)$ scheme is derived from $(\alpha_0$, $G_\mu$, $M_Z)$ using Eq.~(\ref{eq:tuningOS}).
    Similarly, $\seffl$ and $s_{W \,\overline{\rm MS}}^2$ are computed by means of Eqs.~(\ref{eq:tuningSEFF}) and (\ref{eq:swMSfroma0gmu3}).  
  \label{fig:tuned}}
\end{figure}

\begin{figure}
  \includegraphics[width=0.48\textwidth]{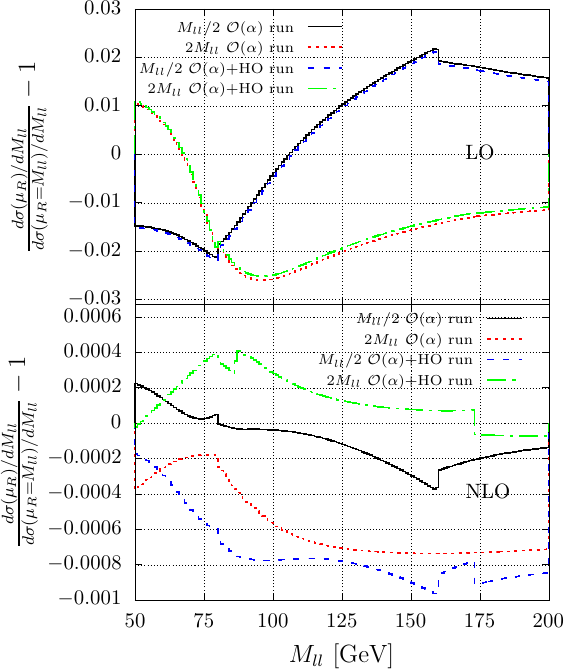}
  \caption{Relative difference of the dilepton invariant mass distribution obtained
    for $\mu_R=2M_{ll}$ (red and green lines) or $\mu_R=M_{ll}/2$ (black and blue curves)
    with respect to the predictions obtained with the default choice $\mu_R=M_{ll}$.
    The $\overline{\rm MS}$ running of $\alpha_{\overline{\rm MS}}(\mu_R^2)$ and
    $s_{W \,\overline{\rm MS}}^2(\mu_R^2)$ is computed at $\mathcal{O}(\alpha)$ in the
    solid and dotted lines, while in the dashed and dot-dashed lines the running
    includes the higher-order effects described in the text. The distributions are
    computed at LO in the upper panel and at NLO in the lower one.
  \label{fig:scvar}}
\end{figure}

\begin{figure}
  \includegraphics[width=0.48\textwidth]{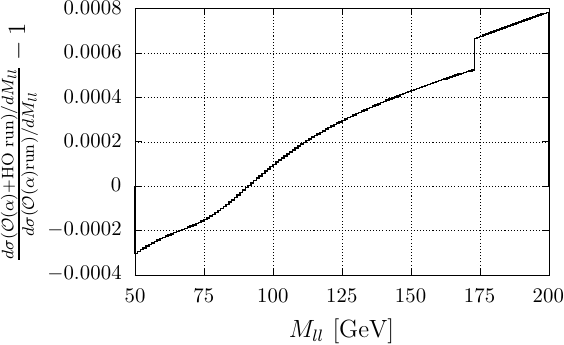}
  \caption{Relative difference between the dilepton invariant mass distributions
    computed in the $\overline{\rm MS}$ scheme with and without including
    the HO corrections to the running of $\alpha_{\overline{\rm MS}}(\mu_R^2)$ and
    $s_{W \,\overline{\rm MS}}^2(\mu_R^2)$ described in the main text.
  \label{fig:homsbar}}
\end{figure}
Concerning the $(\alpha_{\overline{\rm MS}}$, $s_{W \,\overline{\rm MS}}^2$, $M_Z)$ scheme,
it is interesting to analyze the renormalization-scale dependence of the predictions obtained in this scheme.
Fig.~\ref{fig:scvar} shows the ratio
of the dilepton invariant mass cross section distribution computed
with $\mu_R=2M_{ll}$
($\mu_R=M_{ll}/2$)
and the one obtained with the default choice $\mu_R=M_{ll}$ at LO (upper panel) and NLO (lower panel).
Regardless of the accuracy in the matrix element calculation, the running of $\alpha_{\overline{\rm MS}}$ and
$s_{W \,\overline{\rm MS}}^2$  is computed at $\mathcal{O}(\alpha)$ accuracy (solid and dotted lines)
or at $\mathcal{O}(\alpha)$ plus the higher-order corrections taken from Refs.~\cite{Erler:1998sy,Erler:2004in,Erler:2017knj}
(dashed and dot-dashed lines). In the plots, the finite jumps for $M_{ll}=M_W$ ($M_{ll}=2M_W$) are a consequence of the
discontinuity in the $\mathcal{O}(\alpha)$ running of the $\overline{\rm MS}$ parameters for $\mu_R=M_W$ at the denominator
(at the numerator for the choice $\mu_R=M_{ll}/2$). When the HO corrections to the running of
$\alpha_{\overline{\rm MS}}$ and $s_{W \,\overline{\rm MS}}^2$ are included, similar discontinuities appear
also per $M_{ll}=M_{\rm top}$ (and $M_{ll}=2M_{\rm top}$ if $\mu_R=M_{ll}/2$).
At the LO, scale-variation effects are of order $\pm 2\%$ and the size of the jumps related to the $W$
threshold in the running of the couplings is of about a couple of permille,
while the jumps originated by the top threshold in the dashed and dot-dashed lines are not visible
on the scale of the plot. In the NLO calculation, the renormalization-scale dependence
of $\alpha_{\overline{\rm MS}}$ and $s_{W \,\overline{\rm MS}}^2$ cancels against the one of the renormalization counterterms
in the one-loop amplitude and the residual scale dependence starts at $\mathcal{O}(\alpha^2)$.
As a consequence, on the one hand scale variation effects are strongly suppressed (compared to the ones in the upper panel)
and enter at the sub-permille level and, on the other hand, the jumps at the $W$ threshold are visibly reduced.
This does not happen for the discontinuities at the top threshold, since the matching corrections to the running
formulae are beyond $\mathcal{O}(\alpha)$. Though the HO corrections to the running of the $\overline{\rm MS}$ parameters
are not matched by the $\mathcal{O}(\alpha)$ virtual matrix elements, the size of the renormalization-scale dependence
shown by the dashed and dot-dashed  lines is close to the one in the solid and dotted plots where only the $\mathcal{O}(\alpha)$
running of $\alpha_{\overline{\rm MS}}$ and $s_{W \,\overline{\rm MS}}^2$ is used. It is thus reasonable to take the
numerical impact of the HO contribution to the running of the parameters (some $0.01\%$, as shown in Fig.~\ref{fig:homsbar})
as a rough estimate of the missing higher-order corrections to the matrix elements.

As a general remark, while the difference between theoretical predictions
obtained with different input parameter and renormalization schemes can be
considered as a rough and conservative estimate of the theoretical control
over predictions involving weak corrections, there might be motivations
to prefer one scheme to the others, like, for instance, the parametric
uncertainties connected with the knowledge of the input parameters,
the size of the perturbative corrections, the need of a specific free
parameter in the calculation.

In the following we address some of these additional sources of
theoretical uncertainty, in particular in Sect.~\ref{sec:parametric}
we focus on the main parametric uncertainties, in Sect.~\ref{sect:dahad} 
we discuss the treatment of the light quark contributions, and
in Sect.~\ref{sect:widths} 
we consider different available strategies for the
treatment of the unstable gauge bosons.

\section{Parametric uncertainties}
\label{sec:parametric}
We study in the following the parametric uncertainties
induced on $d\sigma / d{M_{ll}}$
and $A_{FB}(M_{ll})$ by the current experimental errors
affecting some of the relevant input parameters for each of the above
considered schemes. In particular, we treat the scheme
$\left(\alpha_0, G_\mu, M_Z\right)$ as free from parametric uncertainties
due to the input parameters because $\alpha_0$, $G_\mu$ and $M_Z$ are known
with excellent accuracy in high energy physics. 
Therefore, for the other two representative schemes,
$( G_\mu$, $M_W$, $M_Z)$ and
$(G_\mu$, $\sin^2 \theta^l_{eff}$, $M_Z)$, we study the
uncertainties induced by the imperfect knowledge of $M_W$ and
$\sin^2 \theta_{eff}^l$, respectively. 
\begin{figure}
  \includegraphics[width=0.48\textwidth]{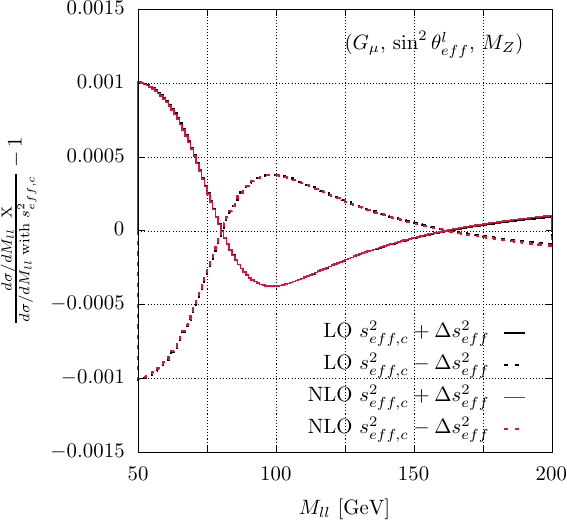}
  \caption{Effects of varying the input parameter
    $\sin^2 \theta^l_{eff} = 0.23154 \pm 0.00016$ in the
    $(G_\mu$, $\sin^2 \theta^l_{eff}$, $M_Z)$ scheme from the central value to
    the upper and lower ones, at leading and next-to-leading order. Here it
    is shown the relative difference between the invariant mass distribution
    obtained with the upper/lower value of $\sin^2 \theta^l_{eff}$ and the
    one with the central value $\sin^2 \theta^l_{eff, \, c} = 0.23154$.
    \label{fig:mv_s2effin}}
\end{figure}
\begin{figure}
  \includegraphics[width=0.48\textwidth]{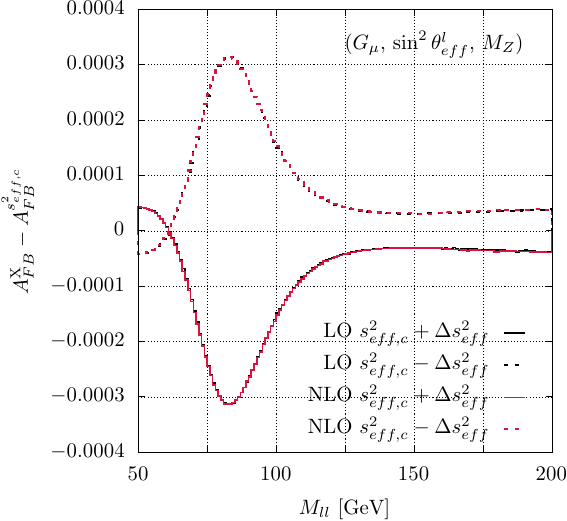}
  \caption{Effects of varying the input parameter
    $\sin^2 \theta^l_{eff} = 0.23154 \pm 0.00016$ in the
    $(G_\mu,\sin^2 \theta^l_{eff}, M_Z)$ scheme from the central value to
    the upper and lower ones, at leading and next-to-leading order.
    In particular it is shown the absolute difference between the asymmetry
    distribution obtained with the upper/lower value of
    $\sin^2 \theta^l_{eff}$ and the one with the central value
    $\sin^2 \theta^l_{eff, \, c} = 0.23154$. \label{fig:afb_s2effin}}
\end{figure}
\begin{figure}
  \includegraphics[width=0.48\textwidth]{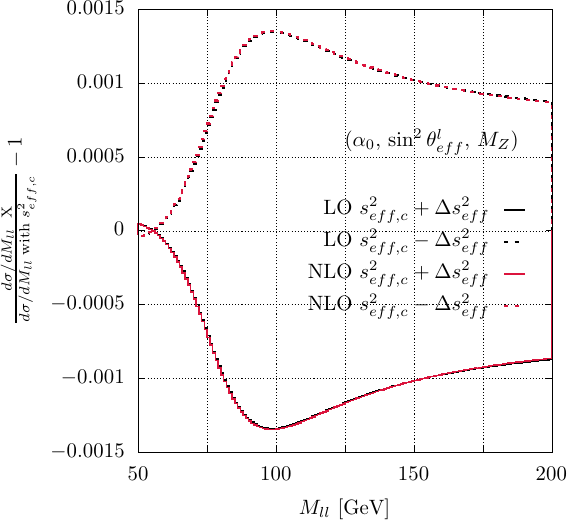}
  \caption{Effects induced on the dilepton invariant mass distribution by a
    variation of the input parameter
    $\sin^2 \theta^l_{eff} = 0.23154 \pm 0.00016$ in the
    $(\alpha_0, \sin^2 \theta^l_{eff}, M_Z)$ scheme from the central value
    to the upper and lower ones, at LO and NLO.
    Same notation and conventions of Fig.~\ref{fig:mv_s2effin}.
    \label{fig:mv_s2effinalpha0}}
\end{figure}
\begin{figure}
  \includegraphics[width=0.48\textwidth]{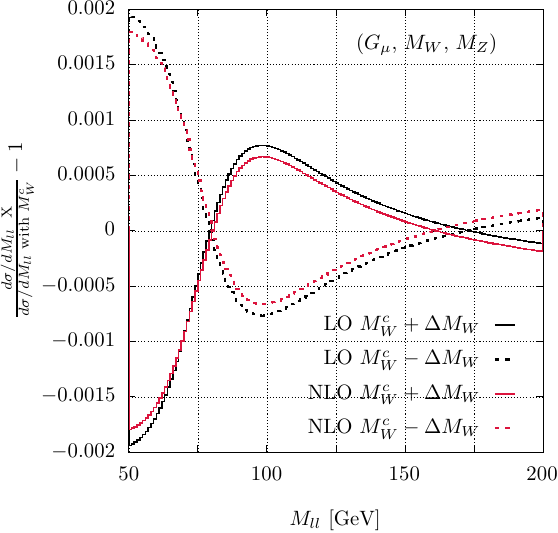}
  \caption{Effects of varying the input parameter
    $M_W = 80.385 \pm 0.015$~GeV in the $(G_\mu, M_W,M_Z)$ scheme from the
    central value to the upper and lower ones, at LO and NLO. Here it is shown the relative difference between the
    invariant mass distribution obtained with the upper/lower value of $M_W$
    and the one with the central value $M_W^c = 80.385$~GeV. \label{fig:mv_mw}}
\end{figure}
\begin{figure}
  \includegraphics[width=0.48\textwidth]{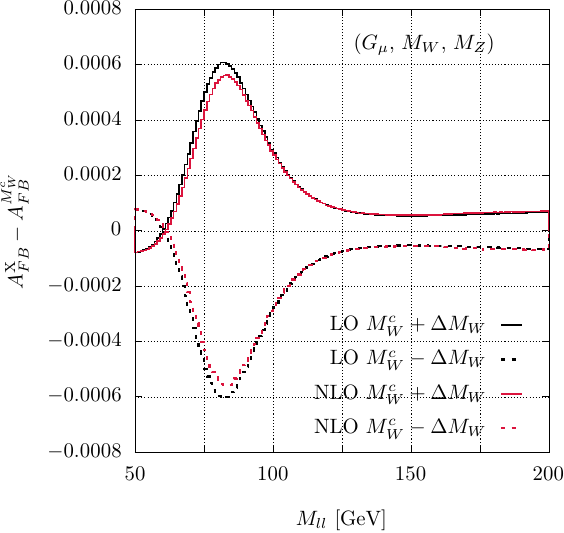}
  \caption{Effects of varying the input parameter
    $M_W = 80.385 \pm 0.015$~GeV in the $(G_\mu, M_W,M_Z)$ scheme from the
    central value to the upper and lower ones, at LO and NLO. In particular it is shown the absolute difference between the
    asymmetry distribution obtained with the upper/lower value of $M_W$ and
    the one with the central value $M_W^c = 80.385$~GeV. \label{fig:afb_mw}}
\end{figure}
Figure~\ref{fig:mv_s2effin} displays the  effect of a variation of
$\sin^2 \theta_{eff}^l$ within the range 
$0.23154 \pm 0.00016$~\cite{ALEPH:2005ab} on the dilepton invariant mass
distribution computed in the
$(G_\mu$, $\sin^2 \theta^l_{eff}$, $M_Z)$ scheme at LO and NLO
accuracy (black and red lines, respectively). 
In particular, the quantity
\begin{equation}\label{eq:delta_param1}  
  \delta^\pm =   \frac{d\sigma}{dM_{ll}^{s^2_{ eff, \, c} \pm \Delta
      s^2_{ eff}}} \Big/  \frac{d\sigma}{dM_{ll}^{s^2_{ eff, \, c}}}
  -1 \, ,
\end{equation}
where $s^2_{ eff, \, c}$ stands for the reference $\sin^2 \theta_{eff}^l$ value
($0.23154$) and $\Delta s^2_{ eff} = 0.00016$, is plotted as a function of
$M_{ll}$. Since the renormalization conditions in the
$(G_\mu$, $\sin^2 \theta^l_{eff}$, $M_Z)$ scheme  
require that $\sin^2 \theta_{eff}^l$ is not affected by radiative
corrections, the variations of $\sin^2 \theta_{eff}^l$ 
have basically the same impact at LO and at NLO. It is worth noticing that
the dependence of $d\sigma / d{M_{ll}}$ 
on $\sin^2 \theta_{eff}^l$ is twofold: on the one hand, it depends on the
$g_V/g_A$ ratio through the $Zf\overline{f}$ vertices 
and, on the other hand, there is an overall dependence coming from the
relation\footnote{When the complex-mass scheme is used, the real part of
  $M_Z^2$ is used.}
\begin{equation}\label{eq:agmufromsw2}  
  \alpha_{G_\mu} = \frac{\sqrt{2}}{\pi} G_\mu \sin^2 \theta_{eff}^l
  \left( 1 -\sin^2 \theta_{eff}^l \right) M_Z^2 \,.
\end{equation}
The latter effect is the source of the enhancement in the low-mass region of
Fig.~\ref{fig:mv_s2effin}, as can be understood 
by comparing the predictions in Fig.~\ref{fig:mv_s2effin} with the ones in
Fig.~\ref{fig:mv_s2effinalpha0}, obtained in the
$\left( \alpha_0, \sin^2 \theta_{eff}^l, M_Z \right)$ scheme. 
As a consequence, in order to assess the sensitivity of the dilepton
invariant mass distribution on the leptonic effective 
weak mixing angle (considered as a measure of the $g_V/g_A$ ratio),
one should consider the normalized 
$d\sigma / dM_{ll}$ distribution, rather than the absolute one. 

In Figure~\ref{fig:afb_s2effin} we plot  the quantity
\begin{equation}\label{eq:delta_param2}
  \Delta^\pm = A_{FB}^{ s^2_{eff, \, c} \pm \Delta s^2_{eff}}(M_{ll}) - A_{FB}^{s^2_{eff, \, c}}(M_{ll})\, ,
\end{equation}
showing the effects on $A_{FB}(M_{ll})$ induced by variations
of $\sin^2 \theta_{eff}^l$ within 
the same range considered in Fig.~\ref{fig:mv_s2effin}. Also in this case,
the dependence on $\sin^2 \theta_{eff}^l$ 
is basically the same at LO and at NLO accuracy. Quantitatively, it amounts
to approximately $\pm 3 \times 10^{-4}$ in the  
resonance region and drops quickly away from the $Z$ peak. As $A_{FB}$ is
defined through a ratio of differential distributions, 
the overall spurious dependence on $\sin^2 \theta_{eff}^l$ related to
Eq.~(\ref{eq:agmufromsw2}) cancels and the results
in Fig.~\ref{fig:afb_s2effin} show the sensitivity of the $A_{FB}$ on the
effective leptonic weak-mixing angle.

In Figures~\ref{fig:mv_mw} and~\ref{fig:afb_mw}, we focus on the parametric
uncertainty coming from
the value of the $W$-boson mass that affects the predictions obtained in the
$(G_\mu, M_W, M_Z)$ scheme.
In particular, we plot the quantities in Eqs.~(\ref{eq:delta_param1})
and~(\ref{eq:delta_param2})
where we replaced $s^2_{ eff, \, c}$ and $\Delta s^2_{eff}$ with the
reference $W$-mass value ($M_W^c=80.385$~GeV)
and its $1\sigma$ error ($\Delta M_W = \pm 15$ MeV~\footnote{This was
  the $1\sigma$ error of 
  Ref.~\cite{ParticleDataGroup:2016lqr}. The current
  Particle Data Group estimate of the uncertainty affecting
  the $M_W$ world average is of $12$~MeV~\cite{Workman:2022ynf},
  excluding the latest CDF measurement~\cite{CDF:2022hxs}.
  Slightly different values have been obtained in Ref.~\cite{Amoroso:2023pey}. 
  For our illustration purposes we can safely stick to
  $\Delta M_W = \pm 15$~MeV.}).
Figures~\ref{fig:mv_mw} (\ref{fig:afb_mw}) and~\ref{fig:mv_s2effin}
(\ref{fig:afb_s2effin}) are very similar.
This can be understood for instance at LO, where the variations of $M_W$ and
$\sin^2 \theta_{eff}^l$ are related by
\begin{equation}\label{eq:dsw2vsdmw2}
  \delta \sin^2 \theta_{eff}^l = - 2 \left( \frac{M_W}{M_Z^2}\right)
  \delta M_W\, ,
\end{equation}
from which we can see that a shift of $15$~MeV in $M_W$ corresponds to a 
shift of $-0.0003$ in $\sin^2 \theta_{eff}^l$, which is  approximately
twice the shift we are considering in Figs.~\ref{fig:mv_s2effin}
and~\ref{fig:afb_s2effin}. The plots show the same pattern also at NLO,
though the relation between
$\sin^2 \theta_{eff}^l$ and $M_W$ beyond LO is more involved (indeed,
at variance with the plots in Figs.~\ref{fig:mv_s2effin}
and~\ref{fig:afb_s2effin},
in Figs.~\ref{fig:mv_mw} and~\ref{fig:afb_mw} the NLO curves do not overlap
with the LO ones).
As in the case of Fig.~\ref{fig:mv_s2effin}, the source of the enhancement
in the low invariant mass
region of Fig.~\ref{fig:mv_mw} is the overall dependence on $M_W$ originating
from the relation
\begin{equation}\label{eq:agmufrommw2}  
  \alpha_{G_\mu} = \frac{\sqrt{2}}{\pi} G_\mu  M_W^2
  \Big( 1-\frac{M_W^2}{M_Z^2}\Big) \,,
\end{equation}
where the real part of the masses is taken in the complex-mass scheme.

\begin{figure}
  \includegraphics[width=0.48\textwidth]{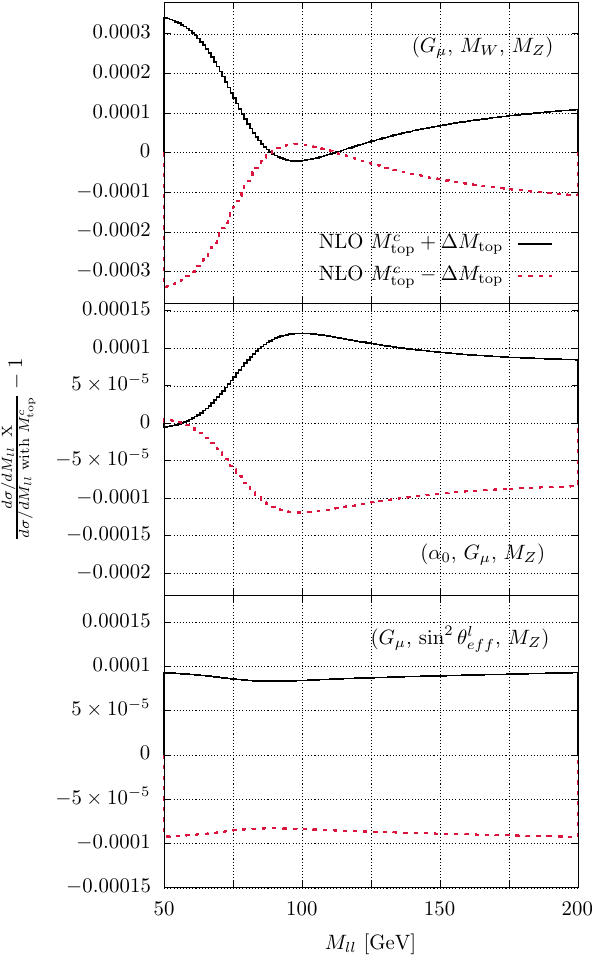}
  \caption{Effects of varying the top-quark mass $M_{\rm top}= 173.0 \pm 0.4$~GeV
    in the three considered schemes, from the central value to the upper and
    lower ones. The top-quark mass enters only the next-to-leading order
    corrections. Here it is shown the relative difference between the
    invariant mass distribution obtained with the upper/lower value of
    $M_{\rm top}$ and the one with the central value $M_{\rm top}^c = 173.0$~GeV.
    \label{fig:mv_mtop}}
\end{figure}
\begin{figure}
  \includegraphics[width=0.48\textwidth]{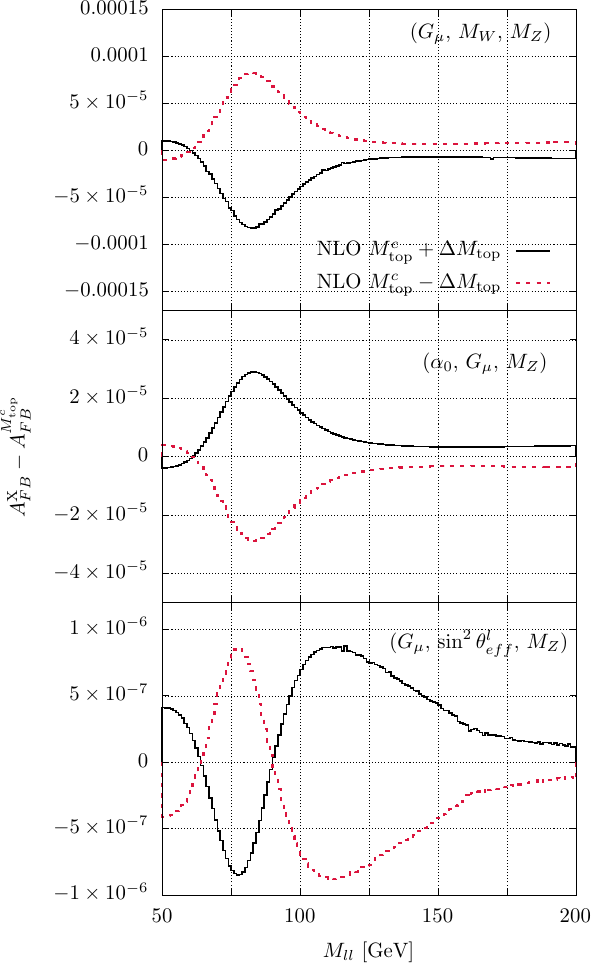}
  \caption{Effects of varying the top-quark mass $M_{\rm top}= 173.0 \pm 0.4$~GeV
    in the three considered schemes, from the central value to the upper and
    lower ones. The top-quark mass enters only the NLO corrections. Here it
    is shown the absolute difference between the asymmetry distribution
    obtained with the upper/lower value of $M_{\rm top}$ and the one with the
    central value $M_{\rm top}^c = 173.0$~GeV. \label{fig:afb_mtop}}
\end{figure}
Figs.~\ref{fig:mv_mtop} and~\ref{fig:afb_mtop} show the sensitivity
of $d\sigma / dM_{ll}$ and $A_{FB}(M_{ll})$ to
variations of $400$~MeV~\footnote{The value
  $M_{\rm top} = 173.0 \pm 0.4$~GeV, which we use in the present simulations,
  corresponds to the 2018 PDG average of
  Ref.~\cite{ParticleDataGroup:2018ovx}, which has been improved to
  $M_{\rm top} = 172.69 \pm 0.30$~GeV in
  Ref.~\cite{Workman:2022ynf}.} in the top-quark mass value,
for the three different input parameter schemes.
Since $M_{\rm top}$ enters parametrically only through
the loop diagrams, we display only the results obtained with the NLO
predictions.  The largest part of the top--quark mass
dependence at $\mathcal{O}(\alpha)$ can be encoded in the $\Delta \rho$ factor
defined in Sect.~\ref{sect:HO}. In the NLO predictions computed in the
$(G_\mu$, $M_W$, $M_Z)$ scheme, $\Delta \rho$ enters in two different ways:
through the overall factor $-2\Delta r$ ($\sim 2 c_W^2/s_W^2$)
and via the counterterm corresponding to $s_W^2$ ($\delta s_W^2 \sim c_W^2 \Delta \rho$).
The former contribution is responsible for the constant shift of about
$\pm 3 \times 10^{-4}$ for $\Delta M_{\rm top} = \pm 0.4$~GeV clearly visible at low
dilepton invariant masses, while the latter is the source of the shape effect
in the upper panel of Fig.~\ref{fig:mv_mtop}. In the $(\alpha_0$, $G_\mu$, $M_Z)$ scheme,
$\Delta \rho$ enters the $\mathcal{O}(\alpha)$ predictions only via the counterterms
$\delta G_\mu$ and $\delta s_W^2$: as a consequence, the $\gamma$-exchange amplitude is not affected
by top-mass variations, as clearly visible in the low dilepton invariant-mass region in the central
panel of Fig.~\ref{fig:mv_mtop}.
In the $(G_\mu$, $\seffl$, $M_Z)$ scheme, $\Delta \rho$ comes from the overall factor $-2 \Delta \tilde r$
($\sim 2 \Delta \rho$) and induces a constant shift approximately three times smaller than
the one coming from $\Delta r$ in the $(G_\mu$, $M_W$, $M_Z)$ scheme (given the different
coefficients multiplying $\Delta \rho$ in the two calculations, namely $c_W^2/s_W^2$ and $1$).
The two lines in the lower panel of Fig.~\ref{fig:mv_mtop} are not completely
flat because, besides the quadratic terms in $M_{\rm top}$ collected in $\Delta \rho$,
there is a residual subleading dependence on the top-quark mass which leads to a tiny
relative effect of order $10^{-6}$.
In the forward-backward asymmetry the overall contributions from $\Delta r$ and $\Delta \tilde r$
largely cancel. As a result, Fig.~\ref{fig:afb_mtop} shows the impact of the $\Delta \rho$ term
in $\delta s_W^2$ (which, in turn, is a shift of the effective $s_W^2$ entering the calculation)
for the $(G_\mu$, $M_W$, $M_Z)$ and $(\alpha_0$, $G_\mu$, $M_Z)$ schemes, while
for the $(G_\mu$, $\seffl$, $M_Z)$ scheme we only see the impact of the non-enhanced $M_{\rm top}$
corrections which is basically two orders of magnitude smaller than the effect observed for the other schemes.

\section{Treatment of $\Delta \alpha^{\rm had}$}
\label{sect:dahad}

The contributions to the running of $\alpha$ coming from the charged leptons and the top quark
can be computed perturbatively in terms of the corresponding contributions to the photon self-energy
and its derivative, namely:
\begin{equation}
  \Delta \alpha^{\rm lept \,(top)}(q^2)=
  -\frac{ {\rm Re} \Sigma_{AA}^{\rm lept \,(top)}(q^2)}{q^2}
  +\frac{ \partial \Sigma_{AA}^{\rm lept \,(top)}(q^2)}{\partial q^2}\Big|_{q^2=0}.
  \label{eq:dapertlt}
\end{equation}
For the light-quark contributions, on the contrary, Eq.~(\ref{eq:dapertlt}) cannot be used because of the
ambiguities related to the definition of the light-quark masses arising from non-perturbative QCD effects.
In the literature, a common strategy to compute the light-quark contribution to $\Delta \alpha$ is the
introduction of light fermion masses as effective parameters which are used to calculate the analogous of
Eq.~(\ref{eq:dapertlt}) for the quark sector:
\begin{equation}
  \Delta \alpha^{\rm had \,pert.}(q^2)=
  -\frac{ {\rm Re} \Sigma_{AA}^{\rm had}(q^2)}{q^2}
  +\frac{ \partial \Sigma_{AA}^{\rm had}(q^2)}{\partial q^2}\Big|_{q^2=0}.
  \label{eq:dapert}
\end{equation}
The light-quark masses are chosen is such a way that the resulting hadronic running of $\alpha$ from 0 to $M_Z^2$
corresponds to the one obtained from the experimental results for inclusive hadron production in $e^+e^-$ collisions
using dispersion relations ($\Delta \alpha^{\rm had \,fit}$), namely:
\begin{equation}
  \Delta \alpha^{\rm had \,pert.}(M_Z^2)=  \Delta \alpha^{\rm had \,fit}(M_Z^2).
  \label{eq:datuning}
\end{equation}
This approach is implemented in {\tt Z\_ew-BMNNPV} and it is used as a default. We stress that the light-quark
masses are only used for the self-energy corrections but they do not enter the vertex and box diagrams. In particular, 
they are not used for the QED corrections, where the light-quarks mass singularities are regularized by means of
dimensional regularization.

Starting from revision 4048, a more accurate treatment of the hadronic
vacuum polarization is available in {\tt Z\_ew\--BMNNPV}.
The code contains an interface to the routines of
Refs.~\cite{Jegerlehner:1985gq,Burkhardt:1989ky,Eidelman:1995ny,Jegerlehner:2003rx,Jegerlehner:2006ju,Jegerlehner:2008rs,Jegerlehner:2011mw,Jegerlehner:2019lxt}
and~\cite{Hagiwara:2003da,Hagiwara:2006jt,Hagiwara:2011af,Keshavarzi:2018mgv}
({\tt HADR5X19.F} and {\tt KNT v3.0.1}, respectively) 
for the calculation of the hadronic running of $\alpha$
based on the experimental data for inclusive $e^+e^-\to$ hadron production at low energies in terms of dispersion relations.
This interface can be activated using the input flag {\tt da\_had\_\-from\_fit=1} and the flag {\tt fit=1,2} can be used to
switch between the two routines for $\Delta \alpha^{\rm had \,fit}$. It is worth noticing that these routines only provide
results in the range $[0,q^2_{\rm max}]$: for larger values of $q^2$, we define 
\begin{eqnarray}
  \Delta \alpha^{\rm had \,fit}( q^2) &=&   \Delta \alpha^{\rm had \,fit}(q^2_{\rm max}) \nonumber \\
  & + & \Delta \alpha^{\rm had \,pert.}(q^2)-\Delta \alpha^{\rm had \,pert.}(q^2_{\rm max}).
  \label{eq:damatch}
\end{eqnarray}
The starting point for the calculation for {\tt da\_had\_from\_fit=1} is the relation
\begin{equation}
  \Delta \alpha^{\rm had \,fit}(q^2)=
  -\frac{ {\rm Re} \Sigma_{AA}^{\rm had}(q^2)}{q^2}
  +\frac{ \partial \Sigma_{AA}^{\rm had\, fit}(q^2)}{\partial q^2}\Big|_{q^2=0}.
  \label{eq:dafit}
\end{equation}
While Eq.~(\ref{eq:dapert}) is a definition of $\Delta \alpha^{\rm had \,pert.}$,
Eq.~(\ref{eq:dafit}) can be considered as a definition of
$\frac{ \partial \Sigma_{AA}^{\rm had\, fit}(q^2)}{\partial q^2}\Big|_{q^2=0}$.
On the one hand, Eq.~(\ref{eq:dafit}) is used in the one-loop corrections to the
photon propagator to replace the combination $\Sigma_{AA}^{\rm had}(s)-s\delta Z_A^{\rm had}$
with $-s \Delta \alpha^{\rm had \,fit}(q^2)+is{\rm Im}\Sigma_{AA}^{\rm had}(s)$
(where the factor $\delta Z_A^{\rm had}=$ $-\frac{ \partial \Sigma_{AA}^{\rm had}(q^2)}{\partial q^2}\Big|_{q^2=0}$
is the light-quark contribution to the photon wave function renormalization counterterm)
and, on the other hand, $\frac{ \partial \Sigma_{AA}^{\rm had\, fit}(q^2)}{\partial q^2}\Big|_{q^2=0}$
is used for the counterterms related to the electric charge and the photon wave function.
More precisely, since the self-energy in Eq.~(\ref{eq:dafit}) can be computed perturbatively
for $q^2$ much larger than $\Lambda^2_{\rm QCD}$,
we take $q^2=M^2_Z$ and tune the quark masses using Eq.~(\ref{eq:datuning}). 
This way, the formal expression of the counterterms is the same as the one used in the default
computation ({\tt da\_had\_from\_fit=0}). The calculations for {\tt da\_had\_from\_fit} set to~0 and~1
are not equivalent: in fact, even though both of them rely on the tuning of the light-quark masses
from Eq.~(\ref{eq:datuning}), the corrections to the photon propagator are different since 
$\Delta \alpha^{\rm had \,fit}(s) \neq  \Delta \alpha^{\rm had \,pert.}(s)$ for $s\neq M_Z^2$.

We notice that the electric-charge and wave function counterterms could also be defined from 
Eq.~(\ref{eq:dafit}) setting the light-quark masses to zero in the photon self-energy diagrams: 
on the one hand, this would lead to differences of order $\alpha \left( \frac{m_q^2}{M_Z^2} \right){\rm log} \left( \frac{m_q^2}{M_Z^2}\right)$
and, on the other hand, setting the light-quark masses to zero would require several modifications
to the routines used for the evaluation of the virtual one-loop corrections.

The impact of the improved treatment of the hadronic running of $\alpha$ is only visible for the
input parameter schemes that use $\alpha_0$ as an independent parameter, since for the other schemes
the terms that depend logarithmically on the light-quark masses cancel.
Fig.~\ref{fig:mv_dalpha1} shows the dependence of
$d\sigma / dM_{ll}$ on the uncertainty $\delta \Delta \alpha^{\rm had \, fit}$
for each of the two adopted parameterizations, with the
$(\alpha_0, M_W, M_Z)$ scheme. The effect of changing
$\Delta \alpha^{\rm had}$ from its central value by a shift of
$\pm \delta \Delta \alpha^{\rm had}$ is at the level of $\pm 0.022$\% for 
{\tt HADR5X19.F} and $\pm 0.027$\% for {\tt KNT v3.0.1}, respectively.
 Variations of  $\Delta \alpha^{\rm had}$ mainly affect the $\delta Z_e$
counterterm, as the light-quark mass logarithms in $\delta Z_e$ have
been traded for $\Delta \alpha^{\rm had}$, leading to an almost constant
shift  in Fig.~\ref{fig:mv_dalpha1}. Changing $\Delta \alpha^{\rm had}$
also affects the NLO corrected $\gamma$ propagator, but the numerical
impact is tiny as the bare self-energy diagrams do not involve logarithmically
enhanced light-quark mass terms: this effect, being only present
for the $\gamma$-mediated amplitude, induces a small shape effect in Fig.~\ref{fig:mv_dalpha1}.
In $A_{FB}$ the contribution from $\delta Z_e$ largely cancels, leading to an absolute
change in $A_{FB}$ at the $10^{-6}$ level as shown in Fig.~\ref{fig:afb_dalpha1}.

\begin{figure}
  \includegraphics[width=0.48\textwidth]{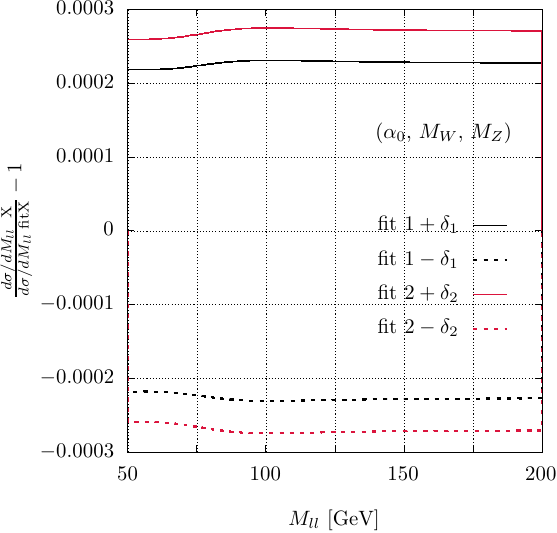}
  \caption{Change in the cross section as a function of the invariant mass if
    one takes the central value for $\Delta \alpha^{\rm had \, fit}$ (that
    here coincides with zero) plus its error
    $\delta \Delta \alpha^{\rm had \, fit}$, here labelled $\delta_1$ for
    results obtained with the flag {\tt fit=1}~\cite{Jegerlehner:1985gq,Burkhardt:1989ky,Eidelman:1995ny,Jegerlehner:2003rx,Jegerlehner:2006ju,Jegerlehner:2008rs,Jegerlehner:2011mw,Jegerlehner:2019lxt}, and $\delta_2$ for results with
    the flag {\tt fit=2}~\cite{Hagiwara:2003da,Hagiwara:2006jt,Hagiwara:2011af,Keshavarzi:2018mgv}. The input scheme used here is $(\alpha_0, M_W, M_Z)$.
    \label{fig:mv_dalpha1} \label{fig:mv_dalpha1}}
\end{figure}

\begin{figure}
  \includegraphics[width=0.48\textwidth]{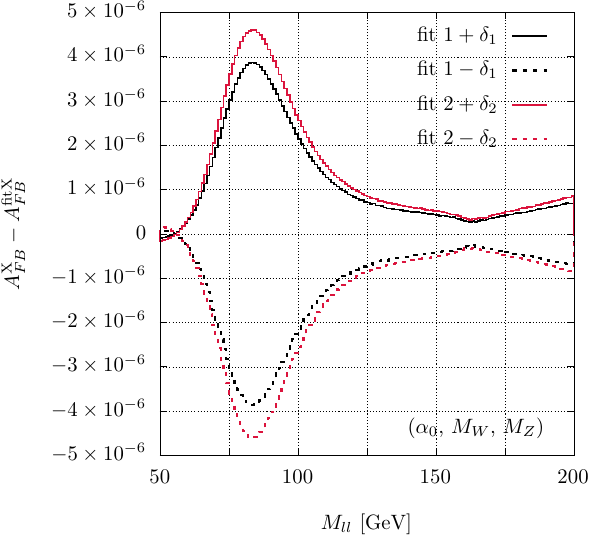}
  \caption{Change in the asymmetry distribution if one takes the central
    value for $\Delta \alpha^{\rm had \, fit}$ plus its uncertainty $\delta \Delta
    \alpha^{\rm had \, fit}$, as in Fig.~\ref{fig:mv_dalpha1}.
    \label{fig:afb_dalpha1}}
\end{figure}

\section{Treatment of the $Z$ width}
\label{sect:widths}
The unstable nature of the $Z$ vector boson is considered 
by default through the complex-mass scheme~\cite{Denner:1999gp,Denner:2005fg,Denner:2006ic}, according to which the squared vector boson masses
are taken as complex quantities
$\mu^2_{V} = M^2_{V} - i \Gamma_{V} M_{V}$, with $(V = W,Z)$, in the
LO and NLO calculation. The input values for $M_V$ and $\Gamma_V$
are assumed to be the on-shell ones, $M_V^{\rm OS}$ and $\Gamma_V^{\rm OS}$, 
and are converted internally in the initialization phase to
the corresponding pole values using the
relations~\cite{Bardin:1988xt,Beenakker:1996kn}
\begin{equation}
  M_V=\frac{M_V^{\rm OS}}{\sqrt{1+\Big(\frac{\Gamma_V^{\rm OS}}{M_V^{\rm OS}}\Big)^2}}, \quad
  \Gamma_V=\frac{\Gamma_V^{\rm OS}}{\sqrt{1+\Big(\frac{\Gamma_V^{\rm OS}}{M_V^{\rm OS}}\Big)^2}}.
  \label{eq:ostopole}
\end{equation}
The pole parameters $M_V$ and $\Gamma_V$ are used throughout the code for the
matrix element calculations.
In the CMS, the couplings that are functions of the gauge-boson masses become necessarily complex quantities.
In particular, in the schemes with $M_W$ and $M_Z$ as input parameters,
the quantities $s_W^2$ and $\delta s_W / s_W$ of Eq.~(\ref{eq:swos})
and Eq.~(\ref{eq:swctos}) are calculated in terms of $\mu_W$ and $\mu_Z$.
Since $\sin^2 \theta_{eff}^l$ is defined through the real part
of the $g_V/g_A$ ratio, it is considered as a real quantity when used as a free parameter.
Similarly, the input parameters $\alpha_0$, $\alpha(M_Z^2)$, and $G_\mu$ are real.
As a consequence, in the input parameter schemes with only one vector boson mass, $M_Z$,
the input couplings are taken as real quantities.

When performing calculations in the input parameter/ renormalization schemes having $G_\mu$ among the free parameters
(with the only exception of the $(\alpha_0$, $G_\mu$, $M_Z)$ one), $G_\mu$ is usually traded for $\alpha_{G_\mu}$
by means of Eqs.~(\ref{eq:agmufromsw2}) or~(\ref{eq:agmufrommw2}). Since the gauge-boson
masses enter in this relations, $\alpha_{G_\mu}$ might in principle acquire an imaginary part if the CMS scheme is employed.
In the code, we follow the standard procedure of taking a real-valued $\alpha_{G_\mu}$ to minimize the
spurious higher-order terms associated with the overall factor $[{\rm Im (\alpha_{G_\mu}}) ]^2$.
More precisely, in Eqs.~(\ref{eq:agmufromsw2}) and~(\ref{eq:agmufrommw2}), we always use the real part
of the gauge-boson masses\footnote{There is some arbitrariness in how to get a real value for $\alpha_{G_\mu}$,
see for instance the discussion in Ref.~\cite{Frederix:2018nkq}.}.

The CMS preserves gauge invariance order by order
in perturbation theory and this feature guarantees also that the
higher-order unitarity violations are not artificially enhanced. For this
reason it is the scheme commonly adopted for multi-particle NLO calculations. 
However, for neutral-current Drell-Yan it is possible to adopt other strategies
for the treatment of the $Z$ resonance. In particular, in the
{\tt Z\_ew-BMNNPV} package, we implemented the so-called \emph{pole} and \emph{factorization}
schemes following Secs.~3.3.ii and~3.3.iii of Ref.~\cite{Dittmaier:2009cr}, respectively
\footnote{In the {\tt W\_ew-BMNNP} package, besides the CMS scheme, the CLA scheme discussed in Ref.~\cite{Dittmaier:2001ay}
is implemented for the treatment of the unstable $W$ boson in charged-current Drell-Yan.}.
These schemes can be switched on by means of the flags {\tt PS\_scheme 1} and
{\tt FS\_scheme 1}, respectively.

\begin{figure}
  \includegraphics[width=0.48\textwidth]{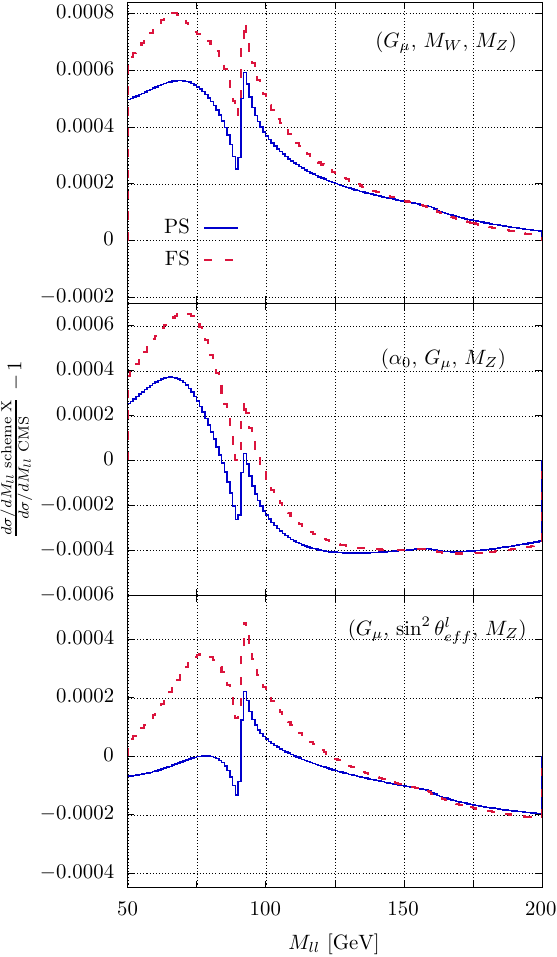}
  \caption{Difference of the pole/factorization scheme with respect to the
    default complex-mass scheme in the dilepton invariant mass cross
    section distribution at NLO. The difference PS-CMS is shown by the solid blue curve,
    the FS-CMS one by the dashed red one. Upper panel:
    $(G_\mu, M_W, M_Z)$; middle panel: $(\alpha_0, G_\mu, M_Z)$; lower panel:
    $(G_\mu, \sin^2 \theta_{eff}^l, M_Z)$. \label{fig:mv_3p_res}}
\end{figure}

In Figs.~\ref{fig:mv_3p_res} we show the relative difference between
the pole (factorization) scheme, blue (red) line, w.r.t. the CMS, for
$d\sigma / dM_{ll}$, considering different input parameter
schemes. A feature common to all
schemes is the oscillation of few 0.01\% of amplitude around the
$Z$ resonance, as already shown separately for $u {\bar u}$
and $d {\bar d}$ partonic initial states in Ref.~\cite{Dittmaier:2009cr}.
Over the whole range $50$~GeV $ < M_{ll} < 200$~GeV, the shapes of
the differences are similar for the three considered input parameter
schemes, with larger differences appearing with the $(G_\mu, \mw, M_Z)$
and $(\alpha_0, G_\mu, M_Z)$ schemes. The structure at the $WW$
threshold present in Fig.~6 of Ref.~\cite{Dittmaier:2009cr} is not
visible, within the available statistical error, because of a partial
cancellation between the contributions of up- and down-type quark
channels.

\begin{figure}
  \includegraphics[width=0.48\textwidth]{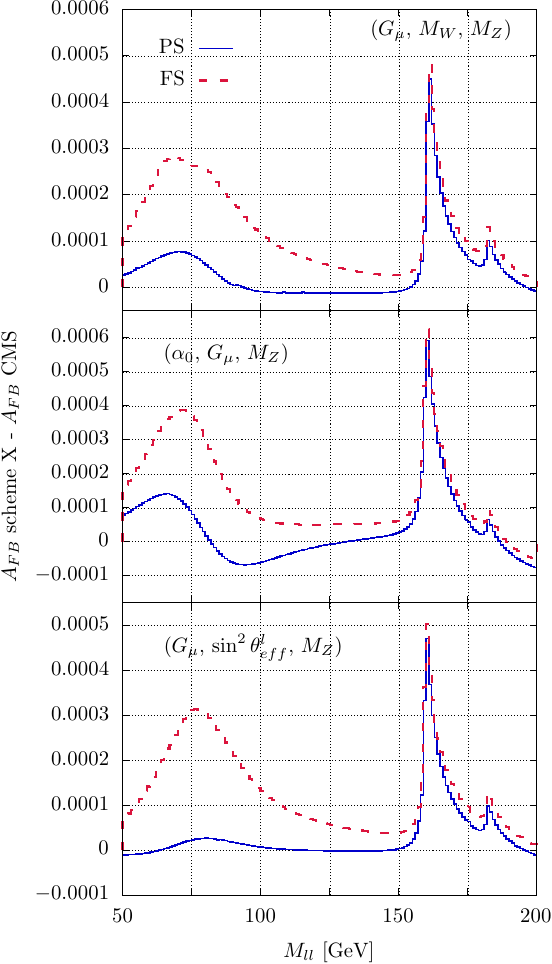}
  \caption{Difference of the pole/factorization scheme with respect to the
    default complex-mass scheme in the forward-backward asymmetry
    distribution at NLO. The difference PS-CMS is shown by the solid blue
    curve, while the FS-CMS one by the dashed red one. Upper
    panel: $(G_\mu, M_W, M_Z)$; middle panel: $(\alpha_0, G_\mu, M_Z)$; lower
    panel: $(G_\mu, \sin^2 \theta_{eff}^l, M_Z)$. \label{fig:afb_3p_res}}
\end{figure}

The same comparison between pole (factorization) scheme and CMS is shown
for $A_{FB}$ as a function of $M_{ll}$ in Fig.~\ref{fig:afb_3p_res}.
In this case we plot the absolute difference instead of the relative
difference with respect to the CMS. For $A_{FB}$ the difference is smooth
around the $Z$ resonance, being of the order of few $10^{-5}$ for the pole
scheme and of the order of $10^{-4}$ for the factorization scheme. 
Contrary to the $d\sigma/dM_{ll}$ case, in Fig~\ref{fig:afb_3p_res}
the $WW$ threshold enhancement, of the order of $5 \times 10^{-4}$, is clearly visible. 

\section{High energy regime}
\label{sec:high-energy}
\begin{figure}
  \includegraphics[width=0.48\textwidth]{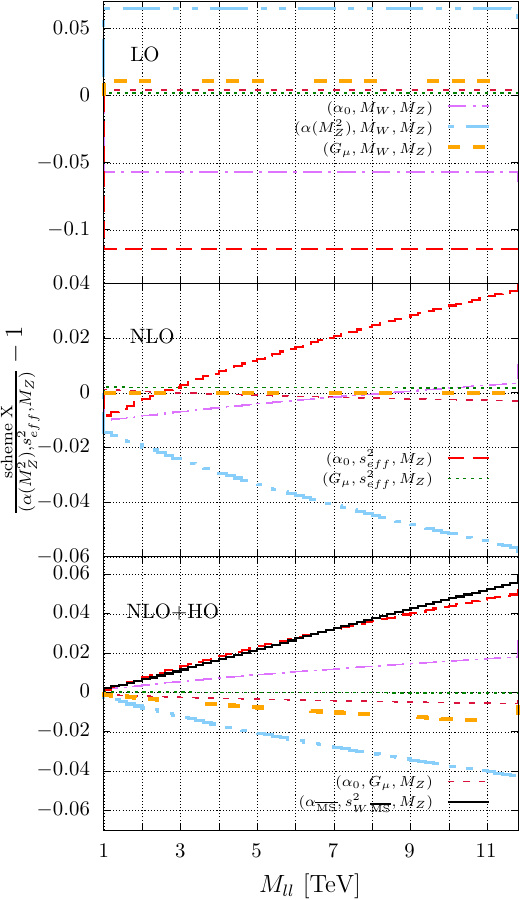}
  \caption{Relative difference of the predictions for the dilepton invariant
    mass cross section distribution at LO (upper panel), NLO (middle panel),
    NLO+HO (lower panel), in the range $1-12$~TeV. The calculation is
    performed with the same inputs of Fig.~\ref{fig:mll_ratio_schemes}.
    \label{fig:mll_highmass}}
\end{figure}
\begin{figure*}
  \includegraphics[width=0.98\textwidth]{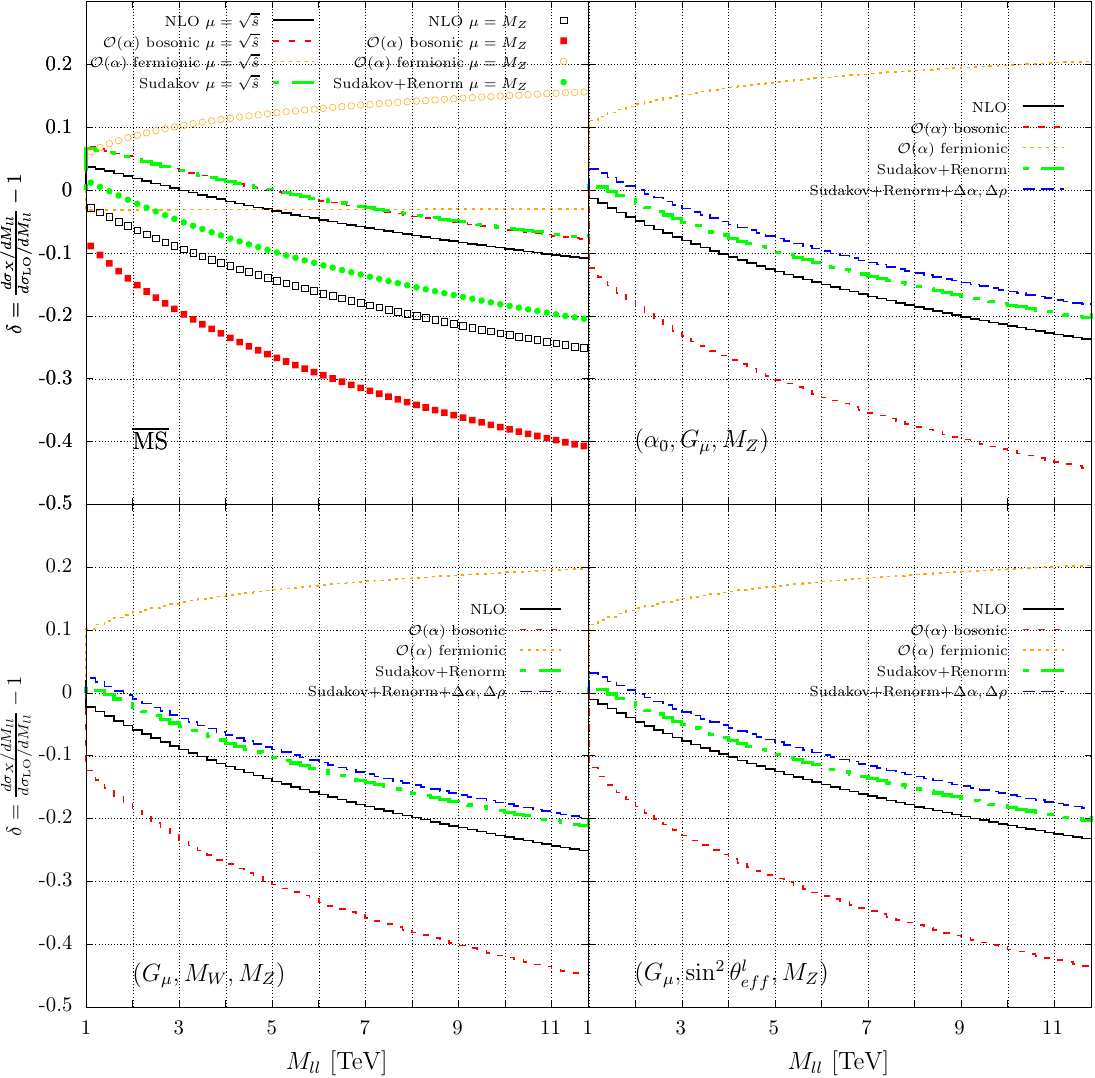}
  \caption{Comparison among the predictions for the dilepton invariant mass
    cross section distribution, obtained by including the full NLO weak
    corrections, the fermionic-only and bosonic-only corrections, or the
    approximation which includes the Sudakov logarithms, the
    parameter-renormalization logs and the leading fermionic corrections
    stemming from $\Delta \alpha$ and $\Delta \rho$. Four schemes are shown:
    the $\overline{\rm MS}$ in both its running- and fixed-scale
    realizations, the latter one with $\mu=\mz$ (top left),
    $(\alpha_0, G_\mu, \mz)$ (top right), $(G_\mu, \mw,\mz)$ (bottom left)
    and $(G_\mu, \seffl,\mz)$ (bottom right). \label{fig:sudakov4}}
\end{figure*}
The analysis presented so far has been focused on the physics at the $Z$
peak. 
To complete this study, we examine now the behaviour of the corrections and
the interplay among different renormalization and input schemes in the
high-energy regime, 
that can be relevant also in view of the upcoming programme of the LHC 
and at future high-energy machines.

In this section, we present the results for a specific initial-state quark
flavour focusing,
for brevity, on $d$-quarks: in this way PDFs contributions exactly cancel
when studying the relative
effect of weak corrections with respect to the LO, as well as in the ratio of
predictions
obtained for different schemes. The main motivation of our choice is the fact
that, at
high dilepton invariant masses, PDFs are poorly constrained and typically
affected by large errors.
If one considers the contribution of all quark flavors at the same time, the
above-mentioned
relative corrections and ratios have a residual dependence on PDFs which
tends to induce,
for high $M_{ll}$ values, quite large unphysical distortions.
The actual size of this effect clearly depends on the specific PDF set used.

In Fig.~\ref{fig:mll_highmass} we repeat the study in
Fig.~\ref{fig:mll_ratio_schemes}
for dilepton invariant masses in the range between 1 and 12~TeV. With respect
to the peak region,
we find a quite different behaviour. First of all, the dilepton invariant
mass ratios at LO
are flat. In the upper panel of Fig.~\ref{fig:mll_ratio_schemes}, the shapes
came from the variations in
the values of $s_W^2$ entering the $g_{Zf\overline{f}}$ couplings in the
$Z$-boson exchange diagram and
the $M_{ll}$ dependence was originated by the different propagators
of the $\gamma$ and the $Z$
as well as by PDFs weighting the different quark flavours: in
Fig.~\ref{fig:mll_highmass} not only
we consider only $d$-quarks, but also the $Z$-boson propagator is effectively
$1/s$ (since $s \gg M_Z^2$)
so that the only $s$ dependence in the differential cross section is the
overall flux factor
(see for instance Eq.~(2.12) of~\cite{Dittmaier:2009cr} with $\chi_Z=1$.)

Moving to the NLO results, the level of agreement between the different input
parameter/renormalization schemes is
of the order of 1\% at the left edge of the plot, but it gets worse as
the partonic center of mass energy increases,
with a 10\% spread at 12~TeV. The inclusion of the fermionic higher-order
effects discussed in the previous sections
improves the picture only around 1~TeV, but it does not reabsorb the
differences among the predictions in
the considered schemes at large $M_{ll}$.

The behaviour shown in Fig.~\ref{fig:mll_highmass} can be understood as
follows. In the considered
dilepton invariant mass range, the bosonic part of the weak NLO corrections
is dominated by the
so-called Sudakov logarithms, which are double and single logarithms  of
kinematic invariants
over the gauge-boson masses. These logs correspond to the infrared limit of
the weak corrections,
where the gauge-boson masses are small compared to the energy scales
involved  and act
as (physical) cutoff for the soft and/or collinear virtual weak
corrections~\cite{Sudakov:1954sw,Beccaria:1998qe,Ciafaloni:1998xg,Beccaria:1999fk,Denner:2000jv,Denner:2001gw,Pozzorini:2001rs,Denner:2003wi,Denner:2004iz,Denner:2006jr,Denner:2008yn,Fadin:1999bq,Ciafaloni:1999ub,Ciafaloni:2000df,Ciafaloni:2000rp,Ciafaloni:2000gm,Ciafaloni:2001vt,Ciafaloni:2001vu,Chiu:2007yn,Chiu:2007dg,Chiu:2008vv,Chiu:2009mg,Becher:2013zua,Manohar:2014vxa,Bauer:2017bnh}.
Besides the Sudakov corrections, there is another class of logarithmic
corrections coming
from parameter renormalization. When using dimensional regularization,
counterterms contain
logarithms of the unphysical mass-dimension scale $\mu_{\rm Dim}$ in the combination 
\be
\frac{1}{\epsilon}-\log\frac{r_{ct}^2}{\mu_{\rm Dim}^2} \, ,
\ee
where $\epsilon=(4-{\rm D})/2$, $D$ being the number of space-time dimensions, and
$r_{ct}$ is related to particle
masses in on-shell based schemes or directly to the renormalization scale
$\mu_R$ in the ${\overline{\rm MS}}$ scheme.
This contribution cancels against similar terms
appearing in the bare loop diagrams (where $r_{ct}$ will be some other
scale, say $r_{bare}$)
leaving contribution of the form $\log (r_{bare}^2 / r_{ct}^2)$. In the
Drell-Yan parameter-renormalization
counterterms only vertex diagrams enter, so the only possible scale
(in particular in the limit
of vanishing gauge-boson masses) is $M_{ll}$. 
The functional form of both the Sudakov and the parameter-renormalization
logarithms is the same in any scheme,
but the coefficients multiplying the logarithms (including the LO-like
amplitude where they appear and the LO
amplitude in interference with it) differ numerically as they are function
of the actual $\alpha$ and $s_W^2$
values used. As a result, the logarithms appearing in the numerator and
denominator in the NLO ratios of Fig.~\ref{fig:mll_highmass}
have different coefficients and they do not cancel, leaving logarithmically
enhanced remnants.
It is worth emphasising that the inclusion of the fermionic higher-order
corrections, by definition
has no impact on the Sudakov corrections (as they are bosonic), but they are
also irrelevant
for the parameter-renormalization logs, as the corrections in
Sect.~\ref{sect:HO} do include an effective running of the
parameters, but only up to the weak scale.

In order to prove the argument above, we implemented a private version of the
{\tt Z\_ew-BMMNPV} including the routines used
in~\cite{Chiesa:2013yma,Mangano:2016jyj} 
for the evaluation of the Sudakov corrections in
{\tt ALPGEN}~\cite{Mangano:2002ea}\footnote{For more
recent implementations of Sudakov logarithms in other frameworks, see
Refs.~\cite{Bothmann:2020sxm,Bothmann:2020sxm,Pagani:2021vyk,Pagani:2023wgc,Lindert:2023fcu}.}.
Though it is true that the Sudakov corrections alone are not a good
approximation for the full NLO weak corrections
to neutral-current Drell-Yan (as pointed-out, for instance,
in~\cite{Denner:2019vbn}),
this is mainly due to the large cancellations between fermionic and bosonic
corrections (as shown in Fig.~\ref{fig:sudakov4})
and to a large UV contribution from parameter-renormalization logarithms.
For the on-shell renormalization based schemes, like the
$(\alpha_0, G_\mu, M_Z)$, $(G_\mu, \sin^2 \theta_{eff}^l, M_Z)$, and
$(G_\mu, M_W, M_Z)$ shown in the plot, the fermionic $\mathcal{O}(\alpha)$
corrections are of the order of $10-20\%$ and mainly come from
the fermionic loops entering parameter renormalization. The sum of the
Sudakov corrections and the logarithms from parameter renormalization
is a reasonable approximation of the NLO corrections, reproducing their
shape with an essentially constant shift of about $5-6\%$.
Similar consideration apply to the $\overline{\rm MS}$ scheme, if fixed
renormalization scale is used ($\mu_R=M_Z$ in Fig.~\ref{fig:sudakov4}).
The picture changes considerably if the $\overline{\rm MS}$ scheme is used
with the renormalization scale set to the dilepton invariant mass:
in this way, the corrections related to parameter renormalization are
reabsorbed in the running couplings and the remaining corrections
are smaller than in the other schemes (the fermionic ones, in particular,
boil-down to an almost flat $-4\%$ effect).
The figure also singles-out the part of the fermionic contributions coming
from the universal enhanced terms
$\Delta \alpha$ and $\Delta \rho$.

\begin{figure}
  \includegraphics[width=0.48\textwidth]{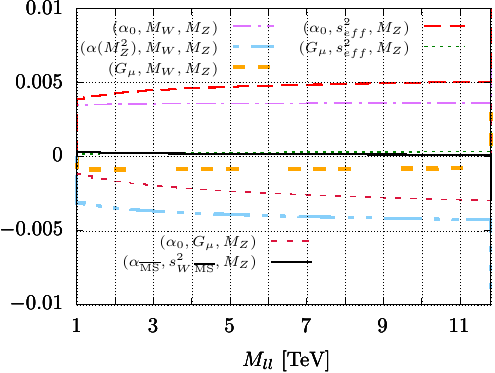}
  \caption{\label{fig:highmassNOlog} Relative difference of the predictions
    for the dilepton invariant mass cross section distribution at NLO+HO,
    after the subtraction of the leading logarithmic corrections (Sudakov
    plus parameter-renormalization logs).}
\end{figure}

To conclude, in Fig.~\ref{fig:highmassNOlog} we repeat the same study of the
lower  panel of Fig.~\ref{fig:mll_highmass} once the leading logarithmic
corrections (Sudakov plus parameter-renormalization logs) have been
subtracted from the NLO+HO predictions. Despite the approximations in the
calculation 
of the logarithmic corrections, it is still possible to read a clear trend
in Fig.~\ref{fig:highmassNOlog}: the ratios fall in the few per mille
range (as in the case of Fig.~\ref{fig:mll_ratio_schemes} for the
near-resonance region) and they tend to be flat\footnote{In the plot, 
  the $\overline{\rm MS}$ scheme is used for a fixed renormalization scale
  equal to $M_Z$, since the running of $\alpha_{\overline{\rm MS}}$ and
  $s_{W \,\overline{\rm MS}}^2$
  would have included part of the logarithms that have been subtracted from
  the denominator in the ratio.}.

As thoroughly discussed in the literature, when electro\-weak Sudakov
logarithms start to become dominant a resummation algorithm needs to
be adopted, in order to obtain reliable predictions.
A very recent discussion on automated resummation algorithms of Sudakov
logarithms in simulation tools can be found in Ref.~\cite{Denner:2024yut}. 
This issue is left to future investigations for the case of DY processes
with {\tt POWHEG-BOX}.

\section{Conclusions}
\label{sec:conclusions}
The precision physics program of the LHC requires flexible
and precise simulation tools to be used for different
purposes. The NC DY process, thanks to its large cross section
and clean signature, plays a particular role in this context
and its NLO electroweak corrections are a mandatory ingredient for
every kind of analysis. In the present paper we have addressed the
issue of the input parameter/renormalization schemes in the gauge sector 
for the electroweak corrections to NC DY. 
In particular, we have provided the relevant expressions for
the counterterms at NLO precision in various realizations of 
on-shell renormalization schemes as well as of an hybrid
$\overline{ \rm MS}$/on-shell renormalization scheme.  
Among the on-shell schemes, we considered explicitly
the following combinations: $(\alpha_0/\alpha(M_Z^2)/G_\mu, M_W, M_Z)$,
$(\alpha_0, G_\mu, M_Z)$ and 
$(\alpha_0/\alpha(M_Z^2)/G_\mu, \sin^2\theta_{eff}^{l}, M_Z)$.
The hybrid scheme we considered, 
containing the input quantities 
$(\alpha_{\overline{\rm MS}}(\mu^2)$,
$s_{W \,\overline{\rm MS}}^2(\mu^2)$, $M_Z)$, 
is of interest for possible future direct determinations of the running 
of the electroweak couplings at high energies. 
In addition to the NLO expressions of the counterterms, we provided, for
each considered scheme, the expressions for the higher-order universal
corrections due to $\Delta \alpha$ and $\Delta \rho$. All the relevant
expressions are presented in a self-contained way, so that they could 
be easily adopted in any simulation tool. For the present phenomenlogical
study, all the discussed input parameter schemes have been implemented
in the code {\tt Z\_ew-BMNNPV} of the {\tt POWHEG-BOX-V2} framework,
which has been used to obtain illustrative
phenomenological results on the differential distributions
$d\sigma / dM_{\mu^+\mu^-}$ and $A_{FB}(M_{\mu^+\mu^-})$, with inclusive acceptance of the leptons. The main features of the various input parameter
schemes have been quantitatively analysed for the two considered observables,
with focus on the invariant mass window which includes the $Z$ peak and
on the high energy region. The latter is characterized by the presence
of Sudakov logarithms, whose impact has been analysed in detail with
a comparison among different schemes. 
In addition to the effects of the electroweak 
corrections, we illustrated the parametric uncertainties on the two considered
observables, associated with the different schemes.
For the schemes with $\alpha_0$ as input parameter, we included the
possibility to calculate $\Delta \alpha$ by means of two different 
parameterizations based on dispersion relations using $e^+ e^-$
collider data. A section is devoted to the discussion of the
improved treatment of the unstable $Z$ boson with respect to the original
version of the code. While the full complex-mass scheme is the new default,
also the pole and the factorization schemes are available as options in the
code. The numerical impact of the different width options on the
considered observables are shown for three representative input
parameter schemes.

\appendix
\section{Input parameters}
\label{appendix:numerical-param}
We list below the values of the EW parameters used for the phenomenological
results presented in the paper:
\bea
  & G_\mu                                           =&   1.1663787 \times 10^{-5}\,{\rm GeV}^2    ,\qquad        \nonumber \\
  & \alpha(M_Z^2)                                   =& 1/128.95072067974729  ,\nonumber \\
  & \alpha_0                                        =&   1/137.0359909956391     ,\qquad                  M_H         = 125        \,{\rm GeV},\nonumber \\
  & M_Z^{\rm OS}                                    =&    91.1876 \,{\rm GeV} ,\qquad                \Gamma_Z^{\rm OS}= 2.4952 \,{\rm GeV} ,\nonumber  \\
  & M_W^{\rm OS}                                    =&    80.385 \,{\rm GeV} ,\qquad                 \Gamma_W^{\rm OS}= 2.085 \,{\rm GeV},  \nonumber \\
  & \sin^2 \theta_{eff}^l                        =&   0.23154 ,\qquad      \nonumber \\
  & s_{W \,\overline{\rm MS}}^2(M_Z^2)              =&   0.23122, \qquad       \nonumber \\
  \label{eq:paramsew}
  \eea
  The numerical values of
  $s_{W \,\overline{\rm MS}}^2$ and
  $\sin^2 \theta_{eff}^l$ correspond to the
  estimates of Ref.~\cite{ParticleDataGroup:2018ovx}. 
The numerical values of the fermionic masses are given below
\begin{eqnarray}
  & m_e             = &     0.51099907 \,{\rm MeV} ,\qquad     m_\mu  =  105.6583 \,{\rm MeV} ,  \nonumber  \\
  & m_\tau          = &  1.7770500  \,{\rm GeV} ,\qquad    \nonumber  \\
  & m_u             = & 69.165474707499480  \, {\rm MeV} ,\qquad   \nonumber \\
  & m_d             = &  69.84 \,{\rm MeV} ,\qquad  \nonumber  \\
  & m_c             = &  1.2    \,{\rm GeV} ,\qquad    m_s  =  150 \,{\rm MeV}   ,  \nonumber  \\
  & M_{\rm top}     = &  173    \,{\rm GeV} ,\qquad    m_b  =  4.7 \,{\rm GeV} . 
  \label{eq:paramsmasses}
\end{eqnarray}
The light-quark masses are used for the calculation of $\Delta \alpha^{\rm had}$, as detailed
in section~\ref{sect:dahad}. 
Depending on the input parameter scheme adopted, we use the values in Eq.~(\ref{eq:paramsew}) only for the independent
parameters, while the other ones are either derived or not used at all. The only exception is $\alpha_0$, which is always
used as input parameter for the calculation of the one-loop QED corrections
(not discussed in the present paper) and enters the loop factor
$\alpha_0/(4\pi)$. 
While the choice of $\alpha_0$ for the photon--fermion coupling is motivated by the physical scale  of the $\gamma f \bar{f}$ splitting and by the
required cancellation of the infrared divergences between virtual
and real contributions, the natural choice of $\alpha$ entering the 
weak loop factor $\alpha / (4\pi)$ is given by $\alpha$ of the
input parameter scheme at hand. However, we leave to the user the
freedom of using the loop factor $\alpha_0 / (4\pi)$ also in
the pure weak corrections by setting to 0 the flag
{\tt a2a0-for-QED-only}.
We stress that the different choices of $\alpha$ for the weak
loop factor introduce differences at $\mathcal{O}(\alpha^2)$.

For the parton distribution functions (PDFs), we use the {\tt NNPDF31\_nlo\_as\_0118\_luxqed}
set~\cite{Manohar:2016nzj,Manohar:2017eqh,Bertone:2017bme} provided by the {\tt LHAPDF-6.2} framework~\cite{Buckley:2014ana}
and set the factorization scale to the invariant mass of the dilepton system.

\section{Input flags}
\label{appendix:flags}

In the following, we briefly describe the input-parameter flags for the {\tt Z\_ew-BMNNPV} package
that have been used to produce the results shown in the main text. These are only a subset of the
available input flags and we refer to the user manual for the complete list of process-specific
input options. The non process-specific input flags can be found in the {\tt POWHEG-BOX-V2} documentation.

\subsubsection*{Options for EW corrections}

{\bf no\_ew}: it is possible to switch off electroweak correction by setting
{\tt no\_ew 1} in the {\tt powheg.input} file. By default {\tt no\_ew}$=0$.

{\noindent
{\bf no\_strong}: allows to switch off the QCD corrections when set to 1.
By default {\tt no\_strong}$=0$.}

{\noindent
{\bf ew\_ho}: the fermionic higher-order corrections
discussed in section~\ref{sect:HO} are included by using the flag 
{\tt ew\_ho 1}. By default {\tt ew\_ho=0}.}

{\noindent
{\bf includer3qcd  }: if set to 1, the expression for $\Delta r$ used in the higher-order
corrections includes the three-loop QCD corrections. Default 0. }

{\noindent
{\bf includer3qcdew}: when equal to 1, the three-loop mixed EW-QCD effects are included in the formula for 
$\Delta r$ used for the fermionic higher-order corrections. Default 0.}

{\noindent
{\bf includer3ew   }: same as the previous flag, but for the three-loop EW corrections. Default 0.}

{\noindent    
    {\bf dalpha\_lep\_2loop  }: if set to 1, $\Delta \alpha$ includes the two-loop leptonic corrections
    from Ref.~\cite{Steinhauser:1998rq} when computing higher-order corrections.
Default 0.}

{\noindent
{\bf QED-only}: for the NC DY, the EW corrections can be split in QED and
pure weak corrections in a gauge invariant way. By setting {\tt QED-only 1}
in the input card, only pure QED corrections are computed. }

{\noindent
{\bf weak-only}: when set to 1, only the virtual weak part of the EW corrections is computed.}

Note that events can be generated without QCD corrections only at LO accuracy
or at LO plus weak (potentially higher-order) corrections: the generation of events
including NLO QED corrections but not the NLO QCD ones is not allowed.

\subsubsection*{Options for unstable particles regularization}

{\bf complexmasses}: if set to 1, the complex-mass scheme of Refs.~\cite{Denner:1999gp,Denner:2005fg,Denner:2006ic} is used.
Default 1. 

{\noindent
{\bf FS\_scheme}: when set to 1, the factorization scheme described in Ref.~\cite{Dittmaier:2009cr} is employed.
Default 0.}

{\noindent
{\bf PS\_scheme}: the calculation is performed in the pole scheme of Ref.~\cite{Dittmaier:2009cr} if {\tt PS\_scheme 1}.
Default 0.}

\subsubsection*{Options for EW input parameter schemes}

{\bf scheme}: this is the main flag for the choice of the EW input scheme.
All the available schemes use the on-shell $Z$ mass $M_Z^{\rm OS}$
({\tt Zmass}) as independent parameter
(internally converted to the corresponding pole value), but differ in the
choice of the remaining two independent parameters. Default 0. 
\begin{itemize}
\item{{\tt scheme 0}: the second EW input parameter is $\alpha_0$.}
\item{{\tt scheme 1}: $\alpha(M_Z^2)$ is taken as free parameter.}
\item{{\tt scheme 2}: $G_{\mu}$ is used as input parameter.}
\item{{\tt scheme 3}: in this scheme, the actual input parameter is $\alpha_0$. However, for each
  phase-space point, the matrix elements are evaluated using the on-shell running of $\alpha$  from $q^2=0$
  to the partonic center of mass of the event (computed in terms of the Born-like momenta {\tt kn\_pborn}).
  The additional factor of $\alpha$ coming from real and virtual QED corrections is always set to $\alpha_0$.
  The loop factor from the virtual weak corrections is set by default to $\alpha_0$,
  but $\alpha(q^2)$ is employed when the flag
  {\tt a2a0-for-QED-only} is equal to~$1$ in the input card. The running is performed by default at NLO accuracy,
  and contains the two-loop leptonic corrections when the flag {dalpha\_lep\_2loop} is set to~$1$.}
\item{{\tt scheme 4}: the $(\alpha_0$, $G_\mu$, $M_Z)$ scheme of Sect.~\ref{sect:agmumz} is used.}
\item{{\tt scheme 5}: the calculation is performed in the $\overline{\rm MS}$ scheme of Sect.~\ref{sect:swms}
  (see below for input flags specific to this scheme choice).}

\end{itemize}
  
When {\tt scheme} is equal to $0$, $1$, $2$, or $3$, the third EW input parameter is 
by default the on-shell $W$ mass $M_W^{\rm OS}$ ({\tt Wmass}) (internally converted
to the corresponding pole value). If the flag {\tt use-s2effin} is different from zero, the third
independent parameter is the effective weak-mixing angle as described in Sect.~\ref{sect:amzsw}.

{\noindent
{\bf use-s2effin}: should be set to the desired value of the effective weak-mixing angle $\seffl$. 
If this flag is present and {\tt scheme} is equal to  $0$, $1$, $2$, and $3$, the
calculation is performed in the $(\alpha_0$, $\seffl$, $M_Z)$,
$(\alpha(M_Z^2)$, $\seffl$, $M_Z)$, $(G_\mu$, $\seffl$, $M_Z)$,
and $(\alpha(q^2)$, $\seffl$, $M_Z)$ schemes, respectively.
This flag is not compatible with the options {\tt scheme}$=4,\,5$.}

{\noindent
{\bf a2a0-for-QED-only}: regardless of the scheme used, the additional loop factor
coming from the virtual weak corrections is set by default to $\alpha_0$. If the flag
is set to 1, the purely weak loop factor is set to the same value used for the LO matrix
element in the selected scheme. Note that the additional factor of $\alpha$ from real
and virtual QED corrections is always equal to $\alpha_0$.}

Besides $M_Z^{\rm OS}$, also the on-shell Z width ({\tt Zwidth}) is a free parameter of
the calculation (internally converted to the corresponding pole value).
The same holds for the on-shell $W$ width  ({\tt Wwidth}), when $M_W^{\rm OS}$ is taken as an EW input parameter,
though this is only relevant when the complex-mass scheme is used.

\subsubsection*{Options for the hadronic running of $\alpha$ and light-quark masses}

{\bf da\_had\_from\_fit}: if set to 1, the calculation of the hadronic
corrections to the photon propagator and its derivative is
based on the experimental data for inclusive $e^+e^-\to$ hadron production
at low energies in terms of dispersion relations. Default 0.

{\noindent
  {\bf fit}: if {\tt da\_had\_from\_fit}=$1$ and {\tt fit}$=1(2)$, the {\tt HADR5X19.F} ({\tt KNT v3.0.1}) routine
  is used for the calculation of the hadronic vacuum polarization. The option {\tt fit}=$0$ is left for cross checks,
  as with this option the quark loops under consideration are computed as in the case {\tt da\_had\_from\_fit}=$0$.
}

{\noindent
{\bf mq\_only\_phot}: if set to 1, the light-quark masses are set to 0 in the $W$, $Z$, and mixed $\gamma Z$
self-energy corrections and in their derivatives, since their light-quark mass
dependence is regular and tiny in the massless quark limit. Default 0. 

The treatment of the hadronic vacuum polarization is critical for the derivative of the photon propagator
and thus for the electric-charge and photon wave-function counterterms in the on-shell scheme. As a consequence,
it is critical for those schemes that use $\alpha_0$ as input parameter, while the impact is minor
when $\alpha(M_Z^2)$, $G_\mu$, or $\alpha(q^2)$ are used.

In the context of the $\overline{\rm MS}$ calculation ({\tt scheme}$=5$), the
electric-charge and photon wave-function counterterms do not depend on the light-quark masses,
while in the hadronic corrections to the bare photon propagator the light-quark mass
dependence is regular and tiny in the limit $m_q \to 0$: as a result, in this scheme
the {\tt da\_had\_from\_fit} flag is not needed. The non perturbative effects in the
$\overline{\rm MS}$ running of $\alpha_{\overline{\rm MS}}$ (and, indirectly, of
$s_{W \,\overline{\rm MS}}^2$) 
are included in the starting values of the evolution $\alpha_{\overline{\rm MS}}(\mu_0^2)$
and $s_{W \,\overline{\rm MS}}^2(\mu_0^2)$, that should correspond to a scale
$\mu_0^2$ sufficiently larger than $4 m_b^2$.

\subsubsection*{Options for the $\overline{\rm MS}$ scheme}

The flags below are only effective when {\tt scheme}$=5$.

{\noindent
{\bf running\_muR\_sw}:
if set to 1, the calculation is performed in the $\overline{\rm MS}$ scheme
with dynamical renormalization scale. For each phase-space point,
the $\overline{\rm MS}$ scale is set to the dilepton invariant mass in the
Born-like kinematic ({\tt kn\_pborn} momenta) and the couplings $\alpha_{\overline{\rm MS}}$ 
and $s_{W \,\overline{\rm MS}}^2$ are evolved accordingly~\cite{Erler:1998sy,Erler:2004in,Erler:2017knj}. 
Default 0.}

{\noindent
  {\bf MSbarmu02}: if {\tt running\_muR\_sw}$=1$, this parameter is the starting
  scale of the evolution of $\alpha_{\overline{\rm MS}}$ 
  and $s_{W \,\overline{\rm MS}}^2$. Otherwise, this is the (constant)
  value of the $\overline{\rm MS}$ renormalization scale. It should be sufficiently larger than $4 m_b^2$.
  By default it is the pole $Z$ mass computed internally from the input parameter $M_Z^{\rm OS}$.
}

{\noindent
  {\bf MSbar\_alpha\_mu02}: is the value of $\alpha_{\overline{\rm MS}}(\mu_0^2)$ for {\tt MSbarmu02} $=\mu_0^2$.
  {\tt MSbar\_alpha\_mu02} has a default value only if {\tt MSbar\-mu02} is not present in the input card
  and corresponds to $\alpha_{\overline{\rm MS}}(M_Z^2)$ computed as a function of $\alpha_0$ according
  to eq. (10.10) of Ref.~\cite{ParticleDataGroup:2020ssz} which includes effects up to
  $\mathcal{O}(\alpha \alpha_S^2)$ (see also Ref.~\cite{Chetyrkin:1996cf}).
  If {\tt excludeHOrun}$=1$, the NLO relation $\alpha_0$ and $\alpha_{\overline{\rm MS}}(M_Z^2)$ is used.

{\noindent
  {\bf MSbar\_sw2\_mu02}: same as {\tt MSbar\_alpha\_mu02}, but for $s_{W \,\overline{\rm MS}}^2$. The default value is
  $s_{W \,\overline{\rm MS}}^2(M_Z^2)=0.23122$.
}

{\noindent
{\bf decouplemtOFF}: if set to 1, switches off the top-quark decoupling. Default 0.}

{\noindent
{\bf decouplemwOFF}: same as {\tt decouplemtOFF}, but for the $W$ decoupling.}

{\noindent
  {\bf MW\_insw2\_thr}: allows to tune the position of the $W$ threshold in
  the $\overline{\rm MS}$ running of $\alpha$ end $s_W^2$ and the corresponding argument of the decoupling logarithms.
  If absent, this parameter is computed as $M_W^{2,\ \rm th}={\rm Re} [ M_Z^2 (1-s_{W \,\overline{\rm MS}}^2(\mu_0^2)) ]$.}

{\noindent
  {\bf OFFthreshcorrs}: when set to 1, the threshold corrections in the $\overline{\rm MS}$ running of $\alpha$ end $s_W^2$
are switched off. Default 0.}

{\noindent
  {\bf excludeHOrun}: if set to 1, the $\overline{\rm MS}$ running of $\alpha$ end $s_W^2$
  is performed at NLO accuracy. Default 0 (i.e. the higher-order effects in
  Refs.~\cite{Erler:1998sy,Erler:2004in,Erler:2017knj} are included).}

{\noindent
{\bf OFFas\_aMS}: when set to 1, the running of $\alpha_{\overline{\rm MS}}$ and 
$s_{W \,\overline{\rm MS}}^2$ does not include the corrections $\mathcal{O}(\alpha \alpha_S)$
and $\mathcal{O}(\alpha \alpha_S^2)$. Default 0.}

{\noindent
{\bf OFFas2\_aMS}: if equal to 1, the running of $\alpha_{\overline{\rm MS}}$ and 
$s_{W \,\overline{\rm MS}}^2$ does not include the corrections $\mathcal{O}(\alpha \alpha_S^2)$
. Default 0.}

{\noindent
{\bf ewmur\_fact}: this entry sets the factor by which the renormalization scale is multiplied. It can be used for studying scale variations, e.g. by setting it to the standard values $2$ or $1/2$. Default 1.}

\subsubsection*{Remaining EW parameters}    

{\bf alphaem   }: $\alpha_0$. Default value: $1/137.0359909956391$. When {\tt scheme}$=0$,
it is used for both the couplings in the LO matrix element and in the NLO corrections.
For other schemes, it is used for the extra power if $\alpha$ in the real and virtual QED
corrections (and in the loop factor for the virtual weak corrections if {\tt a2a0-for-QED-only}
is not set to 1).

\noindent
{\bf alphaem\_z}: $\alpha(M_Z^2)$. Default value: $1/128.95072067974729$. It is only used if {\tt scheme}$=1$.
See above ({\tt alphaem}) for the additional power of $\alpha$ in the NLO corrections.

\noindent
{\bf gmu       }: $G_\mu$. Default value: $1.1663787 \times 10^{-5}$ GeV$^{-2}$. It is only used if {\tt scheme}$=2$.
See above ({\tt alphaem}) for the additional power of $\alpha$ in the NLO corrections.

\noindent
    {\bf azinscheme4 }: if it is positive, the electromagnetic coupling of the $(\alpha_0, G_\mu, \mz)$ scheme
    is set to $\alpha=\alpha_0/(1-\Delta \alpha))$ in the evaluation of the matrix elements (and in the $\alpha/4\pi$ loop factor if {\tt a2a0-for-QED-only}$=1$). Default 0. It is only active when {\tt scheme}=$4$.

\noindent
    {\bf Zmass     }: on-shell $Z$-boson mass $M_Z^{\rm OS}$. Default value: $91.1876$~GeV.
    Internally converted to the corresponding pole value.

\noindent
    {\bf Zwidth    }: on-shell $Z$-boson width $\Gamma_Z^{\rm OS}$. Default value: $2.4952$~GeV.
    Internally converted to the corresponding pole value.    
    
\noindent
    {\bf Wmass     }: on-shell $W$-boson mass $M_W^{\rm OS}$. Default value: $80.385$~GeV.
    Internally converted to the corresponding pole value. This parameter is only used
    if the flag {\tt use-s2effin} is absent when if {\tt scheme}$=0,1,2,3$. Otherwise, it
    is computed from the independent EW parameters.
    
\noindent
{\bf Wwidth    }: on-shell $W$-boson width $\Gamma_W^{\rm OS}$. Default value: $2.085$~GeV.
Internally converted to the corresponding pole value. This parameter is only relevant when
{\tt complexmasses}$=1$ if {\tt scheme} is set $0,1,2,3$ while {\tt use-s2effin} is absent.

\noindent    
{\bf Hmass     }: Higgs-boson mass, only entering weak corrections. Default value: $125$~GeV.

\noindent    
{\bf Tmass     }: top-quark mass. Default value: $173$~GeV.

\noindent    
    {\bf Elmass    }: electron mass. Default value: $0.51099907 $~MeV. Since the calculation is
    performed for massive final-state leptons, this is the parameter used in the phase-space generator
    when running the code for the process $pp\to e^-e^+$.

\noindent    
{\bf Mumass    }: muon mass. Default value: $0.1056583$~GeV. 
    This parameter is also used in the phase-space generator
    when running the code for the process $pp\to \mu^-\mu^+$.

\noindent        
{\bf Taumass   }: tau mass. Default value: $1.777050$~GeV. 
    This parameter is also used in the phase-space generator
    when running the code for the process $pp\to \tau^-\tau^+$.

\noindent    
{\bf Umass     }: up-quark mass. Default value: $0.069165474707499480$~GeV.

\noindent        
{\bf Dmass     }: down-quark mass. Default value: $0.06984$~GeV.

\noindent        
{\bf Cmass     }: charm-quark mass. Default value: $1.2$~GeV.

\noindent        
{\bf Smass     }: strange-quark mass. Default value: $0.15$~GeV.

\noindent    
{\bf Bmass     }: b-quark mass. Default value: $4.7$~GeV.

For a description of the role of light-quark masses in the calculation,
we refer to Sect.~\ref{sect:dahad} and to the flags for the hadronic running
of $\alpha$ and light-quark masses.

\begin{acknowledgements}
  We are grateful to all our colleagues of the LHC Electroweak Working Group,
  Drell-Yan physics and EW precision measurements subgroup,
  for the continuous discussions and pleasant atmosphere. 

  The work of M.C. is partially supported by the Italian Ministero dell'Universit\`a e Ricerca (MUR) and European Union - Next Generation EU through the research grant number 20229KEFAM ``High precision phenomenology at the LHC: combining strong and electroweak corrections in all-order resummation and in Monte Carlo event generators'' under the program PRIN2022. 
\end{acknowledgements}

\bibliographystyle{caps} 
\interlinepenalty=10000
\bibliography{powhegzew_new}   

\end{document}